\documentclass[11pt]{article}
\usepackage[utf8]{inputenc}
\pdfoutput=1

\usepackage{amsmath, amsthm, amssymb, geometry, setspace, graphicx}
\usepackage{graphics}
\usepackage{enumerate}
\usepackage{natbib}
\usepackage{color, multirow, booktabs, bigstrut, makecell, enumitem, caption}
\usepackage{bm}
\usepackage{authblk}

\usepackage{newtxtext}
\usepackage[subscriptcorrection]{newtxmath}
\usepackage{hyperref}
\usepackage[ruled]{algorithm2e}
\def\thickhline{\noalign{\hrule height.8pt}}

\newcommand{\ind}{\overset{ind}{\sim}}
\newcommand{\iid}{\overset{iid}{\sim}}
\newcommand{\LOR}{\textsc{LOR}}

\theoremstyle{plain}

\newtheorem{corollary}{Corollary}
\newtheorem{proposition}{Proposition}

\theoremstyle{definition}

\theoremstyle{remark}

\newgeometry{left=2.8cm,right=2.8cm,bottom=1.5cm,top=1.0cm}
\title{Leveraging External Data for Testing Experimental Therapies with Biomarker Interactions in Randomized Clinical Trials}
\author[1]{Boyu Ren}
\author[2]{Federico Ferrari}
\author[3]{Sandra Fortini}
\author[4]{Steffen Ventz}
\author[5,6]{Lorenzo Trippa\thanks{ltrippa@jimmy.harvard.edu}}
\affil[1]{Laboratory for Psychiatric Biostatistics, McLean Hospital}
\affil[2]{Biostatistics and Research Decision Sciences, Merck \& Co}
\affil[3]{Department of Decision Sciences, Bocconi University}
\affil[4]{Division of Biostatistics, University of Minnesota}
\affil[5]{Department of Data Science, Dana-Farber Cancer Institute}
\affil[6]{Department of Biostatistics, Harvard T.H. Chan School of Public Health}
\date{}

\begin{document}
	
	\maketitle
	
	\begin{abstract}
In oncology the efficacy of novel therapeutics often differs across patient subgroups, and these variations are difficult to predict during the initial phases of the drug development process. The relation between the power of randomized clinical trials and heterogeneous treatment effects has been discussed by several authors. In particular, false negative results are likely to occur when the treatment effects concentrate in a subpopulation but the study design did not account for potential heterogeneous treatment effects. The use of external data from completed clinical studies and electronic health records has the potential to improve decision-making throughout the development of new therapeutics, from early-stage trials to registration. Here we discuss the use of external data to evaluate experimental treatments with potential heterogeneous treatment effects. We introduce a permutation procedure to test, at the completion of a randomized clinical trial, the null hypothesis that the experimental therapy does not improve the primary outcomes in any subpopulation. The permutation test leverages the available external data to increase power. Also, the procedure controls the false positive rate at the desired $\alpha$-level without restrictive assumptions on the external data, for example, in scenarios with unmeasured confounders, different pre-treatment patient profiles in the trial population compared to the external data, and other discrepancies between the trial and the external data. We illustrate that the permutation test is optimal according to an interpretable criteria and discuss examples based on asymptotic results and simulations, followed by a retrospective analysis of individual patient-level data from a collection of glioblastoma clinical trials.
\end{abstract}
	
	\newpage

\section{Introduction}
During the last decade the characterization of oncogenic alterations and resistance mechanisms have been the basis of a rapid increase in the experimental treatments available for clinical testing in cancer research \citep{haslam2021updated}. Novel therapeutics often target patient subpopulations defined by somatic mutations or other biomarkers, and their efficacy can vary across patient subgroups. In the early phases of clinical investigation it is usually unclear and difficult to predict if the experimental therapy improves relevant outcomes, for example survival, and in which subpopulations. Dedicated trial designs \citep{freidlin2010randomized,freidlin2012randomized,ziegler2012personalized} have been proposed to estimate subgroup-specific treatment effects and to improve decision-making, including go/no-go decisions \citep{chu2018blast} and eligibility of registration studies \citep{xu2023machine} accounting for potential heterogeneous treatment effects across subpopulations. Also,  statistical methods for inference on treatment effects and their variations across subgroups have  been studied extensively in the literature. We mention the use of treatment effects pattern plots \citep{bonetti2004patterns}, permutation-based algorithms \citep{wang2015detecting},  randomization-based procedures \citep{ding2016randomization,ding2019decomposing}, Bayesian model averaging to account for multiplicity issues \citep{berger2014bayesian}, and tree-based approaches \citep{wager2018estimation}, among others. Moreover, particular efforts have been made to identify patient subgroups that benefit from the experimental treatment \citep{rigdon2018preventing, wager2018estimation}. For example, \cite{morita2017bayesian} discussed a decision-theoretic solution to identify subgroups, and \cite{rigdon2018preventing} used regression trees to capture heterogeneous treatment effects.

A fundamental requirement for investigating heterogeneous treatment effects in randomized clinical trials (RCTs) is the inclusion of more subjects compared to RCTs that ignores heterogeneous treatment effects \citep{yang2020sample}. However, the sample sizes of most oncology RCTs continue to be chosen for assessing the  average treatment effects and are thus inadequate for accurate inference on interactions between treatments and biomarkers. Additionally, in many settings, such as rare diseases, it is not practical or feasible, due to limited resources, to conduct large RCTs that capture treatment effect's variations across subgroups \citep{nugent2021heterogeneity}. These limitations impact major decisions during the drug development process, including the decision to discontinue the development of the experimental treatment and the choice of the eligibility criteria for registration trials based on previous data from early phase studies. 

To mitigate the outlined limitations, we propose a novel method for the analysis of RCTs with  potential heterogeneous treatment effects,  that integrates external data (ED), including individual patient-level information from completed RCTs or real-world datasets \citep{sherman2016real}, such as electronic health records collected for administrative purposes. This method can improve and accelerate the development of new therapeutics, by leveraging diverse data sources to facilitate the transition from early-phase trials to confirmatory studies,  and supporting the critical decision to continue or terminate the investigation of the experimental treatment. The integration of ED in the design and analysis of clinical trials has received substantial attention in oncology. In particular, recent contributions on the use of external control data  in oncology suggest the potential of accelerating the development of new treatments \citep{rahman2021leveraging,liau2023association}.  Moreover, these contributions might   facilitate  the use of unbalanced randomization ratios (e.g., 1:2 or 1:3 for  the control and experimental arms) because the ED, as well as the control arm of the trial, provide information on the control therapy.

We  discuss a  permutation  procedure   that incorporates a {\it Bayesian working model} and  augments the data from a RCT with ED. 
We focus on a single task: assessing if the RCT data provide evidence of positive treatment effects in some of the patients based on hypothesis testing. Our test controls the false positive rate at the desired $\alpha$-level.  Importantly, the control of false positives does not require any assumption on the ED; it covers scenarios with model misspecification, unmeasured confounders, different pre-treatment patient profiles in the trial population compared to the ED, and other discrepancies between the trial and the ED.
 Our permutation procedure is based on test statistics with straightforward Bayesian interpretations,  including popular summaries  of the evidence of treatment effects. 
 We will provide a decision-theoretic \citep{berger2013statistical} justification of the permutation procedure, illustrating 
 the optimality of the test according to an easy-to-interpret criteria. 
 Through  stylized examples 
 and a data-driven simulation study, we illustrate the  properties of our approach, emphasizing its robustness, the  control of false positive results, and potential  power improvements   compared to popular testing procedures the utilize only the RCT data. 
 
\section{Method}
\subsection{Notation} 

We consider a RCT that randomizes $n$ patients with ratio $1:r$ to the experimental and control arms. We indicate the RCT data ({\it internal data}, ID) with $\mathcal D = (Y, X, A)$, where $Y = (Y_1,\ldots,Y_n)$ 
are the outcomes, $X = (X_1,\ldots,X_n) $ are pre-treatment patient characteristics $(X_i \in \mathbb R^d)$, and $A = (A_1,\ldots, A_n)$ are treatment assignment variables ($A_i = 1$ and $A_i=0$ for the experimental and control treatments). Similarly, $\mathcal D_E = (Y_E, X_E, A_E)$ indicates the ED with $n_E$ patients. To simplify the presentation $X_{E,i}$ will include the same variables as $X_i$, although this assumption can be straightforwardly relaxed. In some cases $\mathcal D_E$ includes only patients who received the control therapy (i.e., $A_{E,i}=0$ for all $i$), for example $\mathcal D_E$ might represent the data from the control arm of a previous RCT that evaluated a different experimental therapy. In other cases $\mathcal D_E$ might include both patients treated with the experimental treatment and 
the control therapy in earlier clinical studies. 

We use $p$ and $p_E$ to indicate the unknown distributions of $\mathcal D$ and $\mathcal D_E$. If the ID and the ED include independent and identically distributed ({\it iid}) replicates, then $p(y_i|x_i,a_i)$ and $p_E(y_{E,i}|x_{E,i},a_{E,i})$ are the distributions of the individual outcome $y_i, y_{E,i}\in\mathbb R$ conditional on treatment $a_i, a_{E,i}\in\{0,1\}$ and pre-treatment profile $x_i, x_{E,i}\in\mathbb R^d$ in the RCT and the ED. These conditional distributions might be different. We assume that the treatment assignment variable $A_i$ and pre-treatment patient characteristics $X_i$ are independent, and that the random variables $A_1,\ldots, A_n$ are independent or exchangeable.

Our goal is to test the effects of the experimental treatment in the RCT population. We will consider the following null hypothesis,
\begin{equation}H_0: p(y, x, a)~\text{is invariant to  permutations of}~a,\;\forall (x,y) \in \mathbb R^{n \times (d+1)} \text{ and } a \in \{0,1\}^n.
	\label{eq:H0}
\end{equation}
In other words, if $H_0$ holds, then for any configuration of outcomes $y = (y_1, \ldots, y_n) \in\mathbb R^n$ and covariates $x = [x_1, \ldots, x_n] \in\mathbb R^{n\times d}$ we have $p(y, x, a) = p(y, x, a')$, for every permutation $a'$ of $a \in \{0,1\}^n$. We use $\mathcal P$ and $\mathcal P_0 \subset \mathcal P$ to indicate the set of potential distributions of $\mathcal D \in \mathbb R^{ n \times (d+1) } \times \{0,1\}^n$ and the subset of distributions concordant with $H_0$. The null hypothesis $H_0$ implies $\mathbb E_p(Y_i|X_i=x_i, A_i=1)= \mathbb E_p(Y_i|X_i=x_i, A_i=0)$ for all $x_i\in \mathbb R^d$. Here $\mathbb E_p$ indicates the expectation with respect to the unknown distribution $p \in \mathcal P$. The alternative hypothesis will be $ p \in \mathcal P \setminus \mathcal P_0$.

To specify the test we define a randomized decision rule \citep{lehmann2005testing} $\phi: \mathbb R^{n\times (d+1)} \times \{0,1\}^n \rightarrow [0,1]$ that rejects $H_0$ with probability $\phi(\mathcal D)\in [0,1]$. 
We consider randomized decisions $\phi$ mainly for analytic convenience.
We will then discuss a similar non-randomized decision rule $\tilde \phi(\mathcal D) \in \{0,1\}$ for practical use. In the next paragraphs $\mathbb E_p[\phi(\mathcal D)] = \int \phi(\mathcal D) dp(\mathcal D)$
is the probability of rejecting $H_0$. Also, for any $\alpha\in [0,1]$, the function $\phi$ is an $\alpha$-level test if $\mathbb E_p[\phi(\mathcal D)]\leq \alpha$ for every $p\in\mathcal P_0$. We summarize the  notation in Table \ref{tab:notation}.

\begin{table}[t]
	\centering
	\scalebox{0.8}{\begin{tabular}{lll}
			\thickhline
			Variable & Notation & Definition\\
			\hline
			& &\\
			\makecell[l]{Outcome, pre-treatment covariates and\\treatment in the RCT} & $Y_i,X_i,A_i$ & $X_i\in\mathbb R^d, A_i\in \{0,1\}$\\
			RCT data (ID) with sample size $n$ & $\mathcal D = (Y, X, A)$ & \makecell[l]{$Y = (Y_1,\ldots,Y_n)$,\\$X = (X_1,\ldots,X_n)$,\\$A = (A_1,\ldots,A_n)$}\\
			\makecell[l]{Outcome, pre-treatment covariates and\\treatment in the ED} & $Y_{E,i}, X_{E,i}, A_{E,i}$ & $X_{E,i}\in\mathbb R^{d_E}$\\
			External data (ED) with sample size $n_E$ & $\mathcal D_E = (Y_E, X_E, A_E)$ & \makecell[l]{$Y_E = (Y_{E,1},\ldots,Y_{E,n_E})$,\\
				$X_E = (X_{E,1},\ldots,X_{E,n_E})$,\\
				$A_E = (A_{E,1},\ldots,A_{E,n_E})$}\\
			Unknown distributions of $\mathcal D$ and $\mathcal D_E$ & $p, p_E$ & \\
			& &\\
			Randomized and non-random decision rules & $\phi(\mathcal D), \tilde\phi(\mathcal D)$ & \makecell[l]{$\phi: \mathbb R^n\times \mathbb R^{n\times d} \times \{0,1\}^n \to [0,1]$\\
			$\tilde\phi: \mathbb R^n\times \mathbb R^{n\times d} \times \{0,1\}^n \to \{0,1\}$}\\
			& &\\
			\makecell[l]{The working model $\mathcal{M}$ specifies conditional \\  distributions  in the RCT and  external group\\  with parameters $\theta$ } & $q_{\theta}, q_{E,\theta}$ & \\
			& &\\
			\makecell[l]{Prior distribution and the conditional distribution, \\given the ED, of the parameters $\theta$}& $\pi(\cdot), \pi(\theta|\mathcal D_E)$ & $\pi(\theta|\mathcal D_E) = q_{E,\theta}(\mathcal D_E)\pi(\theta)/\int_\theta q_{E,\theta}(\mathcal D_E)\pi(\theta)$\\
			& &\\
			Conditional likelihood of the ID given the ED & $m(\mathcal D) $ & $m(\mathcal D) = \int q_\theta(Y|X,A) \pi(\theta|\mathcal D_E)d\theta$\\
            & &\\
			\thickhline
	\end{tabular}}
	\caption{Frequently used notation.}
	\label{tab:notation}
\end{table}

\subsection{An ED Augmented Permutation Test for RCTs with Heterogeneous Treatment Effects} 
We developed our ED augmented permutation test (ED-PT) for the null hypothesis $H_0$ in \eqref{eq:H0} with the following goals:
\begin{enumerate}
	\item The test is tailored to settings with potential heterogeneous treatment effects.
	For instance, several oncology studies enroll patients from various biomarker subgroups, and some groups are more likely to benefit from the experimental treatment than others \citep{dar2021assessment}. Moreover, there are often biological arguments to expect stronger treatment effects in some subgroups than in others \citep{lauko2022cancer}. 
	\item The procedure incorporates information from ED with the aim of increasing the power compared to other testing procedures that do not use ED. 
	\item The control of the false positive rate at level $\alpha$ 
	is robust with respect to model misspecification and potential discrepancies between the distribution of the ID ($p$) and the ED ($p_E$).
	In particular, the control of the type I error rate 
	is preserved when the conditional distributions $p(\cdot|x_i,a_i)$ and $p_E(\cdot|x_{E,i},a_{E,i})$ are different.
\end{enumerate}

\noindent {\it Working Model.} To achieve these aims, we use a working model $\mathcal M$ for the ID and the ED, with outcome distributions conditional on pre-treatment profiles and treatments denoted by $q_\theta(y_i|x_i,a_i)$ and $q_{E,\theta}(y_{E,i}| x_{E,i},a_{E,i})$, respectively.  Throughout the manuscript the {\it true} unknown distributions $(p,p_{E})$ and the conditional densities $(q_\theta, q_{E,\theta})$ of the working model $\mathcal M$ will repeatedly appear together in the same paragraphs. The model $\mathcal M$ is parametrized by $\theta\in \Theta$ and embeds the assumption of conditionally independent outcomes in the RCT and the ED, that is $q_\theta(y|x,a) = \prod_i q_\theta(y_i|x_i,a_i)$ and $q_{E,\theta}(y_E|x_E,a_E) = \prod_i q_{E,\theta}(y_{E,i}|x_{E,i}, a_{E,i})$. The model $\mathcal M$ can incorporate heterogeneous treatment effects. For example, $\mathcal M$ can be a linear regression model including effects of pre-treatment patient characteristics $X_i$ ($X_{E,i}$ for the ED) and treatment $A_i$ ($A_{E,i}$ for the ED) together with interactions. Here the parameters $\theta$  include the regression coefficients and the outcome variance. In this example the unknown distributions  $p$ and  $p_{E}$ might deviate from the linearity assumptions. We can specify the model $\mathcal M$ with identical or distinct regression functions in the ID and ED. Also, we do not require $\mathcal M$ to be a simple parametric model, and allow the use of semi-parametric or non-parametric models. Moreover, $\mathcal M$ can include any individual pre-treatment information, such as the date of diagnosis or the institution where the patient was enrolled. 

We use the Bayesian framework to define the test statistics. 
We first specify a prior distribution $\pi$ on the parameter space $\Theta$ of $\mathcal M$. Then, we summarize the information from the ED through the conditional distribution $\pi(\theta | \mathcal D_E) \propto q_{E,\theta}(Y_E|X_E,A_E) \pi(\theta)$. The model $\mathcal M$ and the prior distribution $\pi$ incorporate prior belief on 
(i) covariate-outcome relationships in the ID and the ED, 
(ii) potential treatment-biomarker interactions, and
(iii) the level of similarity between the regression functions $\mathbb E_{p}(Y_i|X_i,A_i)$ and $\mathbb E_{p_E}(Y_{E,i}|X_{E,i},A_{E,i})$ in the ID and ED. Investigators may use a model $\mathcal M$ with identical conditional outcome distributions 
$q_\theta(\cdot|x,a)=q_{E,\theta}(\cdot|x,a)$ for all $(x, a) \in \mathbb R^d \times \{0,1\}$,
or more flexible solutions, such as Bayesian hierarchical models, to allow for different regression functions for the patients treated with the same therapy in the RCT and the ED. 
Differences between the conditional outcome distributions in  these two groups may arise from several factors, such as variations in measurement technologies and treatment schedules of the therapy \citep{slevin1989randomized}.

\noindent {\it Definition of the test statistics. }
The test statistics is 
\begin{equation}
	m(\mathcal D) = \int q_\theta(Y|X,A) \pi(\theta|\mathcal D_E) d\theta.
	\label{eq:test-statistics}
\end{equation}
 Consider for example a hierarchical Bayesian model $(\mathcal M, \pi)$ that includes separate parameters $\theta=(\theta_{I}, \theta_{E})$ for the RCT and the ED, 
a prior distribution on hyper-parameters $\nu$ and conditionally independent group-specific parameters
$\theta_{I},\theta_{E} | \nu \overset{iid}{\sim} \pi( \cdot \mid \nu).$
The conditional outcome distributions, given treatments and covariates, in the RCT and the ED
are parameterized by $\theta_{I}$ and $\theta_{E}$ respectively. In this case, 
\begin{equation}
	\begin{aligned}
	m(\mathcal D) &= \frac{\int q_{\theta_I}(Y|X,A) q_{E,\theta_E}(Y_E|X_E,A_E) \pi(\theta_I|\nu) \pi(\theta_E|\nu) \pi(\nu) d\theta_I d\theta_E d\nu}{\int q_{E,\theta_E}(Y_E|X_E,A_E) \pi(\theta_E|\nu) \pi(\nu) d\theta_E d\nu} \\
	&= \int q_{\theta_I}(Y|X,A) \pi(\theta_I|\mathcal D_E) d\theta_I.
	\end{aligned}
	\label{eq:test-statistics:Exa}
\end{equation}
ntegrals in \eqref{eq:test-statistics} and \eqref{eq:test-statistics:Exa} are not necessary, and approximation methods (e.g., importance sampling) to compute these quantity can be used. 

\noindent {\it The randomized test $\phi(\mathcal D)$.}
Let $\tau = (\tau_1,\ldots,\tau_n)$ be a permutation of $(1,2,\ldots,n)$, $\mathcal T$ the set of all  permutations, 
 and $A^{(\tau)} = (A_{\tau_1},\ldots,A_{\tau_n})$. 
We denote with $t_\alpha$ the $(1-\alpha)$-quantile 
of the set $\{m(\mathcal D^{(\tau)}); \tau \in \mathcal T\}$, where $\mathcal D^{(\tau)} = (Y,X,A^{(\tau)})$. We define the randomized test $\phi(\mathcal D)$ at level $\alpha$ as 
\begin{equation}
	\phi(\mathcal D) = \left\{
	\begin{array}{cc}
		1 & \text{if } m(\mathcal D) > t_\alpha,\\
		0 & \text{if } m(\mathcal D) < t_\alpha,\\
		\frac{\alpha n! - \sum_{\tau \in \mathcal T} \mathbb I[m(\mathcal D^{(\tau)}) > t_\alpha]}{\sum_{\tau \in \mathcal T} \mathbb I[m(\mathcal D^{(\tau)})=t_\alpha]} & \text{if } m(\mathcal D) = t_\alpha,
	\end{array}
	\right.
	\label{eq:randomized-phi}
\end{equation}
where $\mathbb I(\cdot)$ is the indicator function.

\noindent {\it A decision-theoretic justification of the permutation test.}
%
The Bayesian expected power (BEP; \citealt{brown1987projection, liu2018assessment}) of a test $\phi'$ with respect to $\pi(\theta|\mathcal D_E)$
is 
\begin{equation}
	\text{BEP}(\phi') = \mathbb{E}_{(X,A) \sim p}
	 \left [ 
	\int \left( \int \phi'(Y,X,A) q_\theta(Y|X,A)dY \right) \pi(\theta|\mathcal D_E)d\theta
	\right ]
	.
	\label{eq:BEP}
\end{equation}
The randomized test $\phi$ in expression \eqref{eq:randomized-phi} has maximal BEP.
Proposition \ref{prop:opt1} states the optimality result. A proof is provided 
in Section SM1 in the supplementary materials. 
\begin{proposition}
	The permutation test $\phi(\mathcal D)$ defined in \eqref{eq:randomized-phi} 
	has level $\alpha$ and 
	maximizes the BEP in \eqref{eq:BEP} among all $\alpha$-level tests of $H_0$.
	\label{prop:opt1}
\end{proposition}

\noindent\textit{The non-randomized test $\tilde \phi(\mathcal D)$.} Randomized tests are rarely used in practice and enumeration over the set $\mathcal T$ of all permutations is typically infeasible even for moderate sample sizes $n$.
We slightly modify $\phi(\mathcal D)$ in {\tt Algorithm \ref{alg:app_bayes}} to obtain a practical non-randomized test $\tilde{\phi}(\mathcal D ) \in\{0,1\}$. 
We generate $J$ random permutations $\tau$ from $\mathcal T$,
and use the proportion of these permutations with 
$m(Y,X,A^{(\tau)}) \geq m(\mathcal D)$ 
as a $p$-value approximation. 

\begin{algorithm}
	\caption{Non-randomized permutation test with level $\alpha$.}
	\label{alg:app_bayes}
	\begin{tabbing}
	\qquad {Input:} The number of permutations $J$, the ID $\mathcal D = (Y, X, A)$, the working model $\mathcal M$, and \\
	\qquad\qquad\quad the conditional distribution $\pi(\theta|\mathcal D_E)$\\
	\qquad $m(\mathcal D) \leftarrow \int_{\theta} q_{\theta}(Y|X, A ) \pi(\theta|\mathcal D_E) d\theta$\\
	\qquad {For} $j = 1$ to $j = J$:\\
	\qquad \qquad $\tau~\leftarrow$ a random sample from $\mathcal T$\\
	\qquad \qquad $m_j \leftarrow \int_{\theta} q_{\theta}(Y|X, A^{(\tau)} ) \pi(\theta|\mathcal D_E) d\theta$\\
	\qquad $\tilde \phi(\mathcal D) \leftarrow \mathbb I \left\{ \frac {1+\sum_1^J \mathbb I[m_j\geq m(\mathcal D)]}{1+J}\leq \alpha \right\}$\\
	\qquad {Output:} $\tilde \phi(\mathcal D)$
	\end{tabbing}
\end{algorithm}

The proof of Proposition 1 allows us to notice four properties of the ED-PT:
\begin{enumerate}
	\item[1.] False positives are also controlled at the $\alpha$ level  
	in relevant scenarios where the samples of the ID and/or the ED are not independent and identically distributed, for example, when the  RCT  population varies over time (e.g., \citealp{russo2023inference,kennedy2017subversion}).
	\item[2.] The control of false positive results for $\phi$ at the $\alpha$ level is maintained if the test statistic $m(\mathcal D)$ is computed using approximation methods (e.g., importance sampling).
	\item[3.] The randomized and non-randomized tests (i.e., $\phi \text{ and } \tilde\phi$) have nearly identical type I and type II error rates when $J$ diverges.
	\item[4.] The proposed ED-PT is applicable when the external dataset includes both patients treated with the experimental and control therapies.
\end{enumerate}
We refer to Section SM2 in the supplementary materials for a discussion of these properties.

\subsection{One-sided Testing}
\label{sec:one-sided}
The testing procedure that we introduced in the previous subsection does not distinguish between positive and negative treatment effects. 
In different words, the test rejects $H_0$ with high probability when the experimental therapy is inferior compared  to   the control. 
In many  settings there are strong arguments to exclude the possibility of negative  effects 
of the experimental intervention. We can mention  for example  
an experimental plan, with frequent text and email reminders, to improve adherence 
to the recommendations of a cancer prevention program. 
However, in some  trials the experimental treatment may have negative effects.
For instance the treatment may reduce survival due to treatment-related toxicities.
In such cases our ED-PT may reject $H_0$ due to negative effects.

Several variations of the permutation procedure (\texttt{Algorithm \ref{alg:app_bayes}}) allow the user to 
(i) reject $H_0$ when some of the patients benefit from the experimental treatment and 
(ii) control the likelihood of rejecting $H_0$ when the effects are absent or negative. 
These variations are based on  test statistics with simple Bayesian interpretations.
We describe two modified versions of the testing procedure: 
\begin{enumerate}
	\item[(i)] The first one replaces $m(\mathcal D)$ in \eqref{eq:test-statistics} with 
	\begin{align}
		\label{m1}
		\tilde m_1(\mathcal D) = \int_{\tilde\Theta} \pi( \theta | \mathcal D, \mathcal D_E) d \theta.
	\end{align}	   
		    
	Here we restrict the integral to a subset $\tilde\Theta \subset \Theta$ of the parameter space. For example we can restrict integration to a subset $\tilde\Theta$ with positive and clinically relevant effects for at least one subgroup of patients. We  can modify the permutation procedure in \texttt{Algorithm \ref{alg:app_bayes}}, using the statistics $\tilde m_1(\mathcal D)$ instead of $m(\mathcal D)$.
	
	\item[(ii)] Alternatively, we can define $\tilde m_2(\mathcal D)$ as the expected regret (i.e., the difference in expected utility) between (i) the optimal policy \citep{murphy2003optimal} that treats every patient $i$ with the best available therapy 
	$\arg \max_{a \in \{0,1\} } \mathbb E_{p}(Y_i | X_i, A_i = a)$,
	and (ii) the policy that assigns every patient to the control therapy.
	The Bayesian working model can be used for inference on the optimal policy.  Let  $\tilde a_i (\theta)=\arg \max_{a \in \{0,1\} } \mathbb E_{q_{\theta}}(Y_i | X_i, A_i = a)$ and define the utility
	$ \frac{1}{n} \sum_{i=1}^n \mathbb E_{q_{\theta}}[Y_i| X_i , A_i = \tilde a_i (\theta)]$, an interpretable function of $\theta$ and $X=(X_1,\ldots,X_n)$, which we integrate with respect to the posterior 
	of $\theta$, conditional on the RCT and EC data. 
	The resulting integral is a summary of the efficacy of the optimal policy in the trial population, and it can be compared to the policy that treats every patient with the control therapy
	using the following statistics:
	\begin{align} \label{m2}
		\tilde m_2(\mathcal D) =  \int \frac{1}{n} \sum_{i=1}^n \left\{ \mathbb E_{q_{\theta}}[Y_i| X_i , A_i = \tilde a_i (\theta)]- \mathbb E_{q_{\theta}}(Y_i| X_i , A_i = 0 )\right\} d\pi(\theta \mid \mathcal D, \mathcal D_E). 
	\end{align}
\end{enumerate}

\subsection{ED-PT with Binary Outcomes}
\label{ex:binary}
We consider a RCT with binary outcomes and $1$:$r$ randomization (experimental vs. control). For simplicity we do not include covariates in this example. Additional examples with potential treatment-biomarker interactions and negative treatment effects are discussed in Section SM6 in the supplementary materials. The ED include patients treated with the control therapy, i.e. $A_{E,i}=0$ for all $i$. We first describe the unknown data distributions, $p$ and $p_E$  (ID and ED):
	\begin{equation} 
		\begin{gathered}
			A_i \iid \text{Bernoulli}[1/(1+r)],\quad
			Y_i | A_i \ind \text{Bernoulli}(w_0 + \gamma A_i), \label{eq:prop-model}\\
			Y_{E,i} \iid \text{Bernoulli}(w_0 + \beta_0).
		\end{gathered}
	\end{equation}
	Here $w_0$ is the response rate in the control arm, 
	$\gamma\in [-w_0, 1-w_0]$ and $\beta_0 \in [-w_0, 1-w_0]$ indicate the treatment effect and the mean difference between the outcomes in the 
	internal (i.e., within the RCT) and external control groups. 
	We set $n = 100$, $n_E = 500$, $r = 0.5$ and $w_0 = 0.5$.
	To investigate the type I error rate and power of the test we consider $\gamma = 0$ and $\gamma = 0.25$. We vary $\beta_0$ from -0.1 to 0.1 to examine the robustness of the test procedure 
	with respect to  discrepancies between the control arm in the RCT and the ED.\\
	
	\noindent \textit{Working model $\mathcal M$.} We use a Beta-Bernoulli working model: 
	\begin{equation}
		\begin{gathered}
			Y_i|A_i=a,\theta_1,\theta_0 \overset{ind}{\sim} \text{Bernoulli}(\theta_a),\quad Y_{E,i} | \theta_1,\theta_0\overset{ind}{\sim} \text{Bernoulli}(\theta_0),\\
			\theta_a \overset{iid}{\sim} U[0,1],\;\;a \in\{0,1\}.
		\end{gathered}
	\end{equation}
	The model parameters are $\theta = (\theta_0, \theta_1)$, the response rates in the control and experimental arms. Here $U[0,1]$ indicates the uniform distribution on $[0,1]$, and conditional on  $\theta$, the outcomes are independent. The model $\mathcal M$ assumes identical response rates in the internal and external control groups. $\mathcal M$ is misspecified when $\beta_0\neq 0$.
	Let $n_1 = \sum_i A_i$, $n_0 = n - n_1$, $s_1=\sum_i A_iY_i $, $s_0=\sum_i (1-A_i)Y_i $, $s_E=\sum_i Y_{E,i} $ and $s=s_0+s_1$.
	 The conditional distribution of $\theta_0$ given $\mathcal D_E$
	is a $\text{Beta}(s_E + 1, n_E - s_E + 1)$ distribution, while the conditional distribution of $\theta_1$ given $\mathcal D_E$ remains $U[0,1]$. The conditional likelihood 
	$m(\mathcal D)$, based on standard results on the Beta-Bernoulli model, is
	\begin{equation}
		m(\mathcal D) = \frac{s_1!(n_1 - s_1)!(s_0 + s_E)!(n_0 + n_E - s_0 - s_E)!(n_E + 1)!}{(n_1+1)!(n_E + n_0 + 1)!s_E!(n_E - s_E)!}.
		\label{eq:marg_binary}
	\end{equation}
	
	\noindent {\it Other testing procedures.} In our comparisons we considered: 
	\begin{enumerate}
		\item[Test-A.] A { permutation test without ED}, identical to {\tt Algorithm 1}. We use the same working model for the ID as described above and we do not incorporate ED in the analyses.
		The test statistic is $m'(\mathcal D) = s_1!(n_1 - s_1)!s_0!(n_0 - s_0)!/[(n_1+1)!(n_0+1)!]$,
		which is proportional across permutations to the Bayes factor contrasting the hypotheses $\theta_0 \neq \theta_1$ and $\theta_0 = \theta_1$. 
		\item[Test-B.] A { Wald test based on the ID} with test statistic
		$
		Z = (s_1/n_1 - s_0/n_0)/(s_1(n_1-s_1)/n_1^3 + s_0(n_0-s_0)/n_0^3)^{1/2}$
		and $\alpha$-level rejection region $|Z| > \Phi^{-1}(1-\alpha/2)$, where $\Phi$ is the cumulative distribution function of the standard normal distribution.
		\item[Test-C.] A { Wald test that merges ID and ED} with statistic
		$
		Z = (s_1/n_1 - (s_0+s_E)/(n_0+n_E))/(s_1(n_1-s_1)/n_1^3 + (s_0+s_E)(n_0 + n_E - s_0 - s_E)/(n_0+n_E)^3)^{1/2},
		$
		and rejection region $|Z| > \Phi^{-1}(1-\alpha/2)$.
		%
		\item[Test-D.] An { oracle procedure}. 
		The oracle knows the response probability $w_0$ under the control treatment in the RCT. The test statistic is $Z = (s_1/n_1 - w_0)/[s_1(n_1-s_1)/n_1^3]^{1/2}$ and the rejection region is $|Z| > \Phi^{-1}(1-\alpha/2)$.
	\end{enumerate}
	
\begin{figure}[t]
		\centering
		\includegraphics[width = \textwidth]{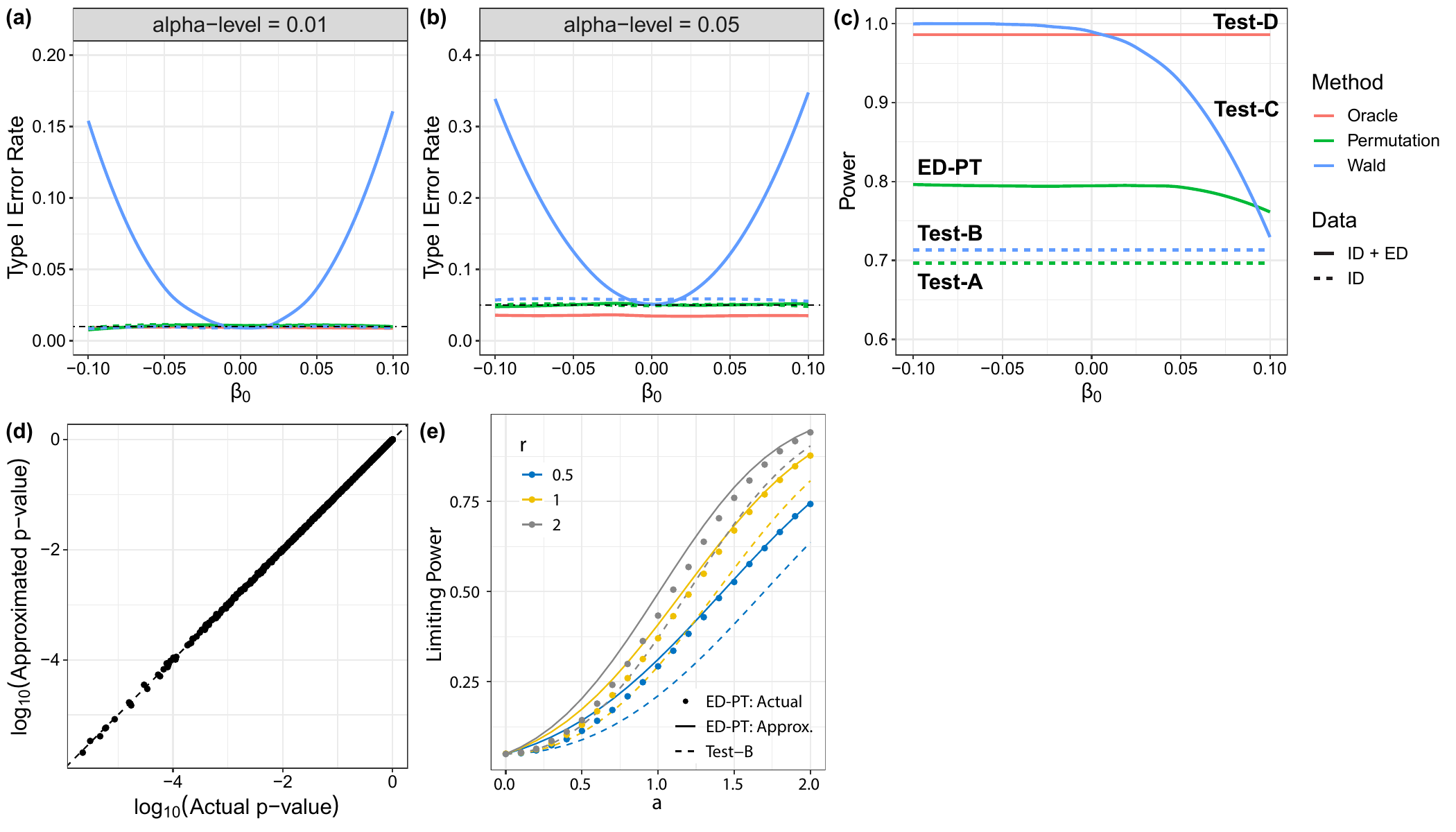}
		\caption{
			ED-PT, an  example with binary outcomes. Panels (a) and (b) illustrate type I error rates at two $\alpha$ levels, while Panel (c) shows the power. All panels (a-c) compare different tests (ED-PT and Test-A to Test-D) with ED (solid lines) and without ED (dashed lines). Here, $\beta_0$, i.e., x-axis in Panels (a) to (c), indicates the difference between the response rates of the control groups in the ED and the ID (see expression \ref{eq:prop-model}). Panel (d) illustrates the comparison between approximate $p$-values in expression \eqref{eq:pval-approx} and the exact $p$-values of $\tilde \phi$. In these simulations, $w_0$ varies between 0.2 and 0.8, $a$ varies between 0 and 2, and $b$ varies between -3 and 3. Panel (e) shows, with dots, for $a$ between 0 and 2 (x-axis), the limiting power function $g(r,r_E,a,b,w_0)$ of ED-PT in \eqref{eq:limiting-power} when $w_0=0.5$, $r_E=5$, $b = 0$ and $r\in\{0.5,1,2\}$. It also includes the approximation in \eqref{eq:limiting-power-approx} with solid lines, and the limiting power function for Test-B (dashed lines).}
		\label{fig:binary_test}
	\end{figure}

	\noindent {\it Simulation results.} 
	We considered scenarios with different $\beta_0$ values, and for each scenario repeated $10,000$ simulations. 
	We estimate the type I error rate for two significance levels $ \alpha = 0.01$ and $0.05$ (Figure \ref{fig:binary_test}, Panels a and b). 
	We also estimate the power for $\alpha = 0.05$ (Figure \ref{fig:binary_test}, Panel c). The graphs show that when $p(y_i| a_i=0) =p_E(y_{E,i}|a_{E,i}=0)$, i.e. $\beta_0 = 0$, both our permutation test and the Wald test (Test-B and Test-C, illustrated in blue) benefit from the use of the ED, which leads to substantial improvements of the power. Importantly, when $p(\cdot| a_i=0) \neq p_E(\cdot|a_{E,i}=0)$
	the  ED-PT (solid green line) controls the type I error rate, while the Wald test that incorporates  ED (Test-C; solid blue line) has an inflated type I error.\\
	
	\noindent \textit{Asymptotic Analysis.} We discuss the asymptotic behavior of our ED-PT. We consider a sequence of $(\mathcal D , \mathcal D_E)$ pairs with increasing sample sizes $n$ and $n_E$, and assume the following:
	
	(A1) { $r = n_0/n_1 > 0$ and $r_E = n_E/n_1 > 0$} are fixed and $n_1\to\infty$,
	
	(A2) $\gamma = a/n_1^{1/2}$, for $a>0$, and $\beta_0 = b/n_1^{1/2},$ with $a$ and $b$ fixed. 
	
	We focus on the test $\tilde\phi$ ({\tt Algorithm \ref{alg:app_bayes}}) when $J$ is large (i.e., $J/n! \rightarrow \infty$). In other words, we consider $\tilde\phi$ and the exact $p$-value, equal to the proportion of permutations $\tau\in\mathcal T$ that satisfy $ m(\mathcal D^{(\tau)})\geq m(\mathcal D)$. We are interested in obtaining the limiting ($n \rightarrow \infty$) power of $\tilde\phi$ under the Assumptions (A1) and (A2). To evaluate the power 
	we utilize the following proposition, which provides a large-sample approximation $\hat{ \texttt{pv}}_{\tilde\phi} $ of the exact $p$-value $\texttt{pv}_{\tilde\phi}$. 
	\begin{proposition}\label{prop:pval-approx}
		Under assumptions (A1) and (A2), 
		$$\frac{\text{\tt pv}_{\tilde\phi} }{ \hat{\text{\tt pv}}_{\tilde\phi} }\overset{p}{\rightarrow} 1,$$
		 when 
		the sample sizes diverge. Here the convergence is in probability, $\text{\tt pv}_{\tilde\phi}$ is the exact $p$-value of the test $\tilde \phi$ and 
		\begin{equation}
				\hat {\text{\tt pv}}_{\tilde\phi} = 1 - 
				\Phi\left(\frac{\max\left[ s_1, \frac{2(s+s_E)}{r+r_E+1} - s_1 \right] - \frac{s}{r+1} }
				{\left\{ \frac{ sr[(r+1)n_1-s]}{ [(r+1)n_1 - 1](r+1)^2} \right\}^{1/2} }\right) + 
				\Phi\left(\frac{\min \left[s_1, \frac{2(s+s_E)}{r+r_E+1} - s_1 \right] - \frac{s}{r+1}}{\left\{ \frac{ sr[(r+1)n_1-s]}{ [(r+1)n_1 - 1](r+1)^2} \right\}^{1/2} }\right).
			\label{eq:pval-approx}
		\end{equation}
	\end{proposition}

	We evaluated the accuracy of the approximation numerically (see Figure \ref{fig:binary_test}, Panel d). In this panel $n_1 = 10,000$, 
	$r = 0.5$, and $r_E = 5$. 
	 We varied $w_0$, $a$ and $b$ using a grid, with $w_0 \in [0.2,0.8]$, $a\in[0,2]$, $b\in[-3,3]$. 
	Figure \ref{fig:binary_test}(d) illustrates 1000 $(\mathcal D , \mathcal D_E)$ simulations, each corresponding to a different combination of $w_0$, $a$ and $b$. It illustrates the accuracy of the approximation $\hat{\texttt{pv}}_{\tilde\phi}$ in Proposition \ref{prop:pval-approx}.
	
	We can now derive the limiting ($n\rightarrow \infty$) power function of $\tilde\phi$.
	\begin{proposition}\label{prop:opt2}
		Under assumptions (A1) and (A2), the limiting power function of the $\alpha$-level test $\tilde\phi$ is
		\begin{equation}
			g(r, r_E, a, b, w_0) =\text{pr}\left\{\Phi[\max(U_1,U_0)] - \Phi[\min(U_1,U_0)] > 1 - \alpha\right\},
			\label{eq:limiting-power}
		\end{equation}
		where $U=(U_1, U_0) \sim N(\mu,\Sigma)$
		has bivariate normal distribution, with parameters 
		\begin{equation}
			\mu =\left[
			\begin{array}{c}
				\frac{ar^{1/2}}{[(r + 1)w_0(1-w_0)]^{1/2}}\\
				\frac{-[r(r + r_E + 1) + 2r_E]a + 2(r+1)r_E b}{(r + r_E + 1)[r(r+1)w_0(1-w_0)]^{1/2}}
			\end{array}\right], 
			\quad
			\Sigma = \left[
			\begin{array}{cc}
				1, & -1 \\
				-1, & 1 + \frac{4r_E}{r(r+r_E+1)}
			\end{array}
			\right]
			\label{eq:y_asym}.
		\end{equation}
	\end{proposition}
	Computing (\ref{eq:limiting-power}) requires numerical integration, for example Monte Carlo simulations. 
	When $a>0$ is large (above 1.5) and $b = 0$, we can approximate \eqref{eq:limiting-power} with a simpler closed-form expression (see Section SM4 in the supplementary materials for the derivation): 
	\begin{equation}
		g(r,r_E,a,b,w_0) \approx \Phi\left[\frac{ar^{1/2}}{[(r+1)w_0(1-w_0)]^{1/2}} - \Phi^{-1}(1-\alpha)\right].
		\label{eq:limiting-power-approx}
	\end{equation}
	Figure \ref{fig:binary_test}(e) illustrates the limiting power function in \eqref{eq:limiting-power} with dots, computed with 100,000 Monte Carlo simulations, the approximation in \eqref{eq:limiting-power-approx} (solid lines) and Test-B with dashed lines, when $a$ varies between 0 and 3. Here $r\in\{1/2,1,2\}$, $w_0 = 0.5$, $n_1 = 10,000$, $r_E = 5$ and $b = 0$. In this example the ED-PT test is asymptotically more powerful than the two-sided Wald test (Test-B) that utilizes only ID. 
	Also, 
	Figure SM5 illustrates large sample rejection regions for the two tests; here
	 $r = 1/2$, $n_1 = 10,000$, $r_E = 5$ and $(n_E, s_E) = (50,000, ~24,995)$.
	
	The power function in Proposition \ref{prop:opt2} allows us to discuss the behavior of the ED-PT test 
	when the user has access to a large external dataset representative of the control therapy
	(i.e., $r_E\rightarrow \infty$). The following corollary provides the limit of $g(r,r_E,a,b,w_0)$ when 
	$r_E $ diverges.
	
	\begin{corollary}\label{prop:opt3}
		 We consider the $\alpha$-level test $\tilde\phi$ with access to  large ED. 
  The limit $\lim_{r_E\to\infty}g(r, r_E, a, b, w_0)$  is equal to the right-hand side of \eqref{eq:limiting-power} with
		$$\Sigma = \left[\begin{array}{cc}
			1 & -1\\
			-1 & 1+4/r
		\end{array}\right], \quad \mu^\intercal =\left[ \frac{ar^{1/2}}{[(r+1)w_0(1-w_0)]^{1/2}}, \frac{-a(r+2) + 2(r+1)b}{[r(r+1)w_0(1-w_0)]^{1/2}} \right].$$
	\end{corollary}
	
\subsection{ED-PT with Normally Distributed Outcomes}
	\label{ex:linear}
	We now consider normally distributed outcomes and include pre-treatment covariates. We  assume again $1:r$ randomization (experimental and control arms) for the RCT, and all patients in the ED are treated with the control therapy, as in Section \ref{ex:binary}. The ID and the ED have the following distributions,
	\begin{equation}
		\begin{gathered}
			A_i \iid \text{Bernoulli}[1/(1+r)],\quad
			Y_i | A_i, X_i \ind N(\eta_0 + \gamma A_i + \beta_1^\intercal X_i + A_i \gamma_1^\intercal X_i, 1),\\ 
			Y_{E,i} | X_{E,i} \ind N[\eta_0 + \beta_{0,0} + (\beta_1 + \beta_{0,1})^\intercal X_{E,i}, 1].
		\end{gathered}
		\label{eq:linear_model}
	\end{equation}
	Recall that $X_i,X_{E,i}\in\mathbb R^d$. Here $\gamma$ is the treatment effect for patients with pre-treatment covariates $X_i$ equal to $0_d$, where $0_d$ is a vector of zeros with length $d$. The terms $A_i \gamma_1^\intercal X_i$ determine variations of the treatment effects across pre-treatment profiles, and the parameters $\beta_0 = (\beta_{0,0},\beta_{0,1})\in\mathbb R^{d+1}$ quantify the discrepancies between ED and the controls within the RCT.\\
	
	\noindent {\it Working Model $\mathcal M$.} 
	We use a linear model,
	\begin{equation}
		\begin{gathered}
			Y_i | A_i, X_i \ind N(\theta_0 + \theta_1^\intercal X_i + \theta_2 A_i + A_i \theta_3^\intercal X_i , 1), \quad
			Y_{E,i} | X_{E,i} \ind N(\theta_0 + \theta_1^\intercal X_{E,i}, 1),
		\end{gathered}
		\label{eq:lm_working}
	\end{equation}
	where $\theta_1,\theta_3\in \mathbb R^{d}$. The Bayesian model uses independent $N(0,\sigma^2)$ prior distributions for the components of $\theta = (\theta_0,\theta_1,\theta_2,\theta_3)$. Based on standard conjugacy results, $\theta|\mathcal D_E \sim N\left(\mu_E, V_E\right)$, where $V_E = \text{diag}\{(\tilde X_E^\intercal\tilde X_E + \sigma^{-2}I_{d+1})^{-1}, \sigma^2I_{d+1}\}$, $\mu_E = [(\tilde X_E^\intercal\tilde X_E + \sigma^{-2}I_{d+1})^{-1}\tilde X_E^\intercal Y_E, 0_{d+1}]$ and $\tilde X_E = (1_{n_E}, X_E)$. Here $1_{n_E}$ is a vector of ones of length $n_E$, and $I_{d+1}$ is the $(d+1) \times (d+1)$ identity matrix. The conditional likelihood of the ID  (expression \ref{eq:test-statistics}) is
	\begin{equation}
		\label{marg_normal}
		m(\mathcal D) \propto \text{exp} \left(\frac{1}{2} \mu^\intercal V^{-1} \mu \right) |V|^{1/2},
	\end{equation}
	where
	$V = (\tilde X^\intercal \tilde X + V_E^{-1})^{-1}$, $\mu = V (\tilde{X}^\intercal Y + V_E^{-1} \mu_E)$, and $\tilde X = [1_n,X,A, (A {1}_n^\intercal) X]$.\\
	
	\noindent {\it Other testing procedures.} Our comparisons include: 
	\begin{enumerate}
		\item[Test-A.] A { permutation test without ED} identical to the ED-PT; the test statistic is $m'(\mathcal D)\propto \exp\left(\frac{1}{2} \mu'^\intercal V'^{-1} \mu' \right) |V'|^{1/2}$, where $V' = (\tilde X^\intercal \tilde X + \sigma^{-2}I_{2(d+1)})^{-1}$ and $\mu' = V'\tilde X^\intercal Y$.
		\item[Test-B.] A { Wald test for $(\theta_2,\theta_3)$ based only on the  ID}, using the model  \eqref{eq:lm_working}. The test statistic is $Z = (\hat\theta_2,\hat\theta_3)^\intercal \Sigma^{-1} (\hat\theta_2, \hat\theta_3)$, where $(\hat\theta_2,\hat\theta_3)$ is the MLE of $(\theta_2,\theta_3)$ under the working model and $\Sigma$ is the 
		submatrix of $\mbox{Cov}(\hat \theta) = (\tilde X^\intercal \tilde X)^{-1} $ 
		that corresponds to the estimates $(\hat\theta_2,\hat\theta_3)$. The rejection region is $Z > \chi^2_{1-\alpha, d + 1}$. Here $\chi^2_{1-\alpha, d + 1}$ is the $1-\alpha$ quantile of a Chi-square distribution. 
		\item[Test-C.] A { Wald test for $(\theta_2,\theta_3)$ based on merged ID and ED}, using the model in \eqref{eq:lm_working}. 
		The test statistic is $Z = (\hat\theta_2,\hat\theta_3)^\intercal \Sigma^{-1} (\hat\theta_2, \hat\theta_3)$,
		as in Test-B, 
		but in this case $(\hat\theta_2,\hat\theta_3)$ and $\Sigma$ are the MLEs and their covariance estimate computed after merging ID and ED. The rejection region is $Z > \chi^2_{1-\alpha, d + 1}$.
		\item[Test-D.] { An oracle procedure.} The oracle knows the regression function of the internal control group. The test is based on the RCT residuals $R_i = Y_i - \eta_0 - \beta_1^\intercal X_i$ for all individuals $i$ that received the experimental treatment (i.e. $A_i=1$). 
		In particular, it is a Wald test for $(\theta_2,\theta_3)$ based on the regression model $R_i|X_i \sim N(\theta_2 + \theta_3^\intercal X_i, 1)$. We compute the MLEs $(\hat \theta_2, \hat \theta_3)$ and their covariance matrix $\Sigma$, including only patients that  received the experimental treatment $(A_i=1)$.  Then, similar to Test-B and Test-C, we use the test statistic $Z$ and the rejection region $Z > \chi^2_{1-\alpha, d + 1}$.
	\end{enumerate}
	
\begin{figure}[t]
		\centering
		\includegraphics[width=0.95\textwidth]{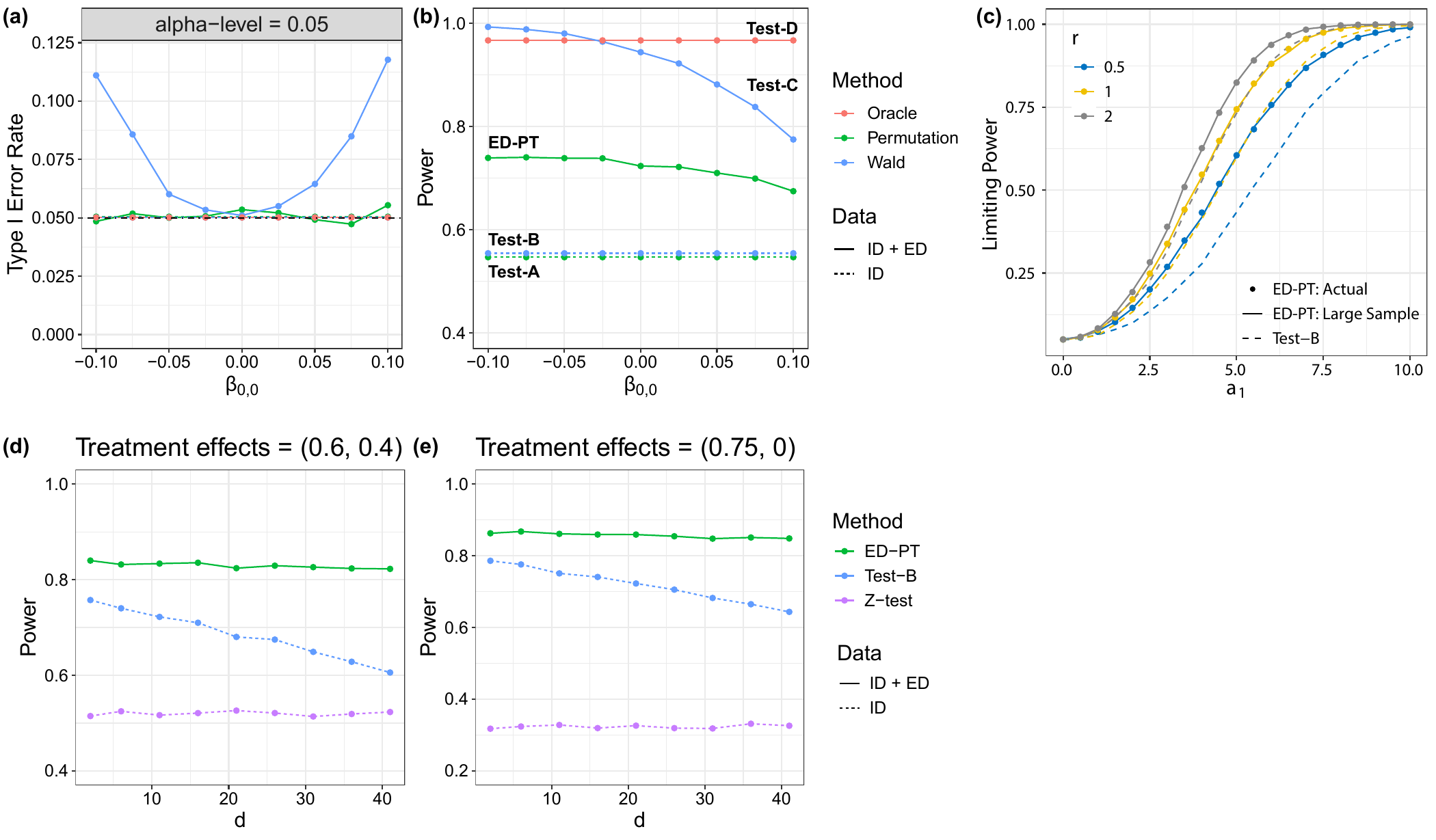}
		\caption{ 
			Power and type I error rates of the ED-PT, an example with  continuous outcomes.
			Panels (a-c) show results when covariates are discrete subgroup indicators. 
			Here $K = 2$ with $\rho = (0.5,0.5)$, $\eta_0 = \beta_{0,1} = 0,$ $\beta_1 = 0.5$, $r_E = 7.5$ and $\sigma^2 = 10$. 
			We used $J = 1\text{,}000$ permutations and 10,000 simulation replicates per scenario.
			Panels (a) and (b): Type I error rates ($\alpha=0.05$) and power for the ED-PT and alternative tests (Test-A to Test-D) with (solid lines) and without (dashed lines) ED. The type I error rate was evaluated with $\gamma = \gamma_1 = 0$ and the power was assessed with $\gamma = 0.5 $ and $ \gamma_1 = -0.2$. The $\beta_{0,0}$ parameter (x-axis) summarizes the discrepancy between the ED and the control arm of the RCT (see also expression \ref{eq:linear_model}). The results were computed with $r = 0.5$ and $n = 150$. Panel (c): Limiting power of the ED-PT, evaluated using expression (S.2) in Section SM5 in the supplementary material with 100,000 Monte Carlo replicates (dots). The results are compared to estimates of the power based on trial simulations, using
			\texttt{Algorithm \ref{alg:app_bayes}} and $n_1 = 10\text{,}000$ (solid lines).  We considered $r\in\{0.5,1,2\}$ and specified $\gamma$, $\gamma_1$, $\beta_{0,0}$, and $\beta_{0,1}$ as in Assumption (B3), with $a_1\in[0,10]$, $a_2 = 0$, and $b = (0,0)$. The panel also illustrates the power of Test-B (dashed lines).
			Panels (d) and (e) show the power of three testing procedures when the covariates include subgroup indicators and continuous pre-treatment variables. The dimension of $X_i$  varied between 2 and 41, with 1 to 40 continuous pre-treatment variables. The details of the simulation are described in the manuscript.
		}
		\label{fig:normal_test}
	\end{figure}

	\noindent \textit{S1,  a scenario with discrete covariates  ($X_i$ and $X_{E,i}$) indicating the patient subgroups.}
Pre-treatment profiles are often summarized by binary and categorical variables.
	We consider a trial with patients partitioned into $K$ subgroups.
In particular 
$X_i$ and $X_{E,i}$ include $K-1$ binary indicators (i.e., $X_{E, i}, X_i\in \{0,1\}^{K-1}$) that point to the individual subgroup. That is, $X_i$ and $X_{E,i}$ can take only $K$ different values, and the $(k-1)$-th element of $X_i$ (or $X_{E,i}$) is equal to one if  patient $i$ belongs to group $k$.
	In Figure \ref{fig:normal_test}(a) and (b) we show simulation results for $K=2$ patient subgroups. Here the proportion of patients in subgroup 1 is 0.5 for the RCT and the ED. 
	Also, $\eta_0 = 0, \beta_1 = 0.5$ and $r = 1/2$. 
	With $K = 2$ the treatment effects in subgroups 1 and 2 are  $\gamma$ and $\gamma + \gamma_1$. 
	We considered $\gamma = \gamma_1 = 0$ to assess the type I error rate of the test, and $(\gamma = 0.5, \gamma_1 = -0.2)$ to evaluate the power. 
	We set $\beta_{0,1} = 0$ and varied $\beta_{0,0}$ from $-0.1$ to $0.1$ to evaluate the robustness of the ED-PT procedure with respect to discrepancies between $p_E(\cdot|x_{E,i},a_{E,i} = 0)$ and $p(\cdot|x_i,a_i = 0)$. 
	
	The results in Figure \ref{fig:normal_test}(a) and (b) 
	were computed with $J=1,000$ permutations, $10,000$ simulations per scenario, $n = 150$, $n_E = 750$, and $\sigma^2 = 10$. If $p_E(\cdot|x_{E,i},a_{E,i} = 0) = p(\cdot|x_i,a_{i} = 0)$ for every pair of
	pre-treatment profiles
	$x_{E,i} = x_i$ (i.e., $\beta_{0,0} = 0$), then the use of ED increases the power of all testing procedures.
	Moreover, when $p_E(\cdot|x_{E,i},a_{E,i} = 0) \neq p(\cdot|x_i,a_{i} = 0)$ for some values of  $x_{E,i} = x_{i}$, our ED-PT controls the type I error rate, while the control of false positives deteriorates for Test-C.
	
	\noindent \textit{Asymptotic analysis.} We investigate the asymptotic behavior of our ED-PT. 
	We focus on the scenarios that we described, 
	with $X_i, X_{E,i} \in \{0,1\}^{K-1}$ indicating subgroups. The unknown outcome distributions ($p$ and $p_E$) are summarized by model \eqref{eq:linear_model}, and the working model by expression \eqref{eq:lm_working}. We derive the limiting power function when the population includes $K>1$ subgroups. Similar to Section \ref{ex:binary}, we consider a sequence of $(\mathcal D, \mathcal D_E)$ pairs with increasing samples sizes $n$ and $n_E$. See Section SM5 in the supplementary materials for details. In Figure \ref{fig:normal_test}(c) we computed the limiting power (see expression S.2 in Section SM5 in the supplementary materials) using 100,000 Monte Carlo simulations (dots). We then compared these results to estimates of the power based on trial simulations (solid lines), with $K = 2$, $\rho = (0.5,0.5)$, $a_1\in [0,10]$, $\eta_0 =\beta_{0,1} = a_2 = b_1 = b_2 = 0$, $\beta_1 = 0.5$, $r\in\{0.5,1,2\}$, $r_E = 7.5$, $n_1 = 10,000$, $J = 1,000$ and 10,000 simulation replicates per scenario. The panel  illustrates that in this scenario the limiting power of the ED-PT is larger than the power of Test-B (dashed lines).\\
	
	\noindent {\it Modified \textit{S1}, negative treatment effects.} We modified Scenario {\it S1} above to showcase the excessive rejection rate of the null hypothesis by the original ED-PT when treatment effects are negative, and illustrate how the modified ED-PT, based on test statistics $\tilde{m}_1(\mathcal D)$ and $\tilde{m}_2(\mathcal D)$, effectively control the rejection rate in such scenarios.
	
	We simulated 10,000 times the pair $(\mathcal D, \mathcal D_E)$ using model \eqref{eq:linear_model} with $K = 2$ patient subgroups. We set $\eta_0=0,$ $\beta_1 = 0.5, \gamma=0$ (the treatment effect in group 1) and $\gamma_1=-1$ (the treatment effect in group 2) in \eqref{eq:linear_model}. 
	Also, $n_1 = 100, r=0.5, r_E=7.5$, $\beta_{0,0}=\beta_{0,1}=0$, and $\alpha = 0.05$. 
	In this modified scenario \texttt{Algorithm \ref{alg:app_bayes}} rejected the null hypothesis with a frequency equal to
	0.85.
	
	To use $\tilde{m}_1$, we specify $\tilde \Theta$ as
	$
	\tilde\Theta = \{ \theta \in \Theta : \theta_2 > \tilde\theta \text{ or } \theta_2 + \theta_3 > \tilde\theta \},
	$
	where $\tilde \theta$ is a threshold that defines clinically relevant treatment effects.
	The modified permutation procedure in \texttt{Algorithm \ref{alg:app_bayes}}, using the statistics $\tilde m_1(\mathcal D)$ instead of $m(\mathcal D)$, rejected $H_0$ with a frequency equal to 0.04 when we set $\tilde\theta = 0$. If we use $\tilde{m}_2(\mathcal D)$ in \texttt{Algorithm \ref{alg:app_bayes}} to replace $m(\mathcal D)$ in this example, we have
	$$
	\tilde m_2(\mathcal D) = \mathbb E [ \rho_1 \theta_2 \mathbb I(\theta_2 > 0) + (1-\rho_1) (\theta_2 + \theta_3) \mathbb I(\theta_2 + \theta_3 >0) \mid \mathcal D, \mathcal D_E].
	$$
	Recall that $\rho_1$ is the prevalence of the first subgroup of patients. In this case, the rejection rate decreased from 0.85 to 0.05.
	
	Figure SM3 in the supplementary materials shows the frequency of rejections of the ED-PT with the test statistic $m(\mathcal D)$ (dashed line) and the  modified versions 
	of the testing procedure 
	with  statistics $\tilde m_1(\mathcal D)$ and $\tilde m_2(\mathcal D)$ (solid blue and red lines, respectively) in Scenario {\it S1} when $\gamma=0$ and $\gamma_1 \in [-1,1].$ It  illustrates that the  modified versions 
	of our ED-PT  in  the presence of negative treatment effects (i.e., $\gamma_1 < 0$) 
	control the frequency of false positive results. Moreover,  with positive treatment effects 
	these modified versions of the ED-PT procedure have power similar to  the ED-PT  with
	test statistics  $m(\mathcal D)$.
	\\
	
	\noindent \textit{S2, a scenario with two subgroups and continuous pre-treatment covariates.} 
	We conclude this subsection by adding continuous pre-treatment covariates to the  Scenario {\it S1}. In particular, we specify pre-treatment profiles $X_{i}, X_{E,i}\in \{0,1\} \times R^{d-1}$, where the first entries of $X_{i} $ and $ X_{E,i}$ are {\it iid}  $\text{Bernoulli}(1/2)$ as in Scenario {\it S1} and the remaining $d-1$ components of  $X_{i} $ and $ X_{E,i}$  have  $N(0_{d-1}, I_{d-1} ) $ distributions. We set $n_1 = 100$, $r=0.5$, $r_E = 7.5$, $\eta_0 = \beta_{0,0} = 0$ and $\beta_{0,1} = 0_{d}$ in the outcome model \eqref{eq:linear_model}. In other words, there are no discrepancies between the outcome distributions $p_E(\cdot|x_{E,i}, a_{E,i}=0)$ and $p(\cdot|x_{i},a_i=0)$ of the ED and the RCT control arm. In Figure \ref{fig:normal_test}(d-e) the dimension $d$ of $X_i$ grows from 2 to 41. The ED-PT is compared to Test-B, and a two-sample Z-test (ID only) that ignores the covariates. In Panel (d) both patient subgroups benefit from the experimental treatment: $\gamma = 0.6$ and $\gamma_1 = (-0.2, 0_{d-1})$. In Panel (e) only one group of patients benefits from the experimental treatment: $\gamma = 0.75$ and $\gamma_1 = (-0.75, 0_{d-1})$. We set $\beta_1 = [0.5, 1_{d-1}/(d-1)^{-1/2}]$, therefore the marginal variability of the outcome $Y_i$ does not vary with the number of continuous covariates ($d-1$). In both panels, the power of ED-PT remains nearly the same as $d$ increases, while for Test-B, which uses only the RCT data, the power decreases. This result suggests that ED-PT successfully leverages the information about $\beta_1$ provided by the ED. Also, the Z-test has lower power compared to the ED-PT and Test-B.
	
\section{Glioblastoma Clinical Trials}
We report the results of retrospective analyses of a collection of Glioblastoma (GBM) datasets. The analyses are based on a resampling schema described in \cite{ventz2022use}. The goal is to  assess   power and type I error rates of our ED-PT and other testing procedures.

\noindent {\it Datasets.} 
The data include  individual patient-level information of newly diagnosed GBM patients that were treated with temozolomide and radiation therapy (TMZ+RT), the current standard of care in GBM \citep{stupp2005radiotherapy}. Pre-treatment patient variables ($X_i$ and $X_{E,i})$ include age, sex, Karnofsky performance status (KPS), MGMT methylation status, and extent of tumor resection (EOR). We use individual patient-level information (i.e., pre-treatment variables and outcomes) of patients treated with TMZ+RT from the AVAGLIO RCT \citep{doi:10.1056/NEJMoa1308345} and the DFCI  electronic health records database. We refer to \cite{10.1158/1078-0432.CCR-22-3524} for further details on these datasets.

{\it Outcome and Subgroups.} The primary outcome is a binary variable that captures survival after 12 months of treatment (OS-12).
 We consider treatment effects that vary across subgroups, defined by patients' KPS ($>90$ and $\leq 90$) and MGMT status (positive and negative). 
 For each patient  treated with TMZ+RT we define $S_i= S(X_i) \in \{1, 2, 3, 4\}$ if the patients' KPS and MGMT variables are 
$[> 90, \text{positive}], $
$[\le 90, \text{positive}],$
$[>90, \text{negative}],$
or 
$[\le 90, \text{negative}]$.
Hypotheses of treatment effects variations modulated by MGMT and KPS in neuro-oncology have been the subject of extensive literature, see for example \cite{chen2018novel}.
The proportions of patients in these groups and the corresponding OS-12 rates in the AVAGLIO and DFCI datasets are provided in Table \ref{tab:proportions}.

\begin{table}[t]
	\resizebox{\textwidth}{!}{
		\begin{tabular}{|r|c|c|c|c|}
			
			\cline{1-5} & KPS$\geq$90, MGMT+ & KPS$<$90, MGMT+ & KPS$\geq$90, MGMT- & KPS$<$90, MGMT- \bigstrut\\
			\hline
			AVAGLIO (ID) N (\%) & 78 (23\%) & 30 (9\%) & 161 (48\%) & 68 (20\%) \bigstrut\\
			 OS-12 & 0.83 & 0.70 & 0.68 & 0.48 \bigstrut\\ 
			\hline
			DFCI (ED) N (\%) & 95 (29.60\%) & 59 (18.38\%) & 92 (28.66\%) & 75 (23.36\%) \bigstrut\\
			 OS-12 & 0.84 & 0.83 & 0.81 & 0.63 \bigstrut\\ 
			\hline
		\end{tabular}%
	}
	\caption{Subgroup sizes and the corresponding OS-12 rates in the AVAGLIO and DFCI datasets.} 
	\label{tab:proportions}%
\end{table}%

\noindent { {\it In silico} RCTs {\it and} ED, $(p,p_E)$.} We use a resampling schema, similar to the one in \cite{ventz2022use}, which allows us to generate {\it in silico} datasets. In particular, we  generate ED and {\it in silico} RCT with pre-treatment characteristics $(X_{E,i}, X_{i} \in \mathbb R^5)$ distributions identical to the empirical distributions in the DFCI and AVAGLIO datasets.
To generate the {\it in silico} RCT and ED, we followed these three steps:
\begin{enumerate}
	\item[(i.a)] {\it In silico ID}. We sample $n$ patients (pre-treatment profiles and OS-12 outcomes) with replacement from the TZM+RT arm of the AVAGLIO study.
	
	\item[(i.b)] {\it In silico ID, treatment assignment}. We randomly assign $n_1= n /(1+r)$ of the $n$ patients to the {\it in silico} experimental arm, and the remaining $n_0=n-n_1$ to the control arm. The experimental and control arms of the {\it in silico} RCT include $n_0$ and $n_1$ {\it iid} replicates with the same joint distribution of pre-treatment variables and outcomes. This joint distribution is identical to the empirical distribution of the TZM+RT group of the AVAGLIO study. In different words, in this {\it in silico} RCT the treatment effects are null. 

	\item[(i.c)] {\it Treatment effects.} To introduce subgroup-specific positive treatment effects  in the {\it in silico } RCT,
	 we randomly relabel with probability $g(S_i)\ge 0$ each negative outcome 
	$(Y_i = 0)$ in the experimental arm (Step i.b) into a positive outcome $(Y_i = 1)$.
	 Similarly, to specify scenarios with negative treatment effects we randomly relabel 
	 each positive outcome $(Y_i = 1)$ into a negative one $(Y_i = 0)$ with probabilities $g(S_i)$ that vary across subgroups.

	\item[(ii)] {\it In silico ED.} 
	We sample $n_E$ patients with replacement from either the TZM+RT arm of  the AVAGLIO study or the DFCI dataset (pre-treatment profiles and OS-12 outcomes) that constitute the {\it in silico} ED.

	\item[(iii)] {\it Hypothesis testing.} We apply our ED-PT in \texttt{Algorithm \ref{alg:app_bayes}} to the {\it in silico} RCT (Step i) and the ED (Step ii).
\end{enumerate}

We set $n = 150, r = 1/2$, $n_E=50, 100, \ldots, 250,$ and consider 
 six scenarios with distinct configurations of the treatment effects 
 (Table \ref{tab:app-scen}). 
  In each scenario the relabeling probabilities $g(s), s\in\{1,2,3,4\}$, which varies across patient subgroups,
  match the desired group-specific log-odds ratio  $\text{LOR}_s$ (treatment vs control).  
 In particular, if $\LOR_s>0$, then $g(s) = 1/\{1+[h_s(e^{\LOR_s}-1)]^{-1}\}$,
 where $h_s$ is the  response rate in subgroup $s$ of the  AVAGLIO dataset. A similar map is used with negative effects ($\LOR_s<0$).

In scenarios 1-5 the treatment effects are null or positive. In contrast, in scenario 6  the experimental treatment has negative effects. We include this scenario to compare the three versions of ED-PT (based on $m$, $\tilde m_1$ and $\tilde m_2$) introduced in Section 2.1 and 2.2  and to examine  the control of false positives.

\begin{table}[t]
	\centering
	\begin{tabular}{|c|l|l|}
		\hline
		Scenario & $\LOR_{s}$ values & Description \\
		\hline
		1& $(0,0,0,0)$ & No effects\\
		2& $(0.5, 1, 1.5, 2)$ & Positive effects all groups\\
		3& $(0,0,3,0)$ & Positive effect for patients with MGMT+ and $\text{KPS}<90$\\
		4& $(5,5,0,0)$ & Positive effects for patients with MGMT+\\
		5& $(2,0,2,0)$ & Positive effects for patients with $\text{KPS}\geq 90$ \\
		6& $(-1,0,0,0)$ & Negative effects for patients with MGMT+ and $\text{KPS}\geq 90$\\
		\hline
	\end{tabular}
	\caption{  Six scenarios  in which the log-odds ratios $\LOR_s, s=1, 2,3, 4$ vary across patient subgroups.}
	\label{tab:app-scen}
\end{table}

\noindent {\it Working model $\mathcal M$.} 
We specify  the working  model, with  conditional distributions $q_\theta$ and $q_{E,\theta}$ parameterized by $\theta=(\theta_{\textsc{\,ID}}, \theta_{\textsc{\,ED}})$.
We use a Bayesian logistic regression model, with six pre-treatment covariates $x_{i}^{1:6}$:
age ($x_{i}^1$), 
sex ($x_{i}^2$),  EOR ($x_{i}^3$),
MGMT ($x_{i}^4$), KPS ($x_{i}^5$), 
and the interaction term  $\text{MGMT}\times \text{KPS}$ ($x_{i}^6$).
The model  includes  additional coefficients  to capture  the treatment effect  and interactions  between
 the treatment and MGMT, KPS, or $\text{MGMT}\times \text{KPS}$.
To summarize, $(x_i,a_i)\rightarrow \text{logit}[q_{\theta}(y_i=1|x_i,a_i)]$ 
 is a linear map of $(1, x_{i}^{1:6}, a_i, a_i \times x_{i}^{4:6} )$ with coefficients 
 $\theta_{\textsc{\,ID}} =(\theta_0, \theta_x, \theta_a, \theta_{I}) \in \mathbb R^{11}$,
  where $\theta_x = (\theta_x^1,\ldots,\theta_x^6)$.
Similarly for  patients treated with the control therapy in the external group,
$ x_{E,i} \rightarrow \text{logit} [q_{E,\theta}(y_{E,i}=1|x_{E,i})] $
 is a linear map  with coefficients 
$\theta_{\textsc{\,ED}} = ( \theta_{E,0}, \theta_{E,x} ) \in \mathbb R^7$.

\noindent {\it Prior model $\pi$.} Based on our previous  analyses  of   GBM RCTs and electronic health records  \citep{ventz2019design, ventz2022use,  10.1158/1078-0432.CCR-22-3524, ventz2022design},
we assume that $\theta_0 = \theta_{E,0}$, $\theta_{E,x}^j= \theta_{x}^j$ for $j \in \{1,2,4,5\}$ and
 $\theta_{E,x}^j = \theta_{x}^j + \theta_{B}^j$ for $j\in\{3,6\}$.  
We use a   normal prior with large variances  
 $\theta = (\theta_0, \theta_x, \theta_a, \theta_I,  \theta_{B}) \sim N( 0_{13}, 100 I_{13} ).$

\noindent {\it Approximation of the test statistics $m(\mathcal D)$.} 
In our ED-PT  the conditional likelihood 
$m(\mathcal D) = \int q_\theta(Y|X,A)\times\allowbreak\pi(\theta|\mathcal D_E) d\theta$ is computed using a Laplace approximation \citep{de1981asymptotic,tierney1986accurate},
\begin{equation}
	m(\mathcal D) \approx (2 \pi ) ^{ 13/2 } | - \hat H |^{-1/2}q_{\hat \theta}(\mathcal D) q_{E, \hat \theta}(\mathcal D_E) \pi(\hat \theta),
	\label{eq:laplace}
\end{equation}
where $\hat\theta$ is the maximum a posteriori (MAP) estimate and $\hat H$ is the Hessian of the log posterior at  $ \hat \theta$.

In \texttt{Algorithm \ref{alg:app_bayes}} we compute the MAP estimate and the Hessian matrix for $J$ permutations $\mathcal D^{(\tau)}$. Similarly,  to compute the modified statistics $\tilde m_1(\mathcal D)$ and $\tilde m_2(\mathcal D)$ in Section \ref{sec:one-sided} we used Laplace approximations.  For $\tilde m_1(\mathcal D)$, we restrict the integration to  $\tilde\Theta \subset \Theta$, the parameter configurations with  positive effects. Also,  to  compute $\tilde m_2(\mathcal D)$, we iteratively sample from  $N[\hat \theta,  (-\hat H)^{-1}]$   to  approximate the  integral in \eqref{m2}.

\noindent {\it Alternative testing procedures.} Our comparisons include several other  testing procedures:
\begin{enumerate}
	\item Test-A-$m$ and $\tilde m$. Permutation tests for ID only using the statistics $m$ (Test-A-$m$) and $\tilde m_j$ (Test-A-$\tilde m_j$), $j=1,2,$  respectively.

	\item Test-B and C. {Wald test for proportions}, as in Section \ref{ex:binary}, using only the ID (Test-B) or merging the ID and the ED (Test-C), without accounting for pre-treatment covariates.
	
	\item Test-LR and LR-ED. Likelihood ratio test based on our working model  (null hypothesis: $[\theta_a,\theta_{I}] =0_4$) using only the  ID  (Test-LR) or  merging the
	ID and the ED (Test-LR-ED).
	
	\item Test-Matching, a matching-based testing procedure. We first apply a matching algorithm to estimate the average  effect in the treated group. Specifically, we match each patient in the RCT's experimental arm to one patient in either the control arm of the RCT or the ED based on propensity scores. We use the R package {\tt MatchIt} \citep{ho2011matchit} to perform matching. We then use g-computation implemented in the R package {\tt marginaleffects} \citep{arel2024interpret} to estimate the treatment effects.

\item Test-IPW, an inverse probability weighting (IPW) procedure. We first estimate $\hat e(x)$, which is
 the conditional probability that a randomly selected patient (i.e., from the RCT or the ED) with pre-treatment characteristics $x$ was enrolled into the RCT, using a logistic regression model. 
We then assign  weights equal to one  to the patients in the ID,  while for the  ED  the individual wights are  $ w_E \hat e(x_{E,i})/[1-\hat e(x_{E,i})]$. Here $w_E\in \mathbb R^{+}$ determines the relative weights 
of the external dataset with respect to the RCT data.  We followed the approach in \cite{li2018balancing} and \cite{wang2023propensity} to obtain average treatment effect estimates in the RCT population. We use the R package \texttt{ipw} \citep{van2011ipw} to compute weights and weighted linear regression to estimate the treatment effects. The R package \texttt{sandwich} \citep{zeileis2020} is used to obtain the robust standard errors \citep{white1980heteroskedasticity} of the estimates.
\end{enumerate}

\noindent {\it Comparative analyses.} Our analyses focus on newly diagnosed 
GBM patients, and are based on two main groups   of   $(\mathcal D, \mathcal D_E)$ {\it in silico} replicates. In the first group  the  distributions  $p(y_i, x_i|a_i=0)$ and $p_E(y_{E,i}, x_{E,i}|a_{E,i}=0)$ are identical  and match the empirical joint distribution of  pre-treatment profiles and the outcomes in the AVAGLIO trial (Step i and ii of our schema). In the second group $p(y_i, x_i|a_i=0)$ and $p_E(y_{E,i}, x_{E,i}|a_{E,i}=0)$ are  different because the ED are generated using a different dataset, the DFCI electronic health records. Treatment effects are included in some of the {\it in silico} RCTs (see step i.c and Table \ref{tab:app-scen}). Then we focus on testing,  using a variety of approaches that differ substantially in several aspects, including the use or exclusion of ED and the potential lack of control of false positives  due to unmeasured confounders or other distortion mechanisms. The aim of the comparative analyses is to identify testing procedures suitable for future GBM trials.

\begin{figure}[t]
 	\centering
 	\includegraphics[width = 0.85\textwidth]{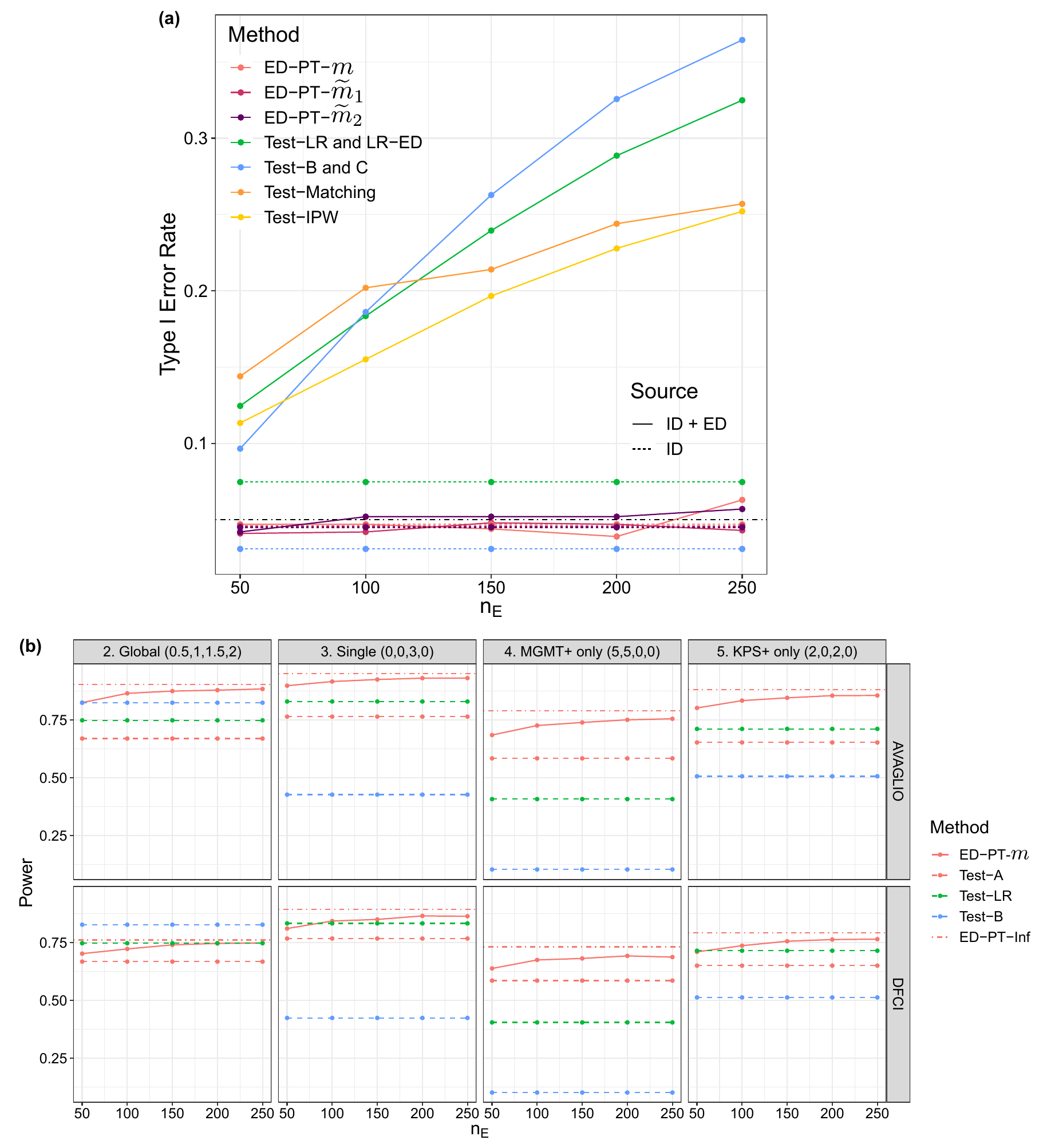}
 	\caption{Comparative analyses with {\it in silico} RCTs and EDs. Type I error rates (a) and power (b) of our ED-PT and alternative testing procedures as $n_E$ increases. In Panel (a), we illustrate the type I error rates of all testing procedures when the ED are generated using the  DFCI electronic health records. The dash-dotted black line indicates the nominal level $\alpha = 0.05$. In Panel (b), the top row illustrates results when the ED are generated using the AVAGLIO trial, while for the bottom row the ED were generated using the DFCI electronic health records.}
 	\label{fig:plot_gbm}
 \end{figure}

\noindent {\it Results.} 
Figure \ref{fig:plot_gbm}(a) shows the type I error rates of all testing procedures in Scenario 1 with ED generated using the DFCI electronic health records. We find that Test-C and Test-LR-ED, which use  the ID and the ED, have inflated type I error rates (Test-C: $0.10$ for $n_E = 50$, and up to $0.36$ when $n_E = 250$; Test-LR-ED: 0.12 for $n_E=50$, and up to $0.32$ when $n_E=250$). Also Test-Matching and Test-IPW  present inflated type I error rates  that are likely due to unmeasured confounding or other types of 
 unadjusted  discrepancies between the conditional distributions $p(\cdot|x_i,a_i = 0)$ and $p_E(\cdot|x_{E,i},a_{E,i} = 0)$. In contrast the permutation tests with (ED-PT-$m$, ED-PT-$\tilde m_1$, and ED-PT-$\tilde m_2$) and without  (Test-A) ED control the type I error rate close to the nominal $\alpha=0.05$ level. 
 
When negative treatment effects are present (Scenario 6 in Table \ref{tab:app-scen}), all testing procedures except ED-PT-$\tilde m_1(\mathcal D)$ and ED-PT-$\tilde m_2(\mathcal D)$ have high rejection probabilities (see Figure SM6). This is expected because the  results (reject $H_0$ or not) of several testing procedures (Test-LR, Test-LR-ED, Test-B and Test-C) do not depend on  the sign of the estimated treatment effects. On the other hand, in  Scenario 6 the ED-PTs based on the modified statistics $\tilde m_1(\cdot )$ and $\tilde m_2(\cdot )$ have a rejection rate close to 0.05. Also, for Test-B and Test-C, the rejection rate becomes smaller than 2.5\% when we modify the rejection region to implement one-sided testing.

Based on the results on the control of false positives, we focused on ED-PT-$m$, Test-A, Test-LR and Test-B for power comparisons. 
Figure \ref{fig:plot_gbm} illustrates the power of these tests in Scenario 2-5. 
The top row illustrates results when we use the AVAGLIO trial to generate both the {\it in silico} RCTs and the ED
while the bottom row reports results when we use the DFCI electronic health records to generate the ED (see point ii of the schema used to generate RCT and external data). 
The top panels show an ideal setting, $p(\cdot|x_i,a_i = 0) = p_E(\cdot|x_{E,i},a_{E,i} = 0)$  when $x_i=x_{E,i}$, whereas the bottom panels  provide more realistic  evaluations of the ED-PT procedure when $p(\cdot|x_i,a_i = 0) \neq p_E(\cdot|x_{E,i},a_{E,i} = 0)$ for some $x_i=x_{E,i}$.
In this second row of panels, with  the ED  generated using the DFCI electronic health records,  we have potential distortion mechanisms \citep{10.1158/1078-0432.CCR-22-3524} such as unmeasured confounding or subtle differences in the definition of the outcomes.

We vary the size of the ED ($n_E$) as indicated by the x-axis to examine its impact on the power of our ED-PT. The solid curve indicates ED-PT-$m$ and the dashed curves correspond to  three alternative tests  (Test-A in red, Test-LR in green, and Test-B in blue). The dash-dotted red curve in Figure \ref{fig:plot_gbm} indicates the power of the ED-PT test when the sample size $n_E$ of the ED diverges ($n_E \rightarrow +\infty$; see Section SM10 in the supplementary materials for details). We did not include the tests that failed to control the type I error rate at the nominal $\alpha$ level.
 
 For all configurations with heterogeneous treatment effects (Scenarios 2 to 5 in Table \ref{tab:app-scen}) we observe a gain in power for the ED-PT-$m$ compared to Test-A. As expected  the  power increase of ED-PT-$m$ compared to Test-A is larger when the ED are generated from the AVAGLIO study and smaller when the ED are generated from the DFCI electronic health records. Except for Scenario 2, where all four subgroups have positive treatment effects, ED-PT-$m$ with a moderate $n_E$ ($n_E>100$) outperforms the other procedures. The improvements in power of ED-PT-$m$ compared to the best performing ID-only approaches in Panel (b) vary across scenarios from 3.7\% (Scenario 3, DFCI  electronic health records used to generate ED) to 29.3\% (Scenario 4, AVAGLIO trial used to generate ED). These increments in power can be attained by the best ID-only testing procedures if the size $n$ of the ID increases between 10\%  (Scenario 3)  and 33\% in (Scenario 4). In Scenario 2  the ED-PT is the most powerful test when the ED are generated using the AVAGLIO trial, but it is 9.6\% less powerful than Test-B when the ED are generated using the DFCI electronic health records. These results  suggest the importance of selecting adequate ED (e.g., previous RCTs or electronic health record), avoiding obsolete data repository and potential distortion mechanisms  \citep{ventz2019design}. For all scenarios, we observe that the power of ED-PT-$m$ with $n_E=250$ is close to the power of ED-PT-$m$ with $n_E \rightarrow \infty$ (ED-PT-Inf).
 
\section{Discussion}
Modified versions of the test statistics $\tilde m_1$ and $\tilde m_2$ can be used to detect negative treatment effects, as illustrated in Sections SM9.1 and SM9.2 of the supplementary materials. When the signs of the treatment effects vary across patient subgroups---positive for some and negative for others---critical decisions must be made throughout drug development. As examples of relevant decisions we mention recommending a phase 3 confirmatory RCT based on phase 2 results, and selecting a phase 3 design with appropriate eligibility criteria and adequate sample size. A comprehensive methodological toolbox is essential to support these decisions. This includes procedures for testing, the primary focus of this manuscript, as well as approaches for estimation and prediction. We also emphasize that context-specific knowledge, such as previous results from preclinical models that characterized the experimental therapy’s mechanism of action, can be fundamental to support key decisions during clinical trials.

\section*{Acknowledgement}
 The authors are grateful to Sergio Bacallado for helpful discussions. Funding was provided by the
National Institutes of Health (NIH) grant R01 LM013352 and Project Data Sphere. Steffen Ventz was supported by the National Cancer
Institute (5P30CA077598-23), a DSI Grant from the Minnesota Supercomputing Institute, and a Medtronic
Faculty Fellowship.

\section*{Supplementary materials}
\label{SM}
Supplementary materials include proofs for all propositions, additional simulations, and additional considerations on the method that we introduced.

\bibliographystyle{chicago}
\bibliography{99BIB}

\begin{thebibliography}{}

\bibitem[\protect\citeauthoryear{Arel-Bundock, Greifer, and Heiss}{Arel-Bundock
  et~al.}{2024}]{arel2024interpret}
Arel-Bundock, V., N.~Greifer, and A.~Heiss (2024).
\newblock How to interpret statistical models using marginaleffects for r and
  python.
\newblock {\em Journal of Statistical Software\/}~{\em 111}, 1--32.

\bibitem[\protect\citeauthoryear{Berger}{Berger}{2013}]{berger2013statistical}
Berger, J.~O. (2013).
\newblock {\em Statistical decision theory and Bayesian analysis}.
\newblock Springer.

\bibitem[\protect\citeauthoryear{Berger, Wang, and Shen}{Berger
  et~al.}{2014}]{berger2014bayesian}
Berger, J.~O., X.~Wang, and L.~Shen (2014).
\newblock A bayesian approach to subgroup identification.
\newblock {\em J Biopharm Stat\/}~{\em 24\/}(1), 110--129.

\bibitem[\protect\citeauthoryear{Bonetti and Gelber}{Bonetti and
  Gelber}{2004}]{bonetti2004patterns}
Bonetti, M. and R.~D. Gelber (2004).
\newblock Patterns of treatment effects in subsets of patients in clinical
  trials.
\newblock {\em Biostatistics\/}~{\em 5\/}(3), 465--481.

\bibitem[\protect\citeauthoryear{Brown, Herson, Atkinson, and Rozell}{Brown
  et~al.}{1987}]{brown1987projection}
Brown, B.~W., J.~Herson, E.~N. Atkinson, and M.~E. Rozell (1987).
\newblock Projection from previous studies: a bayesian and frequentist
  compromise.
\newblock {\em Controlled clinical trials\/}~{\em 8\/}(1), 29--44.

\bibitem[\protect\citeauthoryear{Chen, Zhang, Gan, Wang, Lee, et~al.}{Chen
  et~al.}{2018}]{chen2018novel}
Chen, X., M.~Zhang, H.~Gan, H.~Wang, J.-H. Lee, et~al. (2018).
\newblock A novel enhancer regulates mgmt expression and promotes temozolomide
  resistance in glioblastoma.
\newblock {\em Nat Commun\/}~{\em 9\/}(1), 2949.

\bibitem[\protect\citeauthoryear{Chinot, Wick, Mason, Henriksson, Saran,
  et~al.}{Chinot et~al.}{2014}]{doi:10.1056/NEJMoa1308345}
Chinot, O.~L., W.~Wick, W.~Mason, R.~Henriksson, F.~Saran, et~al. (2014).
\newblock Bevacizumab plus radiotherapy–temozolomide for newly diagnosed
  glioblastoma.
\newblock {\em NEJM\/}~{\em 370\/}(8), 709--722.
\newblock PMID: 24552318.

\bibitem[\protect\citeauthoryear{Chu and Yuan}{Chu and
  Yuan}{2018}]{chu2018blast}
Chu, Y. and Y.~Yuan (2018).
\newblock Blast: Bayesian latent subgroup design for basket trials accounting
  for patient heterogeneity.
\newblock {\em JRSS-C: Applied Statistics\/}~{\em 67\/}(3), 723--740.

\bibitem[\protect\citeauthoryear{Dar, Johansson, Nordenskj{\"o}ld, Iftimi, Yau,
  et~al.}{Dar et~al.}{2021}]{dar2021assessment}
Dar, H., A.~Johansson, A.~Nordenskj{\"o}ld, A.~Iftimi, C.~Yau, et~al. (2021).
\newblock Assessment of 25-year survival of women with estrogen
  receptor--positive/erbb2-negative breast cancer treated with and without
  tamoxifen therapy.
\newblock {\em JAMA Network Open\/}~{\em 4\/}(6), e2114904--e2114904.

\bibitem[\protect\citeauthoryear{De~Bruijn}{De~Bruijn}{1981}]{de1981asymptotic}
De~Bruijn, N.~G. (1981).
\newblock {\em Asymptotic methods in analysis}, Volume~4.
\newblock Courier Corporation.

\bibitem[\protect\citeauthoryear{Ding, Feller, and Miratrix}{Ding
  et~al.}{2016}]{ding2016randomization}
Ding, P., A.~Feller, and L.~Miratrix (2016).
\newblock Randomization inference for treatment effect variation.
\newblock {\em Journal of the Royal Statistical Society Series B: Statistical
  Methodology\/}~{\em 78\/}(3), 655--671.

\bibitem[\protect\citeauthoryear{Ding, Feller, and Miratrix}{Ding
  et~al.}{2019}]{ding2019decomposing}
Ding, P., A.~Feller, and L.~Miratrix (2019).
\newblock Decomposing treatment effect variation.
\newblock {\em Journal of the American Statistical Association\/}~{\em
  114\/}(525), 304--317.

\bibitem[\protect\citeauthoryear{Freidlin, McShane, and Korn}{Freidlin
  et~al.}{2010}]{freidlin2010randomized}
Freidlin, B., L.~M. McShane, and E.~L. Korn (2010).
\newblock Randomized clinical trials with biomarkers: design issues.
\newblock {\em JNCI\/}~{\em 102\/}(3), 152--160.

\bibitem[\protect\citeauthoryear{Freidlin, McShane, Polley, and Korn}{Freidlin
  et~al.}{2012}]{freidlin2012randomized}
Freidlin, B., L.~M. McShane, M.-Y.~C. Polley, and E.~L. Korn (2012).
\newblock Randomized phase ii trial designs with biomarkers.
\newblock {\em JCO\/}~{\em 30\/}(26), 3304.

\bibitem[\protect\citeauthoryear{Haslam, Kim, and Prasad}{Haslam
  et~al.}{2021}]{haslam2021updated}
Haslam, A., M.~Kim, and V.~Prasad (2021).
\newblock Updated estimates of eligibility for and response to genome-targeted
  oncology drugs among us cancer patients, 2006-2020.
\newblock {\em Annals of Oncology\/}~{\em 32\/}(7), 926--932.

\bibitem[\protect\citeauthoryear{Ho, Imai, King, and Stuart}{Ho
  et~al.}{2011}]{ho2011matchit}
Ho, D., K.~Imai, G.~King, and E.~A. Stuart (2011).
\newblock Matchit: nonparametric preprocessing for parametric causal inference.
\newblock {\em Journal of statistical software\/}~{\em 42}, 1--28.

\bibitem[\protect\citeauthoryear{Kennedy, Torgerson, Campbell, and
  Grant}{Kennedy et~al.}{2017}]{kennedy2017subversion}
Kennedy, A.~D., D.~J. Torgerson, M.~K. Campbell, and A.~M. Grant (2017).
\newblock Subversion of allocation concealment in a randomised controlled
  trial: a historical case study.
\newblock {\em Trials\/}~{\em 18\/}(1), 1--6.

\bibitem[\protect\citeauthoryear{Lauko, Lo, Ahluwalia, and Lathia}{Lauko
  et~al.}{2022}]{lauko2022cancer}
Lauko, A., A.~Lo, M.~S. Ahluwalia, and J.~D. Lathia (2022).
\newblock Cancer cell heterogeneity \& plasticity in glioblastoma and brain
  tumors.
\newblock In {\em Seminars in Cancer Biology}, Volume~82, pp.\  162--175.
  Elsevier.

\bibitem[\protect\citeauthoryear{Lehmann and Romano}{Lehmann and
  Romano}{2005}]{lehmann2005testing}
Lehmann, E. and J.~Romano (2005).
\newblock {\em Testing statistical hypotheses}, Volume~3.
\newblock Springer.

\bibitem[\protect\citeauthoryear{Li, Morgan, and Zaslavsky}{Li
  et~al.}{2018}]{li2018balancing}
Li, F., K.~L. Morgan, and A.~M. Zaslavsky (2018).
\newblock Balancing covariates via propensity score weighting.
\newblock {\em JASA\/}~{\em 113\/}(521), 390--400.

\bibitem[\protect\citeauthoryear{Liau, Ashkan, Brem, Campian, and
  Trusheim}{Liau et~al.}{2023}]{liau2023association}
Liau, L.~M., K.~Ashkan, S.~Brem, J.~L. Campian, and J.~E. a.~o. Trusheim
  (2023).
\newblock Association of autologous tumor lysate-loaded dendritic cell
  vaccination with extension of survival among patients with newly diagnosed
  and recurrent glioblastoma.
\newblock {\em JAMA Onc\/}~{\em 9\/}(1), 112--121.

\bibitem[\protect\citeauthoryear{Liu}{Liu}{2018}]{liu2018assessment}
Liu, F. (2018).
\newblock Assessment of bayesian expected power via bayesian bootstrap.
\newblock {\em Stat Med\/}~{\em 37\/}(24), 3471--3485.

\bibitem[\protect\citeauthoryear{Morita and M{\"u}ller}{Morita and
  M{\"u}ller}{2017}]{morita2017bayesian}
Morita, S. and P.~M{\"u}ller (2017).
\newblock Bayesian population finding with biomarkers in a randomized clinical
  trial.
\newblock {\em Biometrics\/}~{\em 73\/}(4), 1355--1365.

\bibitem[\protect\citeauthoryear{Murphy}{Murphy}{2003}]{murphy2003optimal}
Murphy, S.~A. (2003).
\newblock Optimal dynamic treatment regimes.
\newblock {\em JRSS-B: Statistical Methodology\/}~{\em 65\/}(2), 331--355.

\bibitem[\protect\citeauthoryear{Nugent, Madabushi, Buch, Peiris, Crentsil,
  et~al.}{Nugent et~al.}{2021}]{nugent2021heterogeneity}
Nugent, B.~M., R.~Madabushi, B.~Buch, V.~Peiris, V.~Crentsil, et~al. (2021).
\newblock Heterogeneity in treatment effects across diverse populations.
\newblock {\em Pharmaceutical Statistics\/}~{\em 20\/}(5), 929--938.

\bibitem[\protect\citeauthoryear{Rahman, Ventz, McDunn, Louv, Reyes-Rivera,
  et~al.}{Rahman et~al.}{2021}]{rahman2021leveraging}
Rahman, R., S.~Ventz, J.~McDunn, B.~Louv, I.~Reyes-Rivera, et~al. (2021).
\newblock Leveraging external data in the design and analysis of clinical
  trials in neuro-oncology.
\newblock {\em Lancet Onc\/}~{\em 22\/}(10), e456--e465.

\bibitem[\protect\citeauthoryear{Rahman, Ventz, Redd, Cloughesy, Alexander,
  et~al.}{Rahman et~al.}{2023}]{10.1158/1078-0432.CCR-22-3524}
Rahman, R., S.~Ventz, R.~Redd, T.~Cloughesy, B.~M. Alexander, et~al. (2023).
\newblock {Accessible Data Collections for Improved Decision Making in
  Neuro-Oncology Clinical Trials}.
\newblock {\em CCR\/}~{\em 29\/}(12), 2194--2198.

\bibitem[\protect\citeauthoryear{Rigdon, Baiocchi, and Basu}{Rigdon
  et~al.}{2018}]{rigdon2018preventing}
Rigdon, J., M.~Baiocchi, and S.~Basu (2018).
\newblock Preventing false discovery of heterogeneous treatment effect
  subgroups in randomized trials.
\newblock {\em Trials\/}~{\em 19\/}(1), 1--15.

\bibitem[\protect\citeauthoryear{Russo, Ventz, Wang, and Trippa}{Russo
  et~al.}{2023}]{russo2023inference}
Russo, M., S.~Ventz, V.~Wang, and L.~Trippa (2023).
\newblock Inference in response-adaptive clinical trials when the enrolled
  population varies over time.
\newblock {\em Biometrics\/}~{\em 79\/}(1), 381--393.

\bibitem[\protect\citeauthoryear{Sherman, Anderson, Dal~Pan, Gray, Gross,
  et~al.}{Sherman et~al.}{2016}]{sherman2016real}
Sherman, R.~E., S.~A. Anderson, G.~J. Dal~Pan, G.~W. Gray, T.~Gross, et~al.
  (2016).
\newblock Real-world evidence—what is it and what can it tell us.
\newblock {\em NEJM\/}~{\em 375\/}(23), 2293--2297.

\bibitem[\protect\citeauthoryear{Slevin, Clark, Joel, Malik, Osborne,
  et~al.}{Slevin et~al.}{1989}]{slevin1989randomized}
Slevin, M., P.~Clark, S.~Joel, S.~Malik, R.~Osborne, et~al. (1989).
\newblock A randomized trial to evaluate the effect of schedule on the activity
  of etoposide in small-cell lung cancer.
\newblock {\em JCO\/}~{\em 7\/}(9), 1333--1340.

\bibitem[\protect\citeauthoryear{Stupp, Mason, Van Den~Bent, Weller, Fisher,
  et~al.}{Stupp et~al.}{2005}]{stupp2005radiotherapy}
Stupp, R., W.~P. Mason, M.~J. Van Den~Bent, M.~Weller, B.~Fisher, et~al.
  (2005).
\newblock Radiotherapy plus concomitant and adjuvant temozolomide for
  glioblastoma.
\newblock {\em NEJM\/}~{\em 352\/}(10), 987--996.

\bibitem[\protect\citeauthoryear{Tierney and Kadane}{Tierney and
  Kadane}{1986}]{tierney1986accurate}
Tierney, L. and J.~B. Kadane (1986).
\newblock Accurate approximations for posterior moments and marginal densities.
\newblock {\em JASA\/}, 82--86.

\bibitem[\protect\citeauthoryear{van~der Wal and Geskus}{van~der Wal and
  Geskus}{2011}]{van2011ipw}
van~der Wal, W.~M. and R.~B. Geskus (2011).
\newblock ipw: an r package for inverse probability weighting.
\newblock {\em Journal of Statistical Software\/}~{\em 43}, 1--23.

\bibitem[\protect\citeauthoryear{Ventz, Comment, Louv, Rahman, Wen,
  et~al.}{Ventz et~al.}{2022}]{ventz2022use}
Ventz, S., L.~Comment, B.~Louv, R.~Rahman, P.~Y. Wen, et~al. (2022).
\newblock The use of external control data for predictions and futility interim
  analyses in clinical trials.
\newblock {\em Neuro Onc\/}~{\em 24\/}(2), 247--256.

\bibitem[\protect\citeauthoryear{Ventz, Khozin, Louv, Sands, Wen, Rahman,
  Comment, Alexander, and Trippa}{Ventz et~al.}{2022}]{ventz2022design}
Ventz, S., S.~Khozin, B.~Louv, J.~Sands, P.~Y. Wen, R.~Rahman, L.~Comment,
  B.~M. Alexander, and L.~Trippa (2022).
\newblock The design and evaluation of hybrid controlled trials that leverage
  external data and randomization.
\newblock {\em Nat Commun\/}~{\em 13\/}(1), 5783.

\bibitem[\protect\citeauthoryear{Ventz, Lai, Cloughesy, Wen, Trippa, and
  Alexander}{Ventz et~al.}{2019}]{ventz2019design}
Ventz, S., A.~Lai, T.~F. Cloughesy, P.~Y. Wen, L.~Trippa, and B.~M. Alexander
  (2019).
\newblock Design and evaluation of an external control arm using prior clinical
  trials and real-world data.
\newblock {\em CCR\/}~{\em 25\/}(16), 4993--5001.

\bibitem[\protect\citeauthoryear{Wager and Athey}{Wager and
  Athey}{2018}]{wager2018estimation}
Wager, S. and S.~Athey (2018).
\newblock Estimation and inference of heterogeneous treatment effects using
  random forests.
\newblock {\em JASA\/}~{\em 113\/}(523), 1228--1242.

\bibitem[\protect\citeauthoryear{Wang, Zhang, and Tiwari}{Wang
  et~al.}{2023}]{wang2023propensity}
Wang, J., H.~Zhang, and R.~Tiwari (2023).
\newblock A propensity-score integrated approach to bayesian dynamic power
  prior borrowing.
\newblock {\em Stat Biopharm Research\/}, 1--23.

\bibitem[\protect\citeauthoryear{Wang, Schoenfeld, Hoeppner, and Evins}{Wang
  et~al.}{2015}]{wang2015detecting}
Wang, R., D.~A. Schoenfeld, B.~Hoeppner, and A.~E. Evins (2015).
\newblock Detecting treatment-covariate interactions using permutation methods.
\newblock {\em Stat Med\/}~{\em 34\/}(12), 2035--2047.

\bibitem[\protect\citeauthoryear{White}{White}{1980}]{white1980heteroskedasticity}
White, H. (1980).
\newblock A heteroskedasticity-consistent covariance matrix estimator and a
  direct test for heteroskedasticity.
\newblock {\em Econometrica\/}, 817--838.

\bibitem[\protect\citeauthoryear{Xu, Zhang, Zhang, Bian, and Wang}{Xu
  et~al.}{2023}]{xu2023machine}
Xu, J., H.~Zhang, H.~Zhang, J.~Bian, and F.~Wang (2023).
\newblock Machine learning enabled subgroup analysis with real-world data to
  inform clinical trial eligibility criteria design.
\newblock {\em Scientific Reports\/}~{\em 13\/}(1), 613.

\bibitem[\protect\citeauthoryear{Yang, Li, Starks, Hernandez, Mentz,
  et~al.}{Yang et~al.}{2020}]{yang2020sample}
Yang, S., F.~Li, M.~A. Starks, A.~F. Hernandez, R.~J. Mentz, et~al. (2020).
\newblock Sample size requirements for detecting treatment effect heterogeneity
  in cluster randomized trials.
\newblock {\em Stat Med\/}~{\em 39\/}(28), 4218--4237.

\bibitem[\protect\citeauthoryear{Zeileis, K\"oll, and Graham}{Zeileis
  et~al.}{2020}]{zeileis2020}
Zeileis, A., S.~K\"oll, and N.~Graham (2020).
\newblock Various versatile variances: An object-oriented implementation of
  clustered covariances in {R}.
\newblock {\em Journal of Statistical Software\/}~{\em 95\/}(1), 1--36.

\bibitem[\protect\citeauthoryear{Ziegler, Koch, Krockenberger, and
  Gro{\ss}hennig}{Ziegler et~al.}{2012}]{ziegler2012personalized}
Ziegler, A., A.~Koch, K.~Krockenberger, and A.~Gro{\ss}hennig (2012).
\newblock Personalized medicine using dna biomarkers: a review.
\newblock {\em Human genetics\/}~{\em 131}, 1627--1638.

\end{thebibliography}


\begin{thebibliography}{}

\bibitem[\protect\citeauthoryear{Ding}{Ding}{1992}]{ding1992algorithm}
Ding, C.~G. (1992).
\newblock Algorithm as 275: computing the non-central $\chi$ 2 distribution
  function.
\newblock {\em JRSS-C: Applied Statistics\/}~{\em 41\/}(2), 478--482.

\bibitem[\protect\citeauthoryear{Feller}{Feller}{1968}]{FellerW1968AitP}
Feller, W. (1968).
\newblock An introduction to probability theory and its applications. volume 1
  (3rd edition).
\newblock pp.\  525P--525P.

\bibitem[\protect\citeauthoryear{Lehmann and Romano}{Lehmann and
  Romano}{2005}]{lehmann2005testing}
Lehmann, E. and J.~Romano (2005).
\newblock {\em Testing statistical hypotheses}, Volume~3.
\newblock Springer.

\bibitem[\protect\citeauthoryear{Lehmann and Stein}{Lehmann and
  Stein}{1949}]{lehmann1949theory}
Lehmann, E.~L. and C.~Stein (1949).
\newblock On the theory of some non-parametric hypotheses.
\newblock {\em The Annals of Mathematical Statistics\/}~{\em 20\/}(1), 28--45.

\end{thebibliography}

\end{document}


\maketitle
	
	\section{Proof of Proposition 1}
	The control of type I error rate at the nomial level $\alpha$ for the randomized test $\phi$ follows directly from Theorem 15.2.1 in \cite{lehmann2005testing} by noting that $H_0$ is a randomization hypothesis, that is, $p(y,x,a)$ is invariant under permutations of $a$ whenever $p\in\mathcal P_0$.
	
	We prove the optimality of $\phi$ in terms of BEP in two steps:
	\begin{enumerate}
		\item[Step (i).] From Theorem 2 of \cite{lehmann1949theory}, for any $p' \notin  \mathcal P_0$, the randomzied permutation test $\phi'$, which is based on the test statistic $m_{p'}(D)=p'(D)$, is the most powerful $\alpha$-level test for the point alternative $\widetilde H_{1}:  p=p',$ i.e., $ \mathbb E_{p'}[\xi(\mathcal D)] \leq  \mathbb E_{p'}[\phi'(\mathcal D)]$ for any other $\alpha$-level test $\xi$.
		\item[Step (ii).] Since any test $\xi$ of $H_0$ satisfies that
		\begin{align*}
			\text{BEP}(\xi) & =  \mathbb{E}_{(X,A) \sim p}
			\left [ 
			\int \left( \int \xi(\mathcal D) q_\theta(Y|X,A)dY \right) \pi(\theta|\mathcal D_E)d\theta
			\right ]
			\label{eq:BEP}
			= \mathbb{E}_{(X,A) \sim p} [ \mathbb{E}_{m(\mathcal D) }
			[\xi(\mathcal D)  | X,A ]],
		\end{align*}
		it leads to the maximum $\text{BEP}(\xi) $ if $\mathbb{E}_{m(\mathcal D) }\left [  \xi(\mathcal D)  | X,A \right ]$  is maximized
		for all $(X,A)$ on a set of probability one.  
	\end{enumerate} 
	
	By Step (i), the permutation test  $\phi$ based on $m(\mathcal D) = \int  q_\theta(Y|X,A) \pi(\theta|\mathcal D_E) d \theta$ maximizes $ \mathbb{E}_{m(\mathcal D) }	    [\xi(\mathcal D)  | X,A ]$ for all $(X,A)$ on a set of probability one and hence by Step (ii), it also maximizes the $\text{BEP}$. $\square$

	\section{Properties of the ED-PT}
	We discuss four properties of the ED-PT:
\begin{enumerate}
	\item[1.]  \textit{False positives are also controlled at the $\alpha$ level  
	in relevant scenarios where the samples of the ID and/or the ED are not independent and identically distributed.}
	The proof of Proposition 1 builds on Theorem 15.2.1 in
	\cite{lehmann2005testing}.
	This theorem states that, if  the  distribution
	$p(y,x,a)$ is invariant to permutations of $a$,
	then a randomized test $\xi(\mathcal D)$,  defined as in equation (4) of the main manuscript with an arbitrary test statistic $T(\mathcal D)$, satisfies $\mathbb E_p[\xi(\mathcal D)] = \alpha$ for any $p\in \mathcal P_0$. 
	Recall that $\mathcal P_0$ coincides with our $H_0$ in (1). This directly implies that $E_p[\phi(D)] = \alpha$ under the null hypothesis (i.e., when $p\in \mathcal P_0$).
	
	The permutation invariance
	of  $p(y,x,a)$
	holds in the absence of treatment effects when a RCT, say with balanced randomization  $A_i\overset{iid}{\sim} \text{Bernoulli}(1/2)$, collects independent and identically distributed (\textit{iid})  replicates $(X_i,Y_i)$. Here, we emphasize that the same invariance property also holds in relevant scenarios where the triplets $(X_i,Y_i,A_i)$ in the RCT data are not \textit{iid} replicates. We illustrate this through an example  with
	$
	A_i\overset{iid}{\sim} \text{Bernoulli}(1/2),$ and a trend in the  outcomes,
	$$
	Y_i = \beta^\intercal X_i + R_i + \epsilon_i,
	$$
	where $R_i$ is the enrollment time of patient $i$, and $\epsilon_i \overset{iid}{\sim} N(0,1)$. Note  the absence of  treatment effects. In this easy-to-interpret example,  $p(y,x,a)$ is invariant with respect to  permutations of $a$.
	
	\item[2.]   \textit{The control of false positive results for $\phi$ at the $\alpha$ level is maintained if the test statistic $m(\mathcal D)$ is computed using approximation methods.} 
	We consider an approximation $\hat m(\mathcal D)$,  a function of the RCT data $\mathcal D$.   
	Recall that under the null hypothesis $p(y,x,a)$ is invariant to  permutations of $a$. 
	Therefore, based on the same argument used to  demonstrate the control of false positives of $\phi(\mathcal D)$ (Theorem 15.2.1 in
	\citealp{lehmann2005testing}),  the randomized test $\phi(\mathcal D)$ with test statistic $\hat m(\mathcal D)$ controls the false positive rate at level $\alpha$.
	
	\item[3.] \textit{The randomized and non-randomized tests (i.e., $\phi \text{ and } \widetilde\phi$) have nearly identical type I and type II error rates when $J$ diverges.}  
	We condition on the data $(\mathcal D, \mathcal D_E)$ and 
	consider the sequence  of random permutations $\tau_1, \tau_2,\ldots,\tau_J$ in Algorithm 1.
	These permutations are independent and  uniformly distributed over $\mathcal T$.    
	Based on the strong law of large numbers, as $J\to\infty$,
	$$
	\frac{1 + \sum_1^J \mathbb I[m_j \geq m(\mathcal D)]}{1+J} \overset{a.s.}{\to} \frac{\sum_\tau \mathbb I[m(\mathcal D^{(\tau)})\geq m(\mathcal D)]}{n!}.
	$$
	In different words, if $m(\mathcal D) \neq t_\alpha$,  then $\widetilde \phi(\mathcal D)\overset{a.s.}{\to} \phi(\mathcal D)$. Similarly, if
	$m(\mathcal D) = t_\alpha$,
	$$
	\widetilde \phi(\mathcal D) - \phi(\mathcal D) \overset{a.s.}{\to} - \frac{\alpha n! - \sum_{\tau \in \mathcal T}\mathbb I[m(\mathcal D^{(\tau)} > t_\alpha)]}{\sum_{\tau\in \mathcal T} \mathbb I[m(\mathcal D^{(\tau)})=t_\alpha]}.
	$$
	
	In most applications of our permutation procedure and others, it is unlikely to observe ties. That is, $\text{Pr}[m(\mathcal D^{(\tau_j)}) = m(\mathcal D^{(\tau_{j'})})] \approx 0$
	for any $j\neq j'$.
	This implies that the probability of the event $m(\mathcal D) = t_\alpha$ is negligible, which, together  with the above equations, indicates the similarity between the randomized test $\phi(\mathcal D)$ and the more practical non-randomized version $\widetilde\phi(\mathcal D)$. Algorithm 1 allows us to easily 
	evaluate if  
	$\text{Pr}[m(\mathcal D^{(\tau_{j})}) = m(\mathcal D^{(\tau_{j'})})] \approx 0$.
	
	\item[4.] \textit{The proposed ED-PT is applicable when the external dataset includes both patients treated with the experimental and control therapies.} 
	In the ED-PT procedure,  external  information  is  incorporated through the  posterior $\pi(\theta|\mathcal D_E)$, as indicated by the definition of $m(\mathcal D)$ in (2).  In the examples that we discussed, the ED includes only  patients treated with the control therapy. Nonetheless,
	in settings where the    external dataset $\mathcal D_E$  also contains
	pre-treatment covariates and outcomes of  patients treated  with the experimental 
	therapy, the ED-PT procedure can be used with minimal changes.
	Indeed, it is sufficient to  use  the  posterior $\pi(\theta|\mathcal D_E)$, informed by the available data including patients treated with  the experimental ($A_{E,i}=1$) and/or control ($A_{E,i}=0$) therapy, which defines
	the test statistics $m(\mathcal D)$. We do not need   any  other change to apply Algorithm 1. Additionally, we emphasize that the statement in Proposition 1 remains valid, since the proof only involves a generic posterior of the parameters $\pi(\theta|\mathcal D_E)$. In sum,  to  apply the ED-PT, 
	we only need a joint Bayesian model of the RCT data and the available ED.
\end{enumerate}
	
	\section{Proof of Proposition 2}
	\label{sec:prop2}
	Denote the number of responders in experimental and control arms after permuting $A$ by $s_1'$ and $s_0'$ respectively. They satisfy the condition that $s_1' + s_0' = s$. The conditional likelihood $m(\mathcal D^{(\tau)})$ is a function of $s_1', s, n_1$, and $n_0$. However, across the permutations of our testing procedure, the values of  $n_1$, $n_0$, and $s$ remain identical. For this reason, in this proof, we simplify the notation and use $m(s_1')$
	 to indicate the value of $m(D^{(\tau)})$,  which is  expressed as a function of  $s_1'$.
	
	To derive the exact p-value, we first obtain the set $\mathcal N = \{s_1': m(s_1') < m(s_1)\} = (s_\text{min}, s_\text{max})$. It follows that $\texttt{pv} = 1 - \text{Pr}(s_1'\in (s_\text{min}, s_\text{max}))$, where the probability measure corresponds to the permutation distribution of $s_1'$. Note that $m(s_1') = C/h(s_1';n_1, (1 + r + r_E)n_1, s+s_E)$, with $C$ a permutation invariant constant and $h(\cdot;\cdot)$ the probability mass function (pmf) of a hyper-geometric distribution $H(n_1, (1 + r + r_E)n_1, s+s_E)$. As $n_1\to\infty$, $(s+s_E)/n_1 \overset{p}{\to} w_0(1+r+r_E)$ using Assumption (A1) and (A2). Based on standard results of normal approximation of a hyper-geometric distribution (see \citealp[page 194]{FellerW1968AitP}), $h(s_1';n_1, (1 + r + r_E)n_1, s+s_E)$ is asymptotically equivalent to the probability density function (pdf) of a normal distribution with mean $(s+s_E)/(1+r+r_E)$. Since the pdf of a normal distribution is symmetric around its mean, we have $s_\text{min}/ \min(s_1, 2(s+s_E)/(r+r_E+1) - s_1) \overset{p}{\to} 1$ and $s_{\text{max}} / \max(s_1, 2(s+s_E)/(r+r_E+1) - s_1) \overset{p}{\to} 1$.
	
	The permutation distribution of $s_1'$ is a hyper-geometric distribution $H(n_1,(1+r)n_1,s)$ and by the same normal approximation results mentioned above, its cumulative density function (cdf) is asymptotically equivalent to the cdf of a normal distribution. Specifically, $\mathcal H(s_1';n_1,(1+r)n_1,s) / \Phi(s_1';s/(r+1), sr((r+1)n_1-s)/((r+1)n_1 - 1)/(r+1)^2) \overset{p}{\to} 1$, where $\mathcal H$ and $\Phi$ are the cdfs of a hyper-geometric distribution and a normal distribution respectively. The proof is completed by using the asymptotic results on $s_{\text{min}}$ and $s_{\text{max}}$. $\square$
	
	\section{Proof of Proposition 3}
	The limiting power of ED-PT at level $\alpha$ can be written as the limit of the probability that the exact p-value is smaller than $\alpha$ as $n_1\to\infty$. Based on the results of Proposition 2 and Slutsky's theorem, we have
	$$
	g(r,r_E,a,b,w_0) = \lim_{n_1\to\infty} \text{Pr}(\texttt{pv}\leq \alpha) = \lim_{n_1\to\infty} \text{Pr}(\Phi(\max(z_1,z_0)) - \Phi(\min(z_1,z_0) \geq 1-\alpha),
	$$
	where $z_1$ and $z_0$ are
	$$
	z_1 = \frac{s_1 - s/(r+1)}{\sqrt{sr(1-w_0)/(r+1)^2}},~~z_0 = \frac{2(s+s_E)/(r+r_E+1) - s_1 - s/(r+1)}{\sqrt{sr(1-w_0)/(r+1)^2}}.
	$$
	Based on central limit theorem, we have
	\begin{equation}
		\sqrt{n_1}(
		s_1/n_1 - w_0,
		s_0/n_0 - w_0,
		s_E/n_E - w_0) \overset{d}{\to} N\left(
		(a,0,b), w_0(1-w_0)\text{diag}\{1, 1/r, 1/r_E\}.
		\right)
		\label{eq:x-clt}
	\end{equation}
	
	Let $x_1 = s_1/n_1$, $x_0 = s_0/n_0$ and $x_E = s_E/n_E$. We have
	\begin{gather*}
		z_1 = \sqrt{n_1}\cdot \sqrt{\frac{r}{1 - w_0}}\cdot \frac{x_1 - x_0}{\sqrt{x_1 + rx_0}},\\
		z_0 = \sqrt{n_1}\cdot \frac{1}{\sqrt{r(1-w_0)}}\cdot\frac{-(r+\frac{2r_E}{r+r_E+1})x_1 + \frac{r(r+1-r_E)}{r+r_E+1}x_0 + \frac{2r_E(r+1)}{r_E+r+1}x_E}{\sqrt{x_1 + rx_0}}.
	\end{gather*}
	Using \eqref{eq:x-clt} and delta method, we can derive the asymptotic distribution of $(z_1,z_0)$ and the results in Proposition 3 follow. $\square$
	
	\noindent \textbf{Approximation:} to obtain the approximation in (11), note that when $a>0$ is large and $b=0$, $\max(U_1,U_0) = U_1$ with high probability and $\Phi(\min(U_1,U_0))$ is negligible. Therefore, we can approximate $\Phi(\max(U_1,U_0)) - \Phi(\min(U_1,U_0))$ by $\Phi(U_1)$. With (10), we get
	$$
	Pr(\Phi(\max(U_1,U_0)) - \Phi(\min(U_1,U_0))>1-\alpha) \approx \Phi\left(\frac{\sqrt{r}a}{\sqrt{(r+1)w_0(1-w_0)}} - \Phi^{-1}(1-\alpha)\right).
	$$
	
	\section{Asymptotic analysis for normally distributed outcomes}
	We focus on the Scenario S1 in Section 2.5, with $X_i, X_{E,i} \in \{0,1\}^{K-1}$ indicating subgroups.
	The unknown outcome distributions ($p$ and $p_E$) are summarized by model (15), and the working model by expression (16). We derive the limiting power function when the population includes $K>1$ subgroups. 
	Similar to Section 2.4, we consider a sequence of $(\mathcal D, \mathcal D_E)$ pairs with increasing sample sizes 
	$n$ and $n_E$ such that:
	\begin{enumerate}
		\item[(B1)] The proportion $\rho_k$ of patients in subgroup $k = 1,\ldots, K$ is identical in the RCT and the ED populations.
		\item[(B2)] The randomization ratio $r$ is fixed and identical across patient subgroups. Similarly, the ratio $r_E$ between the
		number of 
		external control patients	
		and the number of patients assigned to the experimental arm in the RCT is fixed and does not vary 
		across subgroups.
		\item[(B3)] The subgroup-specific treatment effects are $(\gamma, \gamma 1_{K-1}+\gamma_1) = a/\sqrt{n_1}$ and the discrepancy between ED and RCT controls is captured by $(\beta_{0,0}, \beta_{0,0} 1_{K-1} + \beta_{0,1}) = b/\sqrt{n_1}$. Recall $n_1 = \sum_i A_i$.
		Note that $\gamma_1,\beta_{0,1}\in\mathbb R^{K-1}$, and $a,b\in\mathbb R^{K}$. 
		The parameters $a,b$ will remain the same for the whole sequence of $(\mathcal D, \mathcal D_E)$ pairs.
	\end{enumerate}
	We describe the asymptotic power function of $\widetilde\phi$ in the next Proposition.
	\begin{prop}
		Let $\rho = (\rho_1,\ldots,\rho_K)$, and assume $r$ and $r_E$ remain fixed as $n\to\infty$. Under the assumptions B1, B2, and B3, the asymptotic power function of the $\alpha$ level test $\widetilde\phi$ is
		\begin{equation}
			g(r,r_E,a,b,\rho) = 1 - \int F_{v,a,b}[\widetilde F^{-1}_{v,a,b}(1-\alpha)]\varphi(v)dv,
			\label{eq:asym_normal}
		\end{equation}
		where $\varphi$ denotes the density function of a standard $K-$variate normal distribution, and
		\begin{align*}
			F_{v,a,b}(t) &= \text{Pr}\left\{\sum_{k=1}^K \left[\sqrt{\frac{\rho_kr_E(1+r)}{1+r + r_E}}\left(\frac{r+r_E}{r_E}a_k - b_{k}\right) + v_k + \sqrt{\frac{r(1+r+r_E)}{r_E}}Z_k \right]^2\leq t\right\},\\
			\widetilde F_{v,a,b}(t) &= \text{Pr}\left\{\sum_{k=1}^K \left[\sqrt{\frac{\rho_kr_E(1+r)}{1+r + r_E}}\left(\frac{a_k}{1+r} - b_{k}\right)+ v_k + \sqrt{\frac{r(1+r+r_E)}{r_E}}Z_k
			\right]^2\leq t\right\}.
		\end{align*}
		Here $v=(v_1,\ldots,v_K)\in\mathbb R^K$, $t\in\mathbb R$ and $Z_1,\ldots,Z_K$ are independent standard normal random variables.
		\label{prop:asym_normal}
	\end{prop}
	
	See Section \ref{proof:p4} for a proof of Proposition SM1. We can compute expression \eqref{eq:asym_normal} with Monte Carlo simulations, based on the fact that $F_{v,a,b}$ and $\widetilde F_{v,a,b}$ are cumulative distribution functions of non-central chi-square random variables. These distributions have been studied extensively and algorithms \citep{ding1992algorithm} are available as R functions (\texttt{pchisq} and \texttt{rchisq}).

	\section{Additional simulation scenarios for binary outcomes}
	\subsection{ED-PT for binary outcomes and heterogeneous treatment effects}
	We consider a scenario with $K\geq 2$ subgroups and potential variations of the treatment effects  across subgroups. 
Similar to Section 2.4, the external dataset includes only patients treated with the control therapy. Let $X_i, X_{E,i}\in \{0,1\}^{K-1}$ be subgroup indicators for the RCT and the ED, respectively, as  in Section 2.5 (see Scenario S1). We describe the data distribution:
\begin{equation}
	\begin{aligned}
		A_i &\overset{iid}{\sim} \text{Bernoulli}[1/(1+r)],\\ 
		Y_i|A_i,X_i &\overset{ind}{\sim} \text{Bernoulli}(w_0 + \gamma A_i + \beta_1^\intercal X_i + A_i\gamma_1^\intercal X_i),\\
		Y_{E,i}|X_{E,i} &\overset{ind}{\sim} \text{Bernoulli}(w_0 + \beta_{0,0} + (\beta_1 + \beta_{0,1})^\intercal X_{E,i}).
	\end{aligned}
	\label{eq:binary-HTE}
\end{equation}
The interpretation of the parameters $[\gamma, \beta_0 = (\beta_{0,0}, \beta_{0,1}), \beta_1^\intercal,  \gamma_1]$ here and in Section 2.5 is similar. Also,  $w_0$ is the response probability of a control patient $i$ with $X_i = 0_{K-1}$.

Let $G_i  =  [1:K-1]^\intercal X_i+1$ and $G_{E,i} = [1:K-1]^\intercal X_{E,i}+1$, where $[1:K-1] = (1,2,\ldots,K-1)$. We apply our  permutation  procedure based on the following Bayesian working model: 
\begin{equation}
	\begin{aligned}
		Y_{i}|A_i = a,G_i = k,\theta &\overset{ind}{\sim}\text{Bernoulli}(\theta_{a,k}),\\
		Y_{E,i}|G_{E,i} = k , \theta &\overset{ind}{\sim} \text{Bernoulli}(\theta_{0,k}), \\
		\theta_{a,k}&\overset{iid}{\sim} U[0,1],
	\end{aligned}
	\label{eq:binary-HTE-working}
\end{equation}
where $a \in \{0,1\}$, $\theta=(\theta_{a,k}; a\in\{0,1\},\;\; k\in\{1,\ldots,K\})$ and $\theta_{a,k}$ indicates the probability of response for patients in subgroup $k$ assigned to treatment $a$. The test statistic $m(\mathcal D)$ in this case is
$$
m(\mathcal D) = \prod_{k = 1}^K \frac{s_{1,k}!(n_{1,k} - s_{1,k})!(s_{0,k} + s_{E,k})!(n_{0,k} + n_{E,k} - s_{0,k} - s_{E,k})!(n_{E,k} + 1)!}{(n_{1,k}+1)!(n_{E,k} + n_{0,k} + 1)!s_{E,k}!(n_{E,k} - s_{E,k})!},
$$
where
$n_{1,k}$ and $n_{0,k}$ are the numbers of patients in subgroup $k$ assigned  to  the experimental and control arms of the RCT, and 
$n_{E,k}$ is the size of subgroup $k$ in the ED. Also, $s_{1,k}$ and $s_{0,k}$ are the number of  responders in subgroup $k$ for the experimental and control arms of the RCT, while $s_{E,k}$ is the number of responders in subgroup $k$ in the ED.

We compare our permutation procedure (ED-PT) to four alternatives: mTest-A, mTest-B, mTest-C, and mTest-D, which are modified versions of   Test-A, Test-B, Test-C, and Test-D, respectively, in Section 2.4, to account for the presence of $K$ subgroups. 
 In particular, mTest-B, mTest-C, and mTest-D are based on the following logistic model:
\begin{equation}
\begin{aligned}
	Y_i | A_i, X_i &\sim \text{Bernoulli}[\text{expit}(\theta_0 + \theta_1^\intercal X_i +A_i \theta_2  + A_i\theta_3^\intercal X_i)],\\
	Y_{E,i} | X_{E,i} &\sim \text{Bernoulli}[\text{expit}(\theta_0 + \theta_1^\intercal X_i)].
\end{aligned}
\label{eq:binary-HTE-wald}
\end{equation}
We describe the tests below.
\begin{enumerate}
	\item [mTest-A.] {A permutation test without ED}, similar  to the ED-PT based on the working model in \eqref{eq:binary-HTE-working}, the test statistic is
	$$
	m'(\mathcal D) = \prod_{k = 1}^K \frac{s_{1,k}!(n_{1,k} - s_{1,k})!s_{0,k}!(n_{0,k} - s_{0,k})!}{(n_{1,k}+1)!(n_{0,k} + 1)!}.
	$$
	\item [mTest-B.] {A Wald test for $(\theta_2, \theta_3)$ based only on the ID}, using the saturated logistic model in \eqref{eq:binary-HTE-wald}. The test statistic is $Z = (\hat \theta_2, \hat \theta_3)^\intercal \Sigma^{-1}(\hat \theta_2, \hat \theta_3)$, where $(\hat\theta_2,\hat\theta_3)$ is the MLE of $(\theta_2,\theta_3)$ under the working model \eqref{eq:binary-HTE-wald} and $\Sigma$ is the estimated covariance matrix of $(\hat\theta_2,\hat\theta_3)$. The rejection region is $Z > \chi_{1-\alpha, K}^2$. Here, $\chi^2_{1-\alpha, K}$ is the $1-\alpha$ quantile of a Chi-square distribution with $K$ degrees of freedom.
	\item [mTest-C.] {A Wald test for $(\theta_2, \theta_3)$ based on merged ID and ED,} using the model in \eqref{eq:binary-HTE-wald}. The test statistic has the same form as in mTest-B, but the MLE $(\hat\theta_2,\hat\theta_3)$ and  the covariance matrix estimate are computed after merging the RCT data and the ED.
	\item [mTest-D.] {An oracle procedure.} The oracle knows the response probabilities $(\theta_0, \theta_1)$  in the control arm. 
	The test is based on the same  logistic model in \eqref{eq:binary-HTE-wald}, but with $(\theta_0, \theta_1)$  known and fixed. The only unknown parameters  are $\theta_2$ and $\theta_3$. The procedure is a Wald test, with a test statistic of the same form as in mTest-B. In particular, the MLE $(\hat\theta_2,\hat\theta_3)$ and the  covariance matrix estimate are based on the reduced logistic model with known $\theta_0$ and $ \theta_1$ values. 
	\end{enumerate}

\begin{figure}
	\centering
	\includegraphics[width = 0.95\textwidth]{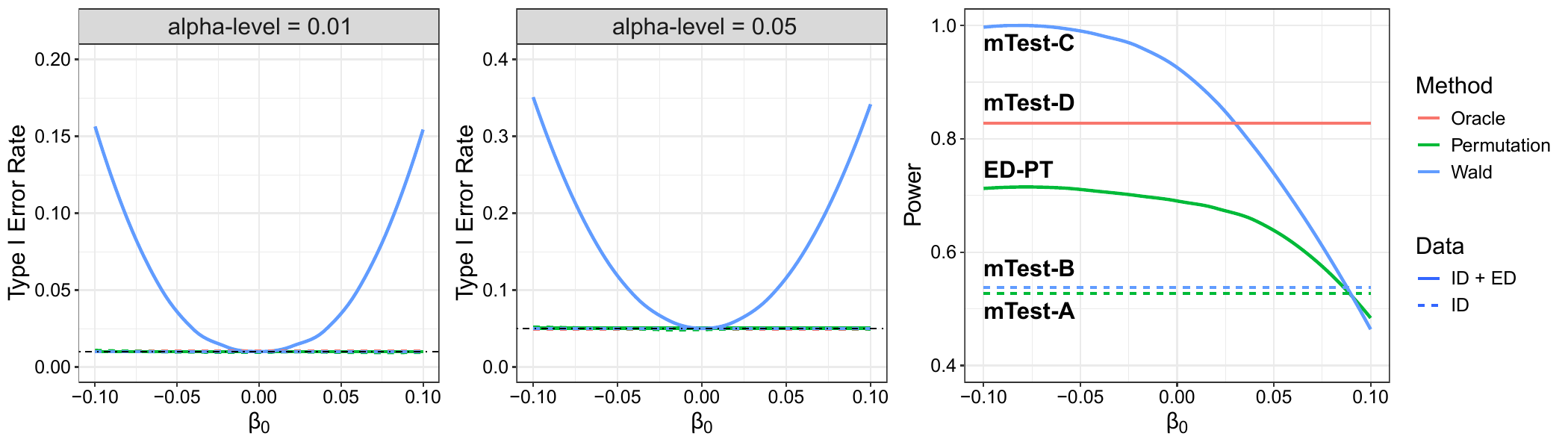}
	\caption{ A simulation study to illustrate the type I error rate and power of five testing procedures. 
	We considered the ED-PT, mTest-A, mTest-B, mTest-C, and mTest-D. The simulation scenarios have been described in the previous paragraphs. In these simulations, $K = 2$, $n_1 = 100$, $r = 1/2$, $r_E = 7.5$, $w_0 = 0.5$, $\beta_1 = 0$ and $\beta_{0,1} = 0$. We use $\gamma = \gamma_1 = 0$ to evaluate the control of false positive results and $(\gamma = 0.25, \gamma_1 = -0.15)$ to evaluate the power. }
	\label{fig:binary-HTE-sim}
\end{figure}

We illustrate the simulation results  based on  a scenario with $K = 2$  subgroups  in Figure \ref{fig:binary-HTE-sim}.
 The prevalence of subgroup 1 is 0.5 in both the RCT and the ED populations. 
 We set $n_1 = 100$, $r = 1/2$, $r_E = 7.5$, $w_0 = 0.5$, $\beta_1 = 0$ and $\beta_{0,1} = 0$. 
 We considered $\gamma = \gamma_1 = 0$ to assess the control of false positive results and $(\gamma = 0.25, \gamma_1 = -0.15)$ to evaluate the power. When evaluating the power, the effect of the experimental treatment, measured by the difference in mean response rates, is 0.25 in subgroup 1 and 0.1 in subgroup 2. We varied $\beta_{0,0}$ from -0.1 to 0.1 to examine the robustness of the testing procedures with respect to discrepancies between the distributions of the ID and the ED. We used $J = 1,000$ permutations and 10,000 simulation replicates. The results are similar to those in Section 2.4 in terms of the relative performances of the ED-PT compared to the other procedures. 

Similar to Proposition 3, we can investigate the asymptotic rejection probability of the ED-PT
  in these  simulation scenarios (see  model \ref{eq:binary-HTE}). We consider a sequence of $(\mathcal D, \mathcal D_E)$ pairs with increasing sample sizes $n$ and $n_E$, such that:
\begin{enumerate}
	\item[(C1)] The proportions $\rho_k >0$ of patients in subgroup $k = 1,\ldots, K$
	  for the RCT and ED populations are identical.
	\item[(C2)] The randomization ratio $r$ is fixed and identical across patient subgroups. 
	Similarly, the ratio $r_E$ between the number of external control patients and the number of patients assigned to the experimental arm of the RCT is fixed and does not vary across subgroups.
	\item[(C3)] The subgroup-specific treatment effects are $(\gamma, \gamma1_{K-1} + \gamma_1) = a/\sqrt{n_1}$ and the discrepancy between ED and RCT controls is captured by $(\beta_{0,0}, \beta_{0,0}1_{K-1} + \beta_{0,1}) = b/\sqrt{n_1}$.
	Note that $\gamma_1,\beta_{0,1} \in \mathbb R^{K-1}$ and $a,b\in\mathbb R^{K}$. The parameters $a,b$ will remain the same for the whole sequence of $(\mathcal D, \mathcal D_E)$ pairs.
\end{enumerate}
We describe the asymptotic power function of $\widetilde\phi$ in Proposition \ref{prop:asym_bernm}. The form of the asymptotic power function is similar to expression \eqref{eq:asym_normal} in Proposition \ref{prop:asym_normal}. The proof is deferred to Section \ref{proof:psm1}.

\begin{prop}
	Let $\rho = (\rho_1,\ldots,\rho_K)$, and assume $r$ and $r_E$ remain fixed as $n\to\infty$. Under the assumptions C1, C2, and C3, the asymptotic power function of the $\alpha$ level test $\widetilde\phi$ is
	\begin{equation}
		g(r,r_E,a,b,\rho) = 1 - \int G_{v,a,b}[\widetilde G^{-1}_{v,a,b}(1-\alpha)]\varphi(v)dv,
		\label{eq:asym_binary}
	\end{equation}
	where $\varphi$ denotes the density function of a standard $K$-variate normal distribution, and
	\begin{align*}
		\resizebox{0.95\textwidth}{!}{$G_{v,a,b}(t) = \text{Pr}\left\{\sum_{k=1}^K \left[\sqrt{\frac{\rho_kr_E(1+r)}{(1+r + r_E)\omega_k(1-\omega_k)}}\left(\frac{r+r_E}{r_E}a_k - b_{k}\right) + v_k + \sqrt{\frac{r(1+r+r_E)}{r_E}}Z_k \right]^2\leq t\right\},$}\\
		\resizebox{0.95\textwidth}{!}{$\widetilde G_{v,a,b}(t) = \text{Pr}\left\{\sum_{k=1}^K \left[\sqrt{\frac{\rho_kr_E(1+r)}{(1+r + r_E)\omega_k(1-\omega_k)}}\left(\frac{a_k}{1+r} - b_{k}\right)+ v_k + \sqrt{\frac{r(1+r+r_E)}{r_E}}Z_k
		\right]^2\leq t\right\}.$}
	\end{align*}
	Here $v=(v_1,\ldots,v_K)\in\mathbb R^K$, $t\in\mathbb R$,  $Z_1,\ldots,Z_K$ are independent standard normal random variables, and $\omega_k$
	 is the  response probability  for  the control arm of the ID within the $k$-th  subgroup.
	\label{prop:asym_bernm}
\end{prop}

\subsection{Scenarios with binary outcomes and negative treatment effects}
We consider the ED-PT and  its two variations, with test statistics $\widetilde m_1$ and $\widetilde m_2$ (defined in Section 2.3), in scenarios with binary outcomes and negative treatment effects. 
We start with a simple scenario without subgroups (i.e., $K = 1$).
 We use the data-generating model in (8) from Section 2.4. 
 We set $\beta_0 = 0$ and vary $\gamma$ from $-0.25$ to $0.25$. All other model parameters remain as specified in Section 2.4. We use the working model in expression (9). The data summary $\widetilde m_1$ is computed using $\widetilde \Theta =\{\theta_1 - \theta_0 > 0\}$, and $
\widetilde m_2(\mathcal D) = \mathbb E[\mathbb I(\theta_1 > \theta_0)(\theta_1 - \theta_0)|\mathcal D, \mathcal D_E].
$

In the second scenario we consider $K = 2$  subgroups and HTEs. 
This is similar to the Modified S1 scenario in Section 2.5, but it focuses on binary outcomes. 
We use the data-generating mechanism in \eqref{eq:binary-HTE}, fixing $\gamma = 0$ while varying $\gamma_1$ from -0.3 to 0.3. The resulting treatment effect for subgroup 1 is null, whereas in subgroup 2, the treatment effect  varies from -0.3 to 0.3.
 We use the working model in \eqref{eq:binary-HTE-working}. Here, $\widetilde m_1$ is computed with $\widetilde\Theta =\{\theta_{1,1} - \theta_{0,1} > 0 \text{ or } \theta_{1,2} - \theta_{0,2} > 0\}$, and $\widetilde m_2$ is given by $\widetilde m_2(\mathcal D) = \mathbb E[\rho_1\mathbb I(\theta_{1,1} > \theta_{0,1})(\theta_{1,1} > \theta_{0,1}) + (1-\rho_1)\mathbb I(\theta_{1,2} > \theta_{0,2})(\theta_{1,2} > \theta_{0,2})|\mathcal D, \mathcal D_E]$.

\begin{figure}
	\centering
	\includegraphics[width=0.95\textwidth]{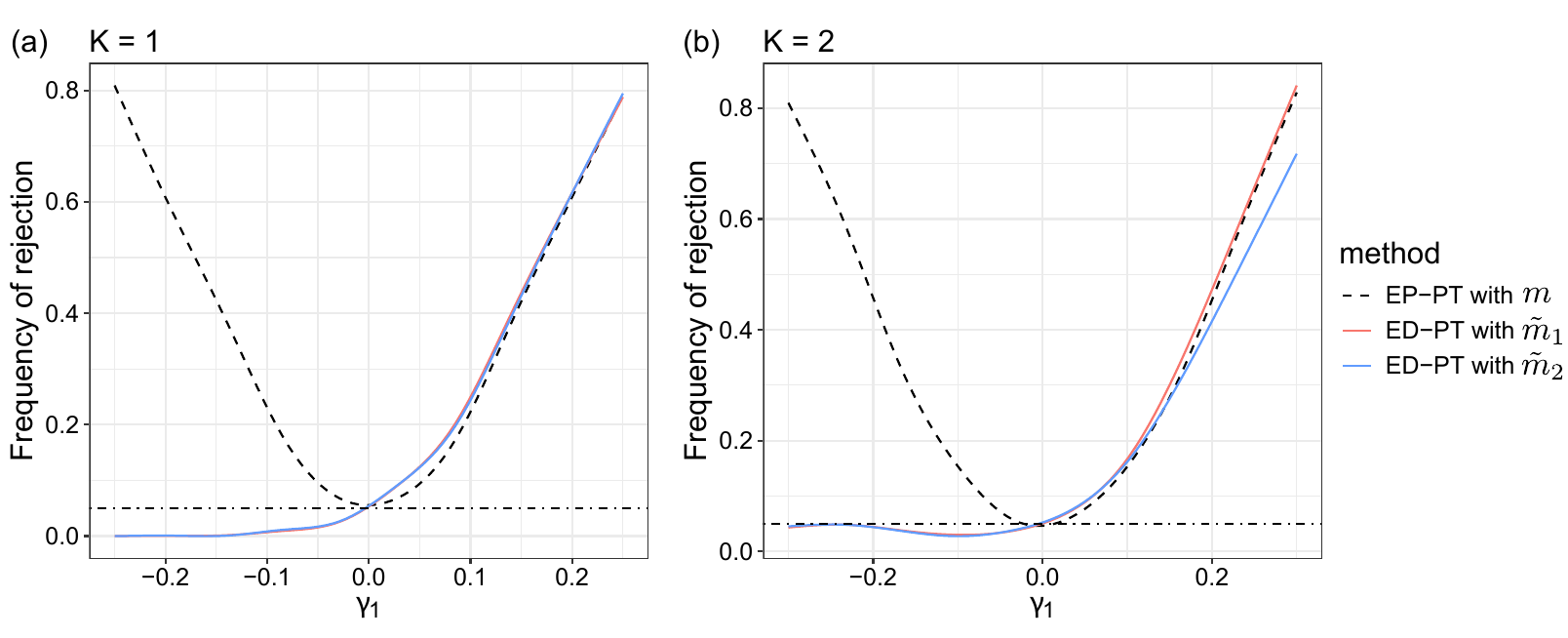}
	\caption{Comparison of ED-PT with test statistics $m$, $\widetilde m_1$, and $\widetilde m_2$.
	We consider a RCT with  binary outcomes in scenarios with (a) one  subgroup ($K = 1$) as described in Section 2.4 with $\beta_0 = 0$ and $\gamma$ varying from $-0.25$ to $0.25$, and (b) $K=2$ subgroups and data generating mechanism summarized in expression \eqref{eq:binary-HTE}, with $\gamma = 0$ and  $\gamma_1$ between $-0.3$ and $0.3$.
 }
 \label{fig:binary-negative}
\end{figure}

We summarize the results in Figure \ref{fig:binary-negative}. In both scenarios, the two variations of ED-PT with test statistics $\widetilde m_1$ and $\widetilde m_2$ control the rates of false positive results in the presence of negative treatment effects. Also, when the treatment effects are positive, the power of ED-PT with $\widetilde m_1$ and $\widetilde m_2$ is nearly identical to the power of ED-PT with $m$.

	\section{Proof of Proposition \ref{prop:asym_normal}}
	\label{proof:p4}
	To simplify the notation, we rewrite the data generating process (12) and the working model (13) as models without intercept and with $X_i$, $X_{E,i}$ as vectors of dimension $K$. More precisely, for $k=1,\dots,K$, the $k$-th coordinate $X_{i,k}$ and $X_{E,i,k}$ of $X_i$ and $X_{E,i}$ are the indicators of patient $i$ being in group $k$ in the internal and external studies, respectively. The parameters $\beta$, $\gamma$ and $\beta_0$ are $K$-dimesional vectors, with $\beta_k$ and $\gamma_k$  representing the control and tratment effects in group $k$ of the internal study, and with  $\beta_{0,k}$ being the discrepancy between control effect in the external and internal studies. According to (B3), $\gamma=a/\sqrt{n_1}$ and $\beta_0=b/\sqrt{n_1  }$. Based on the assumptions and the specification of $\mathcal M$, we can write the conditional likelihood as
	$$
	m(\mathcal D)=p_X(x)p_A(a)\int p(y|x,a,\beta,\gamma)p(\beta|\mathcal D_E)p(\gamma) d\beta d\gamma.
	$$
	The term $p_X(x)p_A(a)$ is invariant to permutations of the $A_i$'s, so that we can omit those terms when writing the test statistic. The posterior distribution of $\beta$ given external data, is made of independent $N(\mu_{E,k},\Sigma_{E,k})$ distributions, with 
	$$\Sigma_{E,k}=(n_{E,k}+\sigma^{-2})^{-1},\quad\quad \mu_{E,k}=n_{E,k}\overline{Y}_{E,k}/(n_{E,k}+\sigma^{-2}),
	$$ where
	$n_{E,k}=\sum_{i=1}^{n_E} X_{E,i,k}$, $\overline{Y}_{E,k}=\sum_{i=1}^{n_E} X_{E,i,k} Y_{E,i}/n_{E,k}.$
	Let us denote by $n_{1,k}$ and $n_{0,k}$ the number of patients in group $k$ that are allocated to treatment and control, respectively, and by $n_{.,k}$ the total number of patients in group $k$ in the internal study. Up to additive terms that are invariant with respect to permutations in $\mathcal T$, the conditional log-likelihood is equivalent ($\equiv$) to
	\begin{align*}
		\log(m(\mathcal D)) &\equiv -\frac 1 2 \sum_{k=1}^K \log \tilde C_k + \sum_{k=1}^K \frac{1}{\tilde C_k} \Bigg[
			(n_{1,k}+\sigma^{-2})(n_{.,k}\overline Y_k+n_{E,k}\overline Y_{E,k})^2 \\
		&\quad \quad + (n_{.,k}+n_{E,k}+\sigma^{-2})n_{1,k}^2\overline{Y}_{1,k}^2 
			- 2n_{1,k}^2\overline{Y}_{1,k}(n_{.,k}\overline{Y}_k+n_{E,k}\overline{Y}_{E,k})
		\Bigg]\\
		& \approx \sum_{k=1}^K \frac{n_{1,k}^2(n_{.,k}+n_{E,k})}{ C_k}
		\left( \overline{Y}_{1,k}- \frac{n_{.,k}\overline{Y}_k+n_{E,k}\overline{Y}_{E,k}}{n_{.,k}+n_{E,k}}  \right)^2
		-\frac 1 2 \sum_{k=1}^K \log  C_k + \mathcal C,
	\end{align*}
	where $\tilde C_k=(n_{.,k}+n_{E,k}+\sigma^{-2})(n_{1,k}+\sigma^{-2})-n_{1,k}^2$, $C_k=n_{1,k}(n_{0,k}+n_{E,k})$, $\overline{Y}_k=\sum_{i=1}^nX_{i,k}Y_i/n_{k}$, $\overline{Y}_{1,k}=\sum_{i=1}^nY_iX_{i,k}A_i/n_{1,k}$, and $\mathcal C$ is a permutation invariant quantity. The approximation follows by noting that $\tilde C_k \approx C_k$ when $n_1$ is large.
	
	Let $\overline{Y}_{0,k}=\sum_{i=1}^nY_iX_{i,k}(1-A_i)/n_{0,k}$, and define
	$$
	Z_{k,1}=\overline Y_{1,k}-\overline Y_{0,k},\quad\quad\quad Z_{k,2}=\overline Y_k=\frac{n_{1,k}\overline Y_{1,k}+n_{0,k}\overline Y_{0,k}}{n_{.,k}}.
	$$
	$ Z_{k,2}$ is invariant under permutations in $\mathcal T$ and
	$
	\overline Y_{1,k}= Z_{k,2}+n_{0,k}Z_{k,1}/n_{.,k}.
	$
	We consider a test statistic $T(Y,X,A)$ that is equivalent to $m(\mathcal D)$ by removing the permutation invariant components from $\log(m(\mathcal D))$:
	$$
	T(Y,X,A)=\sum_{k=1}^K \frac{n_{1,k}^2(n_{.,k}+n_{E,k})}{ C_k}
		\left( 
		\frac{n_{E,k}}{n_{.,k}+n_{E,k}}(Z_{k,2}-
		\overline Y_{E,k})+\frac{n_{0,k}}{n_{.,k}}Z_{k,1}\right)^2
		-\frac 1 2 \sum_{k=1}^K\log \frac{ C_k}{n^2}.
	$$
	The asymptotic power of $\tilde \phi$ is characterized by the limit of the randomized distribution $\hat R(t)$, which is defined as
	$$
	\hat R(t)=\frac{1}{|\mathcal T|}\sum_{\tau\in\mathcal T} \mathbb I(T(Y,X,A^{(\tau)})\leq t),
	$$
	and the $(1-\alpha)$ empirical quantile
	$
	\hat r(1-\alpha)=\inf\{t:\hat R(t)\geq 1-\alpha\}.
	$
	Let $\tau_1$ and $\tau_2$ be independent random permutations in $\mathcal T$. 
	Then, for $j = 1,2$,
	\begin{gather*}
		E(\hat R(t)|(Z_{k,2},n_{1,k},n_{0,k},n_{E,k})_k)=P(T(Y,X,A^{(\tau_j)})\leq t\mid (Z_{k,2},n_{1,k},n_{0,k},n_{E,k})_k),\\
		E(\hat R(t)^2|(Z_{k,2},n_{1,k},n_{0,k},n_{E,k})_k)=P(T(Y,X,A^{(\tau_1)})\leq t,T(Y,X,A^{(\tau_2)})\leq t\mid (Z_{k,2},n_{1,k},n_{0,k},n_{E,k})_k).
	\end{gather*}
	
	For any $j=1,2$,
	\begin{align*}
		T(Y,X,A^{(\tau_j)})=\sum_{k=1}^K \frac{(n_{1,k}^{(j)})^2(n_{.,k}+n_{E,k})}{ C_k^{(j)}}
		\left( 
		\frac{n_{E,k}}{n_{.,k}+n_{E,k}}(Z_{k,2}-
		\overline Y_{E,k})+\frac{n_{0,k}^{(j)}}{n_{.,k}}Z_{k,1}^{(j)}\right)^2
		-\frac 1 2 \sum_{k=1}^K\log \frac{ C_k^{(j)}}{n^2}.
	\end{align*}
	where
	$$
	n_{1,k}^{(j)}=\sum_{i=1}^nX_{i,k}A_i^{(\tau_j)},\quad n_{0,k}^{(j)}=\sum_{i=1}^n X_{i,k}(1-A_i^{(\tau_j)}),\quad C_k^{(j)}=
	n_{1,k}^{(j)}(n_{0,k}^{(j)}+n_{E,k}),
	$$
	$$
	\overline{Y}_{1,k}^{(j)}=\frac{\sum_{i=1}^nY_iX_{i,k}A_i^{(\tau_j)}}{n_{1,k}^{(j)}},\quad \overline{Y}_{0,k}^{(j)}=\frac{\sum_{i=1}^nY_iX_{i,k}(1-A_i^{(\tau_j)})}{n_{0,k}^{(j)}},\quad Z_{k,1}^{(j)}=\overline Y_{1,k}^{(j)}-\overline Y_{0,k}^{(j)}.
	$$
	Note that
	$
	Z_{k,1}^{(j)}=\sum_{i=1}^nW_{i,k}^{(j)}Y_iX_{i,k},
	$
	where
	$
	W_{i,k}^{(j)}=A_i^{(\tau_j)}/n_{1,k}^{(j)}-(1-A_i^{(\tau_j)})/n_{0,k}^{(j)}.
	$
	
Since $A^{(\tau_1)}$ and $A^{(\tau_2)}$ are mutually independent and independent of $Y$, then, for $j=1,2$,
	\begin{align*}
		&E(W_{i,k}^{(j)}\mid (n_{1,k},n_{0,k})_k,(n_{1,k}^{(j^\prime)},n_{0,k}^{(j^\prime)})_{k,j^\prime})=0,\\
		&E(\sum_{i=1}^{n} W_{i,k}^{(j)} Y_iX_{i,k}\mid (n_{1,k},n_{0,k})_k,(n_{1,k}^{(j^\prime)},n_{0,k}^{(j^\prime)})_{k,j^\prime})=0,\\
		&E(\sum_{i=1}^{n} W_{i,k}^{(1)} Y_iX_{i,k}\sum_{i=1}^{n} W_{i,k}^{(2)} Y_iX_{i,k}\mid (n_{1,k},n_{0,k})_k,(n_{1,k}^{(j^\prime)},n_{0,k}^{(j^\prime)})_{k,j^\prime})=0,\\
		&E(\sum_{i=1}^{n} W_{i,k}^{(j)} Y_iX_{i,k} \sum_{i=1}^{n}  Y_iX_{i,k}\mid (n_{1,k},n_{0,k})_k,(n_{1,k}^{(j^\prime)},n_{0,k}^{(j^\prime)})_{k,j^\prime})=0.
	\end{align*}
	By the properties of the normal distribution, $Z_{k,1}^{(1)},Z_{k,1}^{(2)},Z_{k,2}$ are conditionally mutually independent, given $(n_{1,k},n_{0,k})_k$, $(n_{1,k}^{(j^\prime)},n_{0,k}^{(j^\prime)})_{k,j^\prime}$. Hence
	$
	T(Y,X,A^{(\tau_1)})
	$ and $
	T(Y,X,A^{(\tau_2)})
	$
	are conditionally independent, given $(n_{1,k},n_{0,k})_k$, $(n_{1,k}^{(j^\prime)},n_{0,k}^{(j^\prime)})_{k,j^\prime}$ and $Z_{k,2}$.
	
	Let $N_k^{(j)}$ be the cardinality of the set $\{i: X_{i,k}A_iA_i^{(\tau_j)}=1\}$.
	The random variable $N_k^{(j)}$ is conditionally  independent of $(Z_{k,2})_k$, given $ (n_{1,k},n_{0,k})_k$ and $(n_{1,k}^{(j^\prime)},n_{0,k}^{(j^\prime)})_{k,j^\prime}$. Moreover, conditionally on $ (n_{1,k},n_{0,k})_k$ and $(n_{1,k}^{(j^\prime)},n_{0,k}^{(j^\prime)})_{k,j^\prime}$,  $N_k^{(j)}/n_{1,k}^{(j)} \overset{p}{\to}n_{1,k}^{(j)}/n_{.,k}$.
	As $n\rightarrow\infty$,
	\begin{align*}
		& E(\hat R(t)\mid(Z_{k,2})_k,(n_{1,k},n_{0,k},n_{E,k})_k)=
		P(T(Y,X,A^{(\tau_j)})\leq t\mid
		(Z_{k,2})_k,(n_{1,k},n_{0,k},n_{E,k})_k)\\
		&=E\Biggl(P\biggl(
		\sum_{k=1}^K \frac{(n_{1,k}^{(j)})^2(n_{.,k}+n_{E,k})}{ C_k^{(j)}}
		\left( 
		\frac{n_{E,k}}{n_{.,k}+n_{E,k}}(Z_{k,2}-
		\overline Y_{E,k})+\frac{n_{0,k}^{(j)}}{n_{.,k}}Z_{k,1}^{(j)}\right)^2\\
		&\quad\quad-\frac 1 2 \sum_{k=1}^K\log \frac{C_k^{(j)}}{n^2}
		\leq t
		\mid (Z_{k,2},n_{1,k},n_{0,k},n_{E,k})_k),
		(n_{1,k}^{(j^\prime)},n_{0,k}^{(j^\prime)})_{k,j^\prime}
		\biggr)\mid (Z_{k,2},n_{1,k},n_{0,k},n_{E,k})_k
		\Biggr)\\
		&\approx E\Biggl(P\biggl(
		\sum_{k=1}^K \frac{(n_{1,k}^{(j)})^2(n_{.,k}+n_{E,k})}{ n_{1,k}^{(j)}(n_{0,k}^{(j)}+n_{E,k})}
		\Biggl( 
		\frac{n_{E,k}}{n_{.,k}+n_{E,k}}
		\left(\frac{n_{1,k}^{(j)}}{n_{.,k}}\gamma_k -\beta_{0,k}+ \sqrt{
			\frac{n_{.,k}+n_{E,k}}{n_{.,k}n_{E,k}}
		}
		Z_{k,2}^*
		\right)
		\\
		&+\frac{n_{0,k}^{(j)}}{n_{.,k}}
		\left(\left(\frac{N_k^{(j)}}{n_{1,k}^{(j)}}-\frac{n_{1,k}-N_k^{(j)}}{n_{0,k}^{(j)}}\right)\gamma_k+\sqrt{\frac{n_{.,k}}{n_{1,k}^{(j)}n_{0,k}^{(j)}}}{Z_{k,1}^{(j)}}^*\right)
		\Biggr)^2\\
		&\quad\quad-\frac 1 2 \sum_{k=1}^K\log \frac{ C_k^{(j)}}{n^2}
		\leq t
		\mid (Z_{k,2},n_{1,k},n_{0,k},n_{E,k})_k),
		(n_{1,k}^{(j^\prime)},n_{0,k}^{(j^\prime)})_{k,j^\prime}
		\biggr)\mid (Z_{k,2},n_{1,k},n_{0,k},n_{E,k})_k
		\Biggr),
	\end{align*}
	where ${Z_{k,1}^{(1)}}^*$, ${Z_{k,1}^{(2)}}^*$ and $Z_{k,2}^*$ are independent standard normal random variables, independent of $(n_{1,k},n_{0,k},n_{E,k})_k)$,
	$(n_{1,k}^{(j^\prime)},n_{0,k}^{(j^\prime)})_{k,j^\prime}$.

	As $n\rightarrow\infty$, 
	$n_{1,k},n_{1,k}^{(j)}\approx n_1\rho_k,  
	n_{0,k},n_{0,k}^{(j)}\approx n_1 r\rho_k,
	n_{E,k}\approx n_1r_E\rho_k.$ Hence, as $n\rightarrow\infty$,
	\begin{align*}
		& E(\hat R(t)|(Z_{k,2})_k,(n_{1,k},n_{0,k},n_{E,k})_k)=
		P(T(Y,X,A^{(\tau_j)})\leq t\mid
		(Z_{k,2})_k,(n_{1,k},n_{0,k},n_{E,k})_k)\\
		&\approx P\Biggl(
		\sum_{k=1}^K 
		\frac{n_1\rho_k(1+r+r_E)}{r+r_E}\\
		&\Biggl( 
		\frac{ r_E }{1+r+r_E}
		\left(\frac{1}{1+r} \gamma_k -
		\beta_{0,k}+ \sqrt{
			\frac{1+r+r_E}{n_1(1+r)r_E}}
		Z_{k,2}^*
		\right)
		+\frac{r}{1+r}
		\sqrt{\frac{1+r}{n_1\rho_k r}}{Z_{k,1}^{(j)}}^*
		\Biggr)^2\\
		&\quad\quad-\frac 1 2 \sum_{k=1}^K\log (\rho_k^2\frac{r+r_E}{(1+r)^2})
		\leq t
		\mid (Z_{k,2})_k
		\Biggr)\\
		&\approx P\Biggl(\frac{r_E}{(r+r_E)(1+r)}
		\sum_{k=1}^K 
		\Biggl( 
		\sqrt{\rho_k\frac{r_E(1+r)}{1+r+r_E}}
		\left(\frac{a_k}{1+r}-b_k\right)
		+ Z_{k,2}^*
		+\sqrt{\frac{r(1+r+r_E)}{r_E}}{Z_{k,1}^{(j)}}^*
		\Biggr)^2\\   
		&\quad\quad-\frac 1 2 \sum_{k=1}^K\log (\rho_k^2\frac{r+r_E}{(1+r)^2})
		\leq t
		\mid (Z_{k,2})_k
		\Biggr)\\
		&=\tilde F_{(Z_{k,2}^*,a_k,b_k)_k}\left(\frac{(r+r_E)(1+r)}{r_E}(t+\frac 1 2 \sum_{k=1}^K\log (\rho_k^2\frac{r+r_E}{(1+r)^2}))\right)
	\end{align*}
	Analogously, 
	\begin{align*}
		& E(\hat R(t)^2|(Z_{k,2})_k,(n_{1,k},n_{0,k},n_{E,k})_k)=
		P(T(Y,X,A^{(\tau_1)})\leq t,T(Y,X,A^{(\tau_2)})\leq t \mid
		(Z_{k,2})_k,(n_{1,k},n_{0,k},n_{E,k})_k)\\
		&\approx \left(\tilde F_{(Z_{k,2}^*,a_k,b_k)_k}\left(\frac{(r+r_E)(1+r)}{r_E}(t+\frac 1 2 \sum_{k=1}^K\log (\rho_k^2\frac{r+r_E}{(1+r)^2}))\right)\right)^2.
	\end{align*}
	It follows that, as $n\rightarrow\infty$,
	\begin{align*}
		E\left(\left(\hat R(t)- \tilde F_{(Z_{k,2}^*,a_k,b_k)_k}\left(\frac{(r+r_E)(1+r)}{r_E}(t+\frac 1 2 \sum_{k=1}^K\log (\rho_k^2\frac{r+r_E}{(1+r)^2}))\right)\right)^2\mid (Z_{k,2}^*,n_{1,k},n_{0,k},n_{E,k})_k\right)\stackrel{p}{\rightarrow}0,
	\end{align*}
	which entails
	\begin{align*}
		P\left(\left|\hat R(t)-\tilde F_{(Z_{k,2}^*,a_k,b_k)_k}\left(\frac{(r+r_E)(1+r)}{r_E}(t+\frac 1 2 \sum_{k=1}^K\log (\rho_k^2\frac{r+r_E}{(1+r)^2}))\right)\right|>\epsilon\mid (Z_{k,2}^*,n_{1,k},n_{0,k},n_{E,k})_k\right)\stackrel{p}{\rightarrow}0,
	\end{align*}
	for every $\epsilon>0$, as $n\rightarrow\infty$. Thus, the $(1-\alpha)$ quantile $\hat r(1-\alpha)$ of $\hat R(t)$ satisfies
	\begin{align*}
		P\left(\left|\hat r(1-\alpha)-q_{(Z_{k,2}^*,a_k,b_{0,k})_k}(1-\alpha)
		\right|>\epsilon\mid (Z_{k,2}^*,n_{1,k},n_{0,k},n_{E,k})_k\right)\stackrel{P}{\rightarrow}0,
	\end{align*}
	as $n\rightarrow\infty$, where
	$$
	\tilde F_{(Z_{k,2}^*,a_k,b_k)_k}\left(\frac{(r+r_E)(1+r)}{r_E}(q_{(Z_{k,2}^*,a_k,b_{0,k})_k}(1-\alpha)+\frac 1 2 \sum_{k=1}^K\log (\rho_k^2\frac{r+r_E}{(1+r)^2}))\right)=1-\alpha.
	$$
	In other words
	$$
	q_{(Z_{k,2}^*,a_k,b_{0,k})_k}(1-\alpha)=\frac{r_E}{(r+r_E)(1+r)}\tilde F_{(Z_{k,2}^*,a_k,b_{k})_k}^{-1}(1-\alpha)-\frac 1 2 \sum_{k=1}^K\log (\rho_k^2\frac{r+r_E}{(1+r)^2}).
	$$
	
	Now we find the asymptotic conditional distribution of distribution of $T(Y,X,A)$, given $((Z_{k,2},n_{1,k},n_{0,k},n_{E,k})_k$. 
	\begin{align*}
		&P(T(Y,X,A)\leq t\mid (Z_{k,2},n_{1,k},n_{0,k},n_{E,k})_k)\\
		& \approx P\Biggl(
		\sum_{k=1}^K \frac{(n_{1,k})^2(n_{.,k}+n_{E,k})}{ n_{1,k}(n_{0,k}+n_{E,k})}
		\Biggl( 
		\frac{n_{E,k}}{n_{.,k}+n_{E,k}}
		\left(\frac{n_{1,k}}{n_{.,k}}\gamma_k -\beta_{0,k}+ \sqrt{
			\frac{n_{.,k}+n_{E,k}}{n_{.,k}n_{E,k}}
		}
		Z_{k,2}^*
		\right)
		+\frac{n_{0,k}}{n_{.,k}}
		\left(
		\gamma_k+\sqrt{\frac{n_{.,k}}{n_{1,k}n_{0,k}}}{Z_{k,1}}^*\right)
		\Biggr)^2\\
		&\quad\quad-\frac 1 2 \sum_{k=1}^K\log \frac{C_k}{n^2}
		\leq t
		\mid (Z_{k,2},n_{1,k},n_{0,k},n_{E,k})_k
		\Biggr)\\
		&\approx P\Biggl(\frac{r_E}{(r+r_E)(1+r)}
		\sum_{k=1}^K 
		\Biggl( 
		\sqrt{\rho_k\frac{r_E(1+r)}{1+r+r_E}}
		\left(\frac{r+r_E}{r_E}a_k-b_k\right)
		+ Z_{k,2}^*
		+\sqrt{\frac{r(1+r+r_E)}{r_E}}{Z_{k,1}^{(j)}}^*
		\Biggr)^2\\   
		&\quad\quad-\frac 1 2 \sum_{k=1}^K\log (\rho_k^2\frac{r+r_E}{(1+r)^2})
		\leq t
		\mid (Z_{k,2})_k
		\Biggr)\\
		&= F_{(Z_{k,2}^*,a_k,b_k)_k}\left(\frac{(r+r_E)(1+r)}{r_E}(t+\frac 1 2 \sum_{k=1}^K\log (\rho_k^2\frac{r+r_E}{(1+r)^2}))\right).
	\end{align*}
	Notice that 
	\begin{align*}
		&P(T(Y,X,A)\leq q_{(Z_{k,2},a_k,b_{k})_k}(1-\alpha)\mid (Z_{k,2},n_{1,k},n_{0,k},n_{E,k})_k)\\
		&\approx 
		F_{(Z_{k,2}^*,a_k,b_k)_k}\left(
		\frac{(r+r_E)(1+r)\left(q_{(Z_{k,2},a_k,b_{k})_k}(1-\alpha)+\frac 1 2 \sum_{k=1}^K\log (\rho_k^2\frac{r+r_E}{(1+r)^2})\right)}{r_E}\right)\\
		&=F_{(Z_{k,2}^*,a_k,b_{k})_k}
		(\tilde F_{(Z_{k,2}^*,a_k,b_{k})_k}^{-1}(1-\alpha)).
	\end{align*}
	It follows that 
	\begin{align*}
		&P\left(T(Y,X,A)>\hat r(1-\alpha)\mid (n_{1,k},n_{0,k},n_{E,k})_k\right)\\
		&=E\left(
		P\left(T(Y,X,A)>\hat r(1-\alpha)\mid (Z_{k,2},n_{1,k},n_{0,k},n_{E,k})_k\right)\mid (n_{1,k},n_{0,k},n_{E,k})_k
		\right)\\
		&\approx E\left(E\left(
		1-F_{(Z_{k,2}^*,a_k,b_{k})_k}(\tilde F_{(Z_{k,2}^*,a_k,b_{k})_k}^{-1}(1-\alpha))\mid (Z_{k,2}^*)_k
		\right)\right)\\
		&=1-\int F_{v,a,b}(\tilde F_{v,a,b}^{-1}(1-\alpha)) \varphi({ v})d{ v}.
	\end{align*}

\section{Proof of Proposition \ref{prop:asym_bernm}}
\label{proof:psm1}
We use the same technique in Section \ref{proof:p4} to prove the proposition. Ignoring the terms that are invariant to permutations, we can write equivalent ($\equiv$) expressions of $m(\mathcal D)$:
\begin{equation*}
		m(\mathcal D) \equiv  \prod_{k=1}^K
  \frac{
{{n_{.,k}+n_{E,k}}\choose{s_k+s_{E,k}}}
}
{
{{n_{1,k}}\choose{s_{1,k}}}
{{n_{E,k}+n_{0,k}}\choose{s_{E,k}+s_{E,k}}}
}=
 \prod_{k=1}^K\frac{1}{h(s_{1,k};n_{1,k}, n_{.,k} + n_{E,k}, s_k+s_{E,k})},
			\end{equation*}
   where $h$ is the pmf of a hypergeometric distribution (see Section \ref{sec:prop2}) and $n_{.,k} = n_{1,k} + n_{0,k}$. As $n_1\rightarrow\infty$, $(s_k+s_{E,k})/n_{1,k}\stackrel{p}{\rightarrow }w_{0,k}/(1+r+r_E)$.

   By the normal approximation of the hypergeometric distribution, $h(s_{1,k};n_{1,k}, n_{.,k} + n_{E,k}, s_k+s_{E,k})$ is asymptotically equivalent in probability to the probability density function of a normal distribution with mean $n_{1,k}m_{E,k}$ and variance $m_{E,k}(1-m_{E,k})n_{1,k}(1-n_{1,k}/(n_{.,k}+n_{E,k}))$, where $m_{E,k}=(s_k+s_{E,k})/(n_{.,k} +n_{E,k})$. Thus, the $\tilde \phi$ is asymptotically equivalent to a permutation test with test statistic
   \begin{align*}
   T_1(\mathcal D):=\frac 1 2\sum_{k=1}^K \log(n_{1,k}(1-n_{1,k}/(n_{.,k}+n_{1,k})))+
   \sum_{k=1}^K\frac{(s_{1,k}-m_{E,k}n_{1,k})^2}{2m_{E,k}(1-m_{E,k})n_{1,k}(1-\frac{n_{1,k}}{n_{.,k}+n_{E,k}})}
\end{align*}
Denote by $A^{(\tau_1)}$ and $A^{(\tau_2)}$ two independent permutations of $A$, and define
\begin{align*}
	n_{1,k}^{(j)}=\sum_{i=1}^nX_{i,k}A_i^{(\tau_j)},&\quad n_{0,k}^{(j)}=\sum_{i=1}^n X_{i,k}(1-A_i^{(\tau_j)}),
	\\
	s_{1,k}^{(j)}=\sum_{i=1}^nY_iX_{i,k}A_i^{(\tau_j)},&\quad s_{0,k}^{(j)}=\sum_{i=1}^nY_iX_{i,k}(1-A_i^{(\tau_j)}).
\end{align*}
We can then define an equivalent test statistic to $T_1$ as follows.
$$
T_1(\mathcal D^{(\tau_j)}) \equiv  T_2(\mathcal D^{(\tau_j)}):=\frac 1 2\sum_{k=1}^K \log\left(\frac{n_{1,k}^{(j)}}{n_1}\left(1-\frac{n_{1,k}^{(j)}}{n_{.,k}+n_{E,k}}\right)\right)+
   \sum_{k=1}^K\frac{\left(\sqrt{n_{1,k}^{(j)}}
   \left(\frac{s_{1,k}^{(j)}}{n_{1,k}^{(j)}}-m_{E,k}\right)\right)^2}{2m_{E,k}(1-m_{E,k})\left(1-\frac{n_{1,k}^{(j)}}{n_{.,k}+n_{E,k}}\right)}\\
$$
The random vector $(n_{1,1}^{(j)},\dots,n_{1,K}^{(j)})$ has multivariate hypergeometric distribution with parameters
$(n_{.,1},\dots,n_{.,k},n_1)$, that can be approximated, as $n\rightarrow\infty$ by a multivariate normal distribution with mean vector $$\left(\frac{n_{.,1}}{n} n_1,\dots, \frac{n_{.,K}}{n}n_1\right)$$ and variance covariance matrix satisfying
$$
\text{Var}(n_{1,k}^{(j)})=n_1\frac{n_{.,k}}{n}\left( 1-\frac{n_{.,k}}{n}\right) \frac{n_0}{n}
,\quad 
\text{Cov}(n_{1,k}^{(j)},n_{1,i}^{(j)})=-n_1\frac{n_{.,k}n_{.,i}}{n^2}\frac{n_0}{n}.
$$
It follows that, as $n\rightarrow\infty$, $n_{1,k}^{(j)}=n_{1,k}+o_P(n_{1,k})$.
\\
	Let us introduce the notation:
	$$
 \overline{Y}_{1,k}^{(j)}=\frac{1}{n_{1,k}^{(j)}}\sum_{i=1}^nY_iX_{i,k}A_i^{(\tau_j)},\quad 
	\overline{Y}_{0,k}^{(j)}=\frac{1}{n_{0,k}^{(j)}}\sum_{i=1}^nY_iX_{i,k}(1-A_i^{(\tau_j)}),
	$$
	$$
	Z_{k,1}^{(j)}=\overline Y_{1,k}^{(j)}-\overline Y_{0,k}^{(j)},\quad\quad\quad Z_{k,2}=\overline Y_k=\frac{n_{1,k}^{(j)}\overline Y_{1,k}^{(j)}+n_{0,k}^{(j)}\overline Y_{0,k}^{(j)}}{n_{.,k}}
	$$
Notice that $ Z_{k,2}$ is invariant to permutations in $\mathcal T$. Moreover,
	$$
	\overline Y_{1,k}^{(j)}= Z_{k,2}+\frac{n_{0,k}^{(j)}}{n_{.,k}}Z_{k,1}^{(j)}.
	$$
	We can write that
\begin{align*}
&T_2(\mathcal D^{(j)})
   \approx \frac 1 2\sum_{k=1}^K \log\left(\rho_k\left(1-\frac{1}{1+r+r_E}\right)\right)+
   \sum_{k=1}^K\frac{\left(\sqrt{n_{1,k}^{(j)}}
   \left(\frac{s_{1,k}^{(j)}}{n_{1,k}^{(j)}}-m_{E,k}\right)\right)^2}{2m_{E,k}(1-m_{E,k})\left(1-\frac{n_{1,k}^{(j)}}{n_{.,k}+n_{E,k}}\right)}.\end{align*}
   Thus, the permutation test is asymptotically equivalent to a permutation test with test statistic
   \begin{align*}
          &T(\mathcal D^{(j)}):= 
        \sum_{k=1}^K
   \frac{\left(\sqrt{n_{1,k}^{(j)}}
   \left(
   \frac{n_{E,k}}{n_{.,k}+n_{E,k}}(Z_{k,2}-\overline Y_{E,k})+\frac{n_{0,k}^{(j)}}{n_{.,k}}Z_{k,1}^{(j)}
     \right)\right)^2}{m_{E,k}(1-m_{E,k})\left(1-\frac{n_{1,k}^{(j)}}{n_{.,k}+n_{E,k}}\right)}.
   \end{align*}
   As in Section \ref{proof:p4}, we characterize the asymptotic behavior of
	\begin{align*}
		\hat R(t)=\frac{1}{|\mathcal T|}\sum_{A'\in\mathcal T} \mathbb I(T(Y,X,A')\leq t),
	\end{align*}
	and its empirical quantile $\hat r(1-\alpha)=\inf\{t:\hat R(t)\geq 1-\alpha\}.$ 
	
	Using the same argument in Section \ref{proof:p4}, we conclude that 
	$
	T(Y,X,A^{(\tau_1)})
	$ and $
	T(Y,X,A^{(\tau_2)})
	$
	are asymptotically conditionally independent, given $(n_{1,k},n_{0,k})_k$, $(n_{1,k}^{(j^\prime)},n_{0,k}^{(j^\prime)})_{k,j^\prime}$ and $Z_{k,2}$. For $j=1,2$, let $N_k^{(j)}$ be the cardinality of the set $\{i: X_{i,k}A_iA_i^{(\tau_j)}=1\}$.
	The random variable $N_k^{(j)}$ is conditionally  independent of $(Z_{k,2})_k$, given $ (n_{1,k},n_{0,k})_k$ and $(n_{1,k}^{(j^\prime)},n_{0,k}^{(j^\prime)})_{k,j^\prime}$. Moreover, conditionally on $ (n_{1,k},n_{0,k})_k$ and $(n_{1,k}^{(j^\prime)},n_{0,k}^{(j^\prime)})_{k,j^\prime}$,  $N_k^{(j)}/n_{1,k}^{(j)} \overset{p}{\to} n_{1,k}^{(j)}/n_{.,k}$. 
	As $n\rightarrow\infty$, 
	\begin{align}\label{eq:rhat}
		&  E(\hat R(t)|(Z_{k,2})_k,(n_{1,k},n_{0,k},n_{E,k})_k)=
		P(T(Y,X,A^{(\tau_j)})\leq t\mid
		(Z_{k,2})_k,(n_{1,k},n_{0,k},n_{E,k})_k) \nonumber\\
				&\approx E\Biggl(P\biggl(
		\sum_{k=1}^K \frac{n_{1,k}^{(j)}(n_{.,k}+n_{E,k})}{(n_{0,k}^{(j)}+n_{E,k})m_{E,k}(1-m_{E,k})}
		\Biggl( 
		\frac{n_{E,k}}{n_{.,k}+n_{E,k}}
		\left(\frac{n_{1,k}^{(j)}}{n_{.,k}}\gamma_k -\beta_{0,k}+ \sqrt{
			\frac{n_{.,k}+n_{E,k}}{n_{.,k}n_{E,k}}w_k(1-w_k)
		}
		Z_{k,2}^*
		\right)
		\nonumber\\
		&+\frac{n_{0,k}^{(j)}}{n_{.,k}}
		\left(\left(\frac{N_k^{(j)}}{n_{1,k}^{(j)}}-\frac{n_{1,k}-N_k^{(j)}}{n_{0,k}^{(j)}}\right)\gamma_k+\sqrt{\frac{n_{.,k}}{n_{1,k}^{(j)}n_{0,k}^{(j)}}w_k(1-w_k)}{Z_{k,1}^{(j)}}^*\right)
		\Biggr)^2\leq t\mid (Z_{k,2},n_{1,k},n_{0,k},n_{E,k})_k
		\Biggr),
	\end{align}
	where ${Z_{k,1}^{(1)}}^*$, ${Z_{k,1}^{(2)}}^*$ and $Z_{k,2}^*$ are independent standard normal random variables, independent of $(n_{1,k},n_{0,k},n_{E,k})_k)$,
	$(n_{1,k}^{(j^\prime)},n_{0,k}^{(j^\prime)})_{k,j^\prime}$. As $n\rightarrow\infty$, $n_{1,k}/n_1 \approx n_{1,k}^{(j)}/n_1 \approx \rho_k$, $n_{0,k}/n_1 \approx n_{0,k}^{(j)}/n_1 \approx r\rho_k$, and $n_{E,k}/n_1\approx r_E\rho_k$. Hence, as $n\rightarrow\infty$,
	\begin{align*}
		& E(\hat R(t)|(Z_{k,2},n_{1,k},n_{0,k},n_{E,k})_k)=
		P(T(Y,X,A^{(\tau_j)})\leq t\mid
		(Z_{k,2})_k,(n_{1,k},n_{0,k},n_{E,k})_k)\\
		&\approx P\Biggl(\frac{r_E}{(r+r_E)(1+r)}
		\sum_{k=1}^K 
		\Biggl( 
		\sqrt{\rho_k\frac{r_E(1+r)}{1+r+r_E}}
		\left(\frac{a_k}{(1+r)\sqrt{w_k(1-w_k)}}-\frac{b_k}{\sqrt{w_k(1-w_k)}}\right)\\
		&+ Z_{k,2}^*
		+\sqrt{\frac{r(1+r+r_E)}{r_E}}{Z_{k,1}^{(j)}}^*
		\Biggr)^2
		\leq t
		\mid (Z_{k,2},n_{1,k},n_{0,k},n_{E,k})_k)
		\Biggr)\\
		&=\tilde G_{(Z_{k,2}^*,a_k,b_k)_k}\left(\frac{(r+r_E)(1+r)}{r_E}t\right)
	\end{align*}
	Analogously, 
	\begin{align*}
		& E(\hat R(t)^2|(Z_{k,2})_k,(n_{1,k},n_{0,k},n_{E,k})_k)=
		P(T(Y,X,A^{(1)})\leq t,T(Y,X,A^{(2)})\leq t \mid
		(Z_{k,2})_k,(n_{1,k},n_{0,k},n_{E,k})_k)\\
		&\approx \left(\tilde G_{(Z_{k,2}^*,a_k,b_k)_k}\left(\frac{(r+r_E)(1+r)}{r_E}t\right)\right)^2.
	\end{align*}
	It follows that, as $n\rightarrow\infty$,
	\begin{align*}
		E\left(\left(\hat R(t)- \tilde G_{(Z_{k,2}^*,a_k,b_k)_k}\left(\frac{(r+r_E)(1+r)}{r_E}t\right)\right)^2\mid (Z_{k,2}^*,n_{1,k},n_{0,k},n_{E,k})_k\right)\stackrel{P}{\rightarrow}0,
	\end{align*}
	which entails
	\begin{align*}
		P\left(\left|\hat R(t)-\tilde G_{(Z_{k,2}^*,a_k,b_k)_k}\left(\frac{(r+r_E)(1+r)}{r_E}t\right)\right|>\epsilon\mid (Z_{k,2}^*,n_{1,k},n_{0,k},n_{E,k})_k\right)\stackrel{P}{\rightarrow}0,
	\end{align*}
	for every $\epsilon>0$, as $n\rightarrow\infty$. Thus, the $(1-\alpha)$ quantile $\hat r(1-\alpha)$ of $\hat R(t)$ satisfies
	\begin{align*}
		P\left(\left|\hat r(1-\alpha)-q_{(Z_{k,2}^*,a_k,b_{k})_k}(1-\alpha)
		\right|>\epsilon\mid (Z_{k,2}^*,n_{1,k},n_{0,k},n_{E,k})_k\right)\stackrel{P}{\rightarrow}0,
	\end{align*}
	as $n\rightarrow\infty$, where
	$$
	\tilde G_{(Z_{k,2}^*,a_k,b_k)_k}\left(\frac{(r+r_E)(1+r)}{r_E}q_{(Z_{k,2}^*,a_k,b_{k})_k}(1-\alpha)\right)=1-\alpha.
	$$
	In other words
	$$
	q_{(Z_{k,2}^*,a_k,b_{k})_k}(1-\alpha)=\frac{r_E}{(r+r_E)(1+r)}\tilde G_{(Z_{k,2}^*,a_k,b_{k})_k}^{-1}(1-\alpha).
	$$
	Now we find the asymptotic conditional distribution of distribution of $T(Y,X,A)$, given $((Z_{k,2},n_{1,k},n_{0,k},n_{E,k})_k$. Let 
	$$
 \overline{Y}_{1,k}=\frac{1}{n_{1,k}^{(j)}}\sum_{i=1}^nY_iX_{i,k}A_i,\quad 
	\overline{Y}_{0,k}=\frac{1}{n_{0,k}^{(j)}}\sum_{i=1}^nY_iX_{i,k}(1-A_i),
	$$
	$$
	Z_{k,1}=\overline Y_{1,k}-\overline Y_{0,k},\quad\quad\quad Z_{k,2}=\overline Y_k=\frac{n_{1,k}\overline Y_{1,k}+n_{0,k}\overline Y_{0,k}}{n_{.,k}}
	$$
  Then,
	$
	\overline Y_{1,k}= Z_{k,2}+\frac{n_{0,k}}{n_{.,k}}Z_{k,1},
	$
  and
	\begin{align*}
		&P(T(Y,X,A)\leq t\mid (Z_{k,2},n_{1,k},n_{0,k},n_{E,k})_k)\\
		&\approx P\Biggl(\frac{r_E}{(r+r_E)(1+r)}
		\sum_{k=1}^K 
		\Biggl( 
		\sqrt{\rho_k\frac{r_E(1+r)}{1+r+r_E}}
		\left(\frac{r+r_E}{r_E}\frac{a_k}{\sqrt{w_k(1-w_k)}}-\frac{b_k}{\sqrt{w_k(1-w_k)}}\right)\\&
		+ Z_{k,2}^*
		+\sqrt{\frac{r(1+r+r_E)}{r_E}}{Z_{k,1}^{(j)}}^*
		\Biggr)^2		\leq t
		\mid (Z_{k,2},n_{1,k},n_{0,k},n_{E,k})_k
		\Biggr)\\
		&= G_{(Z_{k,2}^*,a_k,b_k)_k}\left(\frac{(r+r_E)(1+r)}{r_E}t\right),
	\end{align*}
    where $Z_{k,2}^*$ is defined as is \eqref{eq:rhat}, and $Z_{k,1}^*$ has standard normal distribution and is conditionally independent of $Z_{k,2}^*$ given $(Z_{k,2},n_{1,k},n_{0,k},n_{E,k})_k$.
	Note that 
	\begin{align*}
		&P(T(Y,X,A)\leq q_{(Z_{k,2},a_k,b_{k})_k}(1-\alpha)\mid (Z_{k,2},n_{1,k},n_{0,k},n_{E,k})_k)\\
		&\approx 
		G_{(Z_{k,2}^*,a_k,b_k)_k}\left(
		\frac{(r+r_E)(1+r)q_{(Z_{k,2},a_k,b_{k})_k}(1-\alpha)}{r_E}\right)\\
		&=G_{(Z_{k,2}^*,a_k,b_{k})_k}
		(\tilde G_{(Z_{k,2}^*,a_k,b_{k})_k}^{-1}(1-\alpha)).
	\end{align*}
	It follows that 
	\begin{align*}
		&P\left(T(Y,X,A)>\hat r(1-\alpha)\mid (n_{1,k},n_{0,k},n_{E,k})_k\right)\\
		&=E\left(
		P\left(T(Y,X,A)>\hat r(1-\alpha)\mid (Z_{k,2},n_{1,k},n_{0,k},n_{E,k})_k\right)\mid (n_{1,k},n_{0,k},n_{E,k})_k
		\right)\\
		&\approx E\left(E\left(
		1-G_{(Z_{k,2}^*,a_k,b_{k})_k}(\tilde G_{(Z_{k,2}^*,a_k,b_{k})_k}^{-1}(1-\alpha))\mid (Z_{k,2}^*)_k
		\right)\right)\\
		&=1-\int G_{v,a,b}(\tilde G_{v,a,b}^{-1}(1-\alpha)) \varphi({ v})d{ v}.
	\end{align*}
    $\square$

\bigskip

\noindent    {\em Remark:}
The power of the test for $k=1$ in Proposition 3 can be retrieved as follows from the proof of Proposition \ref{prop:asym_bernm}. Let
\begin{align*}
&A=\sqrt{\frac{r(1+r+r_E)}{r_E}},\\
&B=\sqrt{\frac{r_E(1+r)}{(1+r+r_E)w(1-w)}}\left(\frac{a}{1+r}-b\right), B'=\sqrt{\frac{r_E(1+r)}{(1+r+r_E)w(1-w)}}\left(\frac{r+r_E}{r_E}a-b\right).
\end{align*}
Note,
$
P\left(T(\mathcal D^{(\tau_j)})\leq t   \right)
\approx E(\tilde G_{Z_2^*,a,b}(t)).
$
Hence
$
\texttt{pv}\approx 1-E(\tilde G_{Z_2^*,a,b}(T(\mathcal D))\mid T(\mathcal D)).
$
        It follows that
        \begin{align*}
        g_B(r,r_E,a,b)&=P(\texttt{pv}<\alpha)
        \approx E[\mathbb I (E\{\tilde G_{Z_2^*,a,b}[T(\mathcal D)]\mid T(\mathcal D)\}>1-\alpha)]\\
        &= P\{\tilde G_{Z_{2}^*,a,b}[T(\mathcal D)]>1-\alpha\}.
        \end{align*}
 On the other hand,
 \begin{align*}
     &\tilde G_{v,a,b}(t)=P\left(\left(B+v+AZ_1^*\right)^2\leq t\right)=P\left(
\frac{-B - v -\sqrt t}{A}\leq Z_1^*\leq \frac{-B - v +\sqrt t}{A}
     \right)\\
     &=\Phi\left(\frac{-B - v +\sqrt t}{A}\right)-\Phi\left(\frac{-B - v -\sqrt t}{A}\right).
 \end{align*}
Furthermore,
$$
T(\mathcal D)\stackrel{d}{\approx}
(B'+Z_{2}^*+AZ_1^*)^2,
$$
which entails
$$
\sqrt{T(\mathcal D)}\stackrel{d}{\approx}
|B'+Z_2^*+AZ_1^*|.
$$
Thus, when $n\to\infty$,
\begin{align*}
g_B(r,r_e,a,b)=
&P\Biggl(\Phi\left(\max\left(\frac{B'-B}{A}+Z_1^*,-\frac{B'+B+2Z_2^*}{A}-Z_1^*\right)\right)\\
&
-\Phi\left(\min\left(\frac{B'-B}{A}+Z_1^*,-\frac{B'+B+2Z_2^*}{A}-Z_1^*\right)\right)>1-\alpha\Biggr).
\end{align*}
Recall that $U_0=\frac{B'-B}{A}+Z_1^*$ and $U_1=-\frac{B'+B+2Z_2^*}{A}-Z_1^*$, we obtain
$$
g_B(r,r_e,a,b) = P(\Phi(\max(U_0,U_1))-\phi(\min(U_0,U_1))>1-\alpha),
$$
and
$$
\left[\begin{array}{c}
U_0\\
\\U_1
\end{array}
\right]\sim \mathcal N\left(\left[
\begin{array}{c}
\frac{\sqrt r a}{\sqrt{(1+r)w(1-w)}}
\\
\frac{
-(r(1+r+r_E)+2r_E)a+2(1+r)r_Eb
}
{(1+r+r_E)\sqrt{r(1+r)w(1-w)}}
\end{array}
\right]
,
\left[
\begin{array}{cc}
1&-1\\
\\
-1&1+\frac{4r_E}{r(1+r+r_E)}
\end{array}
\right]
\right).
$$

\section{Additional simulation scenarios for continuous outcomes}

\subsection{Scenarios with continous outcomes and negative treatment effects}

\begin{figure}[!htp]
		\centering
		\includegraphics[scale = 0.6]{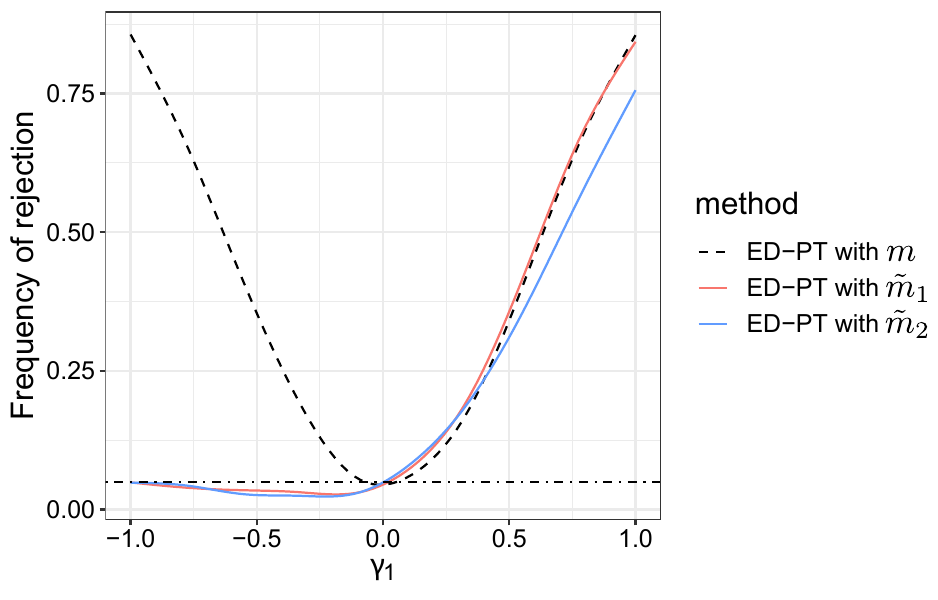}
		\caption{Comparisons of the ED-PT with the test statistic $m$ (black dashed line)  and  two variations of it with test statistics $\widetilde m_1$ (red line) and $\widetilde m_2$ (blue line) when the treatment effects $\gamma_1$ varies. The dash-dotted line indicates frequency of 0.05.}\label{fig:power_mod_norm}
	\end{figure}

Figure \ref{fig:power_mod_norm} summarizes the simulation results for the Modified S1 scenario in Section 2.5 when the treatment effects varies from negative to positive. The results confirm that the ED-PT with test statistics $\tilde m_1(\mathcal D)$ and $\tilde m_2(\mathcal D)$ defined in Section 2.3 can be used for the one-sided test of positive treatment effects. Both versions of ED-PT control the rejection rates when the treatment effect is negative ($\gamma_1 < 0$) and have similar power as the two-sided ED-PT (i.e., ED-PT with the test statistic $m$) when the treatment effect is positive.

	We also examine the ED-PT procedure with test statistics $\widetilde m_1$ and $\widetilde m_2$  
	defined in Section 2.3 for detecting positive effects  when a subgroup of patients has negative effects. We considered the modified S1 scenario  (see Section 2.5) with $K = 3$ subgroups,
	and changed the parameterization of this scenario in two different ways.
	In the first case, we set $\gamma = 0.75$ and $\gamma_1 = (0,-0.75)$, corresponding to treatment effects of 0.75, 0.75, and 0 for subgroups 1, 2, and 3, respectively.
	 In the second case, we have $\gamma = 0.75$ and $\gamma_1 = (0,-1.5)$, resulting in treatment effects equal to 0.75, 0.75, and -0.75 for the three subgroups. 
	 In both cases, we set $\rho = (1/3,1/3,1/3)$ and keep all other model parameters the same as in S1. 
	 In the first case, 
	 the ED-PT with test statistics $\widetilde m_1$ and $\widetilde m_2$ (significance level $\alpha = 0.05$ and  $J = 1\text{,}000$ permutations), rejects the null hypothesis with frequencies equal to 0.865 and 0.876   
	 over $10\text{,}000$ simulations.
	 These frequencies in the second case become 0.894 and 0.888. In the outlined simulations, the power of the ED-PT using test statistics $\widetilde m_1$ and $\widetilde m_2$ remains similar when the third subgroup exhibits a null or a negative effect.

 \subsection{ED-PT for detecting negative treatment effects}
  
\begin{figure}[!htp]
	 \centering
	 \includegraphics[scale = 0.6]{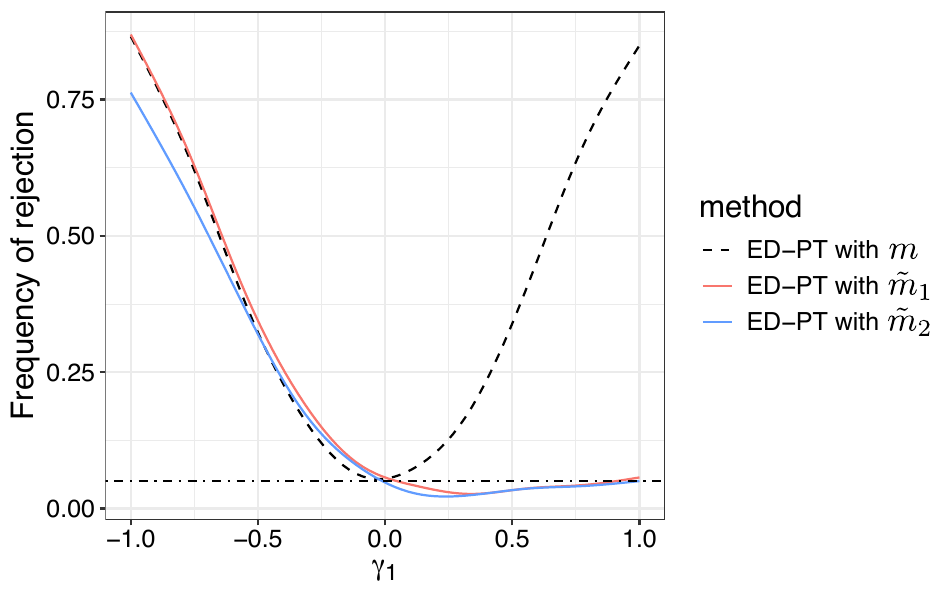}
	 \caption{A comparison of three versions of the ED-PT. We consider the modified  $\widetilde m_1$ (red line) and $\widetilde m_2$ (blue line) statistics for detecting negative treatment effects. 
	 We  also include the ED-PT with test statistics $m$ (black line).
	 We use  the  Modified S1 scenario in Section 2.5 and illustrate the control of false positive results (when $\gamma_1\geq 0$) and power (when $\gamma_1<0$) of the  tests. The dashed line indicates the nominal level $\alpha = 0.05$.
	 }
	 \label{fig:rej-neg}
 \end{figure}

 The statistics $\widetilde m_1$ and $\widetilde m_2$ can be adapted to detect negative treatment effects. In particular $\widetilde{m}_1(\mathcal{D})$ can be adapted by defining $\widetilde{\Theta}$ in (6) as the set of parameter values that imply negative effects in some of the patients. Similarly, by changing the definition of $\widetilde a_i(\theta)$ to indicate the worst available therapy, that is, $\widetilde a_i(\theta) = \argmin_{a \in\{0,1\}}E_p(Y_i|X_i,A_i = a)$, and modifying the definition of $\widetilde m_2(\mathcal D)$  in (7) to
 $$
 \widetilde m_2(\mathcal D) = \frac{1}{n}\int\sum_{i = 1}^n\left\{\mathbb E_{q_\theta}(Y_i|X_i, A_i = 0) - \mathbb E_{q_\theta}[Y_i|X_i, A_i = \widetilde a_i(\theta)]\right\}d\pi(\theta|\mathcal D, \mathcal D_E),
 $$
 the ED-PT can detect negative treatment effects. To illustrate an example, we consider the Modified S1
   scenario in Section 2.5. Figure \ref{fig:rej-neg} summarizes the rejection frequencies  after redefining the test statistics to detect negative effects. 
	To compute  power approximations  in Figure \ref{fig:rej-neg},  we use
	 $10\text{,000}$ simulations per scenario,   the significance level is    $\alpha = 0.05$, and the number of permutations used for testing  is
	  $J = 1\text{,}000$. In these simulations, the treatment effect of subgroup 1 remains null and the treatment effect of subgroup 2 ($\gamma_1$) varies from -1 to 1. We also include the ED-PT with test statistics $m$ for comparison. The results indicate similar power of 
   the ED-PT procedure with the modified versions of $\widetilde m_1$ and $\widetilde m_2$ that we defined
	to detect negative effects. Also, 
	in  the presence of negative treatment effects,
	the power of these two procedures  is comparable to 
	the power of the ED-PT with  statistic $m$.
 
	\section{Limiting power of ED-PT with fixed $n_1$ and $n_E\to\infty$ in Application}
	When $n_E\to\infty$, the posterior distribution of the model parameters $(\theta_{E,0},\theta_{E,x})$ concentrate on their respective asymptotic values. Denote these values by $\tilde\theta = (\tilde\theta_{E,0},\tilde\theta_{E,x})$. Since we assume in $\mathcal M$ that $\theta_0 = \theta_{0,E}$, $\theta_{x}^j = \theta_{E,x}^j$ for $j\in\{1,2,4,5\}$ and $\theta_x^j = \theta_{E,x}^j + \theta_B^j$ for $j\in\{3,6\}$, it follows that the expression of the conditional likelihood $m(\mathcal D)$ is
	$$
	m(\mathcal D) \propto \int_{\theta_B,\theta_a,\theta_I}\tilde p_i^{y_i}(1-\tilde p_i)^{1-y_i}\pi(\theta_B,\theta_a,\theta_I)d\theta_Bd\theta_ad\theta_I,
	$$
	where $\tilde p_i = \text{expit}[\tilde\theta_{E,0} + \theta_{E,x}^\intercal x_i + (\theta_{B},\theta_a,\theta_I)^\intercal (x_i^3,x_i^6,a_i,a_i\times x_i^{4:6})]$. To evaluate this conditional likelihood, we can apply a similar Laplace approximation as in (18) by introducing an offset term $\tilde\theta_{E,0} + \theta_{E,x}^\intercal x_i$ into the logistic regression and reducing the design matrix to that formed by $(x_i^3,x_i^6,a_i,a_i\times x_i^{4:6})$. We can then use this asymptotic test statistic $m(\mathcal D)$ to compute the statistical power of our procedure when $n$ is fixed and finite but $n_E\to\infty$. This provides an upper bound for the power for our ED-PT. Of note, since we generate the {\it in silico} IDs and EDs with bootstrap, $\tilde\theta$ can be computed exactly by fitting a logistic regression model without any terms involving treatment assignment to the entire dataset (AVAGLIO trial or DFCI EHRs) from which ED is sampled. The two modified versions of $m(\mathcal D)$ can be computed in a similar manner.

	\section{Supplementary Figures}

	\subsection{Large sample rejection regions of the ED-PT and Test-B for the example in Section 2.4}
	\begin{figure}[htp]
		\centering
		\includegraphics[width=0.8\textwidth]{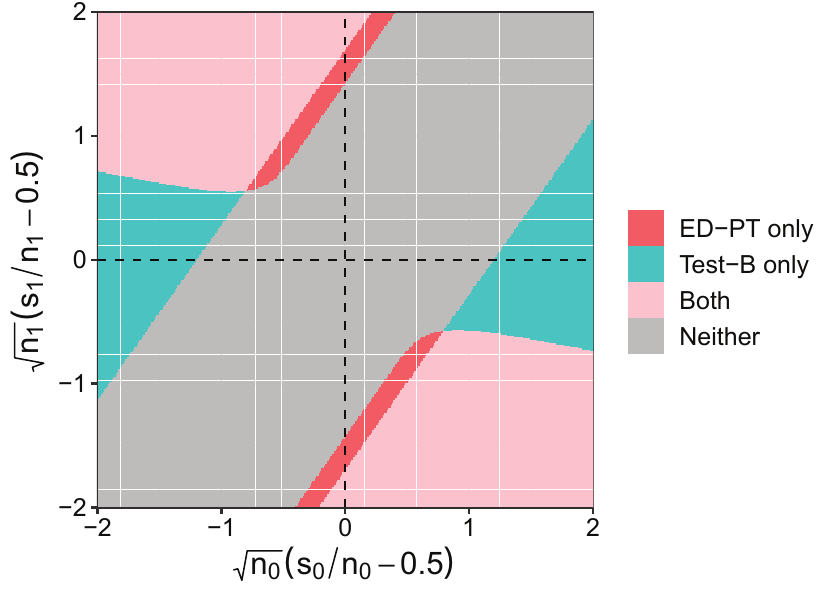}
		\caption{Rejection regions of the ED-PT and Test-B when $n_1 = 10\text{,}000$, $r = 1/2$, $r_E = 5$, $b = 0$ and $(n_E,s_E) = (50\text{,}000, 24\text{,}995)$.  The scenario with binary outcome has been described in Section 2.4.
		}
	\end{figure}

	\subsection{Rejection frequency of all testing procedures under Scenario 6}
	\begin{figure}[h]
		\centering
		\includegraphics[scale = 0.8]{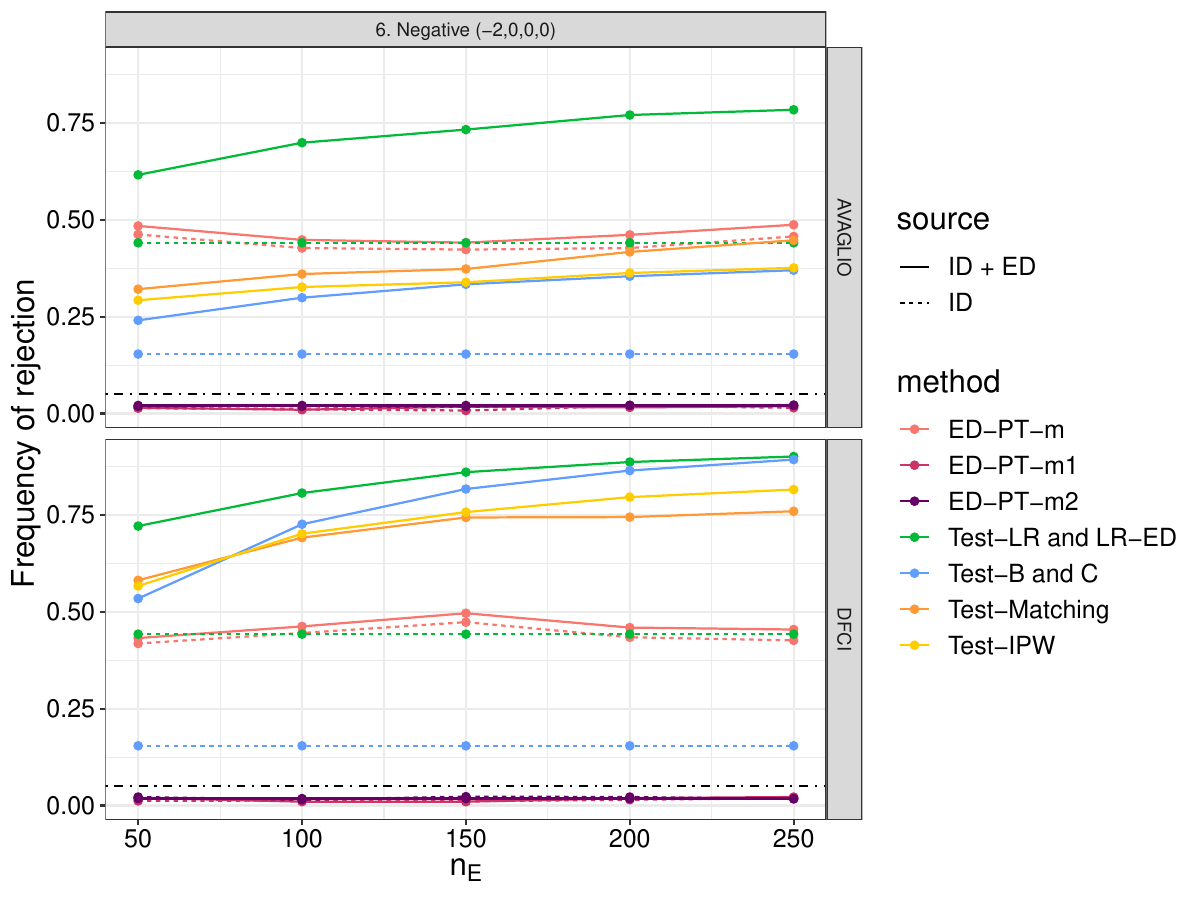}
		\caption{Frequency of rejection of all testing procedures in the scenario with negative treatment effect only (Scenario 6) when the ED is generated from the AVAGLIO trial and the DFCI EHRs. Note that the frequencies for ED-PT-$\widetilde m_1$ and ED-PT-$\widetilde m_2$ are nearly identical and the dots for their results overlap partially. The dash-dotted line indicates frequency of 0.05.}
		\label{fig:power_app_s6}
	\end{figure}
	
	\bibliography{99BIB}
	\bibliographystyle{chicago}